\documentclass[pra,twocolumn,showpacs,superscriptaddress,notitlepage,floatfix,nofootinbib]{revtex4-1}
\usepackage{makeidx}
\usepackage[utf8]{inputenc}
\usepackage{graphicx}
\usepackage{mathrsfs,amsmath,amssymb,amsfonts,amsthm,pifont}
\usepackage{graphicx}
\usepackage{braket}
\usepackage[subrefformat=parens,labelformat=parens,caption=false]{subfig}
\usepackage{amsfonts}
\usepackage{dcolumn}
\usepackage{bm}
\usepackage{bbm}
\usepackage{color}
\usepackage[dvipsnames]{xcolor}
\usepackage{esint}
\usepackage{lineno}
\usepackage{verbatim}
\usepackage{cancel}
\usepackage{multirow}
\usepackage{longtable}
\usepackage{exscale}
\usepackage[breaklinks,
            colorlinks,
            urlcolor=blue,
            linkcolor=blue,
            anchorcolor=blue,
            citecolor=blue]{hyperref}
           
\usepackage{soul}

\newcommand{\beq}{\begin{equation}} 
\newcommand{\eeq}{\end{equation}}
\newcommand{\bqa}{\begin{eqnarray}} 
\newcommand{\eqa}{\end{eqnarray}}
\newcommand{\nn}{\nonumber}

\newcommand{\dg}{^\dagger}
\newcommand{\rt}[1]{\sqrt{#1}\,}

\newcommand{\an}[1]{\langle{#1}\rangle}

\newcommand{\sch}{Schr\"odinger} 
\newcommand{\hei}{Heisenberg}

\newcommand{\vdp}{van~der~Pol}
\newcommand{\ray}{Rayleigh}

\newcommand{\ls}{Stuart--Landau}

\newcommand{\el}{Euler--Lagrange}

\newcommand{\fn}{Fitzhugh--Nagumo}
\newcommand{\lie}{Li\'{e}nard}

\newcommand{\ito}{It\^o} 
\newcommand{\str}{Stratonovich}

\newcommand{\Tr}{{\rm Tr}}

\newcommand{\tbo}[2]{\left(\begin{array}{c} {#1} \\ {#2} \end{array}\right)}

\newcommand{\ann}{\hat{a}}
\newcommand{\adg}{\hat{a}\dg}

\newcommand{\Hhat}{\hat{H}}

\newcommand{\Dcal}{{\cal D}}
\newcommand{\Lcal}{{\cal L}}

\newcommand{\wo}{\omega_0}

\newcommand{\xhat}{\hat{x}}
\newcommand{\yhat}{\hat{y}}

\newcommand{\Ncal}{{\cal N}}
\newcommand{\Scal}{{\cal S}}

\newcommand{\Kcal}{{\cal K}}

\newcommand{\alphastar}{\alpha^*}

\newcommand{\qhat}{\hat{q}}
\newcommand{\phat}{\hat{p}}

\newcommand{\supopdot}{\text{\Large $\cdot$}}

\newcommand{\normord}[1]{:\,\mathrel{#1}\,:}
\def\nLongleftarrow{/\mkern-25mu\Longleftarrow}

\newcommand{\Lin}{{\cal L}^{\in}}
\newcommand{\Lnot}{{\cal L}^{\notin}}

\newcommand{\fxpt}{\text{\tiny $\bullet$}}

\begin{document}

\title{Quantization of nonlinear non-Hamiltonian systems}

\author{Andy Chia}
\email{cqtcah@nus.edu.sg}
\affiliation{Centre for Quantum Technologies, National University of Singapore}

\author{Wai-Keong Mok}
\email{darielmok@caltech.edu}
\affiliation{Institute for Quantum Information and Matter, California Institute of Technology, Pasadena, California 91125, USA}

\author{Leong-Chuan Kwek}
\affiliation{Centre for Quantum Technologies, National University of Singapore}
\affiliation{National Institute of Education, Nanyang Technological University, Singapore}

\author{Changsuk Noh}
\email{cnoh@knu.ac.kr}
\affiliation{Department of Physics, Kyungpook National University, Daegu, South Korea}

\date{\today}

\begin{abstract}

Several important dynamical systems are in $\mathbb{R}^2$, defined by the pair of differential equations $(x',y')=(f(x,y),g(x,y))$. A question of fundamental importance is how such systems might behave quantum mechanically. In developing quantum theory, Dirac and others realized that classical Hamiltonian systems can be mapped to their quantum counterparts via canonical quantization. The resulting quantum dynamics is always physical, characterized by completely-positive and trace-preserving evolutions in the Schr\"{o}dinger picture. However, whether non-Hamiltonian systems can be quantized systematically while respecting the same physical requirements has remained a long-standing problem. Here we resolve this question when $f(x,y)$ and $g(x,y)$ are arbitrary polynomials. By leveraging open-systems theory, we prove constructively that every polynomial system admits a physical generator of time evolution in the form of a Lindbladian. We call our method cascade quantization, and demonstrate its power by analyzing several paradigmatic examples of nonlinear dynamics such as bifurcations, noise-activated spiking, and Li\'{e}nard systems. In effect, our method can quantize any classical system whose $f(x,y)$ and $g(x,y)$ are analytic with arbitrary precision. More importantly, cascade quantization is exact. This means restrictive system properties usually assumed in the literature to facilitate quantization, such as weak nonlinearity, rotational symmetry, or semiclassical dynamics, can all be dispensed with by cascade quantization. We also highlight the advantages of cascade quantization over existing proposals, by weighing it against examples from the variational paradigm using Lagrangians, as well as non-variational approaches.

\end{abstract}

\maketitle
\tableofcontents

\section{Introduction}
\label{Introduction}

Quantization, the process by which we ascribe a quantum-mechanical description to a given classical theory has enabled profound advances in physics \cite{GR01,AS10,Lou00,Mil19,GAF10,Dut05,Mah00,GY19,Sno20,VD17,BGGW21,Rip22}. An interesting and important problem is the quantization of general classical dynamical systems. That is, how to sensibly map classical equations of motion to a valid quantum-mechanical evolution.\footnote{This problem should not be confused with the assignment of an operator to a scalar function on classical phase space, also commonly called a quantization problem \cite{Car22}. An example of such an assignment from a function to an operator is the Weyl transform, or Weyl quantization \cite{AE05,Hal13}. Our problem on the other hand, is to map classical \emph{dynamics} to some analogous version in quantum theory.}

We consider here classical dynamical systems in $\mathbb{R}^2$, specified by 
\begin{align}
\label{GenDynSys}
	x' = f(x,y) \; ,  \quad   y' = g(x,y)   \; , 
\end{align}
where a prime denotes a time derivative, and we have assumed the system to be autonomous for simplicity. Equation \eqref{GenDynSys} is defined to be a Hamiltonian system when there exists a real-valued function $H(x,y)$ (called the Hamiltonian), such that $f(x,y)=\partial H/\partial y$ and $g(x,y)=-\partial H/\partial x$ \cite{Str15,JS07}. A necessary and sufficient condition for \eqref{GenDynSys} to be a Hamiltonian system is \cite{JS07}
\begin{align}
\label{HamiltonianCondition}
	\frac{\partial f}{\partial x} + \frac{\partial g}{\partial y} = 0      \; .
\end{align}
Condition \eqref{HamiltonianCondition} also defines a system whose phase-space area remains constant under \eqref{GenDynSys}. Such systems are also called conservative. Hence, Hamiltonian systems are equivalent to conservative systems. For us, the most important attribute of Hamiltonian systems is that they may be quantized effectively by first finding $H(x,y)$, and then turning it into a self-adjoint operator $\Hhat$.\footnote{The necessity and sufficiency of \eqref{HamiltonianCondition} can be proved by construction. That is, if we know the system to be Hamiltonian, then we also know how to find $H(x,y)$.} This approach is predicated on the structural similarities between quantum theory and the variational formulation of mechanics by the likes of Euler, Lagrange, and Hamilton \cite{Dir25,Dir58,Dir63}. As such, it came to be known as canonical quantization, and is now a staple in the armory of most physicists \cite{BD05}.

However, most often, dynamical systems are not Hamiltonian, which prevents canonical quantization from being applicable. If the system is also nonlinear, as is generally the case \cite{Str15}, quantization becomes highly nontrivial. In this work we tackle the quantization of systems that are both nonlinear and non-Hamiltonian. Despite the abundance of such systems, their quantization has hitherto been an open problem. It is not known how, or if it is even possible, to engineer analytically a quantum-mechanical generator of time evolution given an arbitrary nonlinear non-Hamiltonian system. Here we solve this problem for the system in \eqref{GenDynSys} with $f(x,y),g(x,y)\in\mathbb{P}_m$, where $\mathbb{P}_m$ is the set of all bivariate polynomials of degree $m$. Numerous important dynamical systems are of this form, such as those featuring limit cycles \cite{Jen13}, excitability \cite{LGONSG04}, and a range of bifurcations \cite{Str15,JS07}. It also includes chaotic systems when \eqref{GenDynSys} becomes non-autonomous, as in the case of driven nonlinear oscillators \cite{LM96}. In fact, our method allows any system whose $f(x,y)$ and $g(x,y)$ are analytic to be quantized with arbitrary precision by truncating their power-series representation at arbitrarily high orders (keeping in mind of course that some analytic functions may have a finite radius of convergence).

To define a quantum generator of time evolution corresponding to \eqref{GenDynSys}, we first turn $x$ and $y$ into operators $\xhat$ and $\yhat$ satisfying $[\xhat,\yhat]=i\hat{1}$, where we have set $\hbar=1$. Under any suitable choice of operator ordering between $\xhat$ and $\yhat$, the functions $f(x,y)$ and $g(x,y)$ can then be mapped to their quantum analogs, say $\bar{f}(\xhat,\yhat)$ and $\bar{g}(\xhat,\yhat)$.\footnote{We require $\bar{f}(\xhat,\yhat)$ and $\bar{g}(\xhat,\yhat)$ to have any \emph{suitable} operator ordering as opposed to simply \emph{any} ordering because they should be Hermitian. For example, if $f(x,y)=xy^2$, then $\bar{f}(\xhat,\yhat)=(\xhat \yhat^2 + \yhat^2 \xhat)/2$ and $\bar{f}(\xhat,\yhat)=(\xhat \yhat^2 + \yhat \xhat \yhat + \yhat^2 \xhat)/3$ are both suitable mappings of $f(x,y)$ (the latter being Weyl ordering \cite{AE05,Hal13}), but not $\bar{f}(\xhat,\yhat)=\xhat\yhat^2$.} We can then generalize \eqref{GenDynSys} to quantum mechanics by demanding its evolution be mimicked in expectation values (a generalized Ehrenfest theorem):
\begin{align}
\label{<GenDynSys>}
	\an{\xhat}' = \an{\bar{f}(\xhat,\yhat)}   \;,   \quad    \an{\yhat}' = \an{\bar{g}(\xhat,\yhat)}   \; ,
\end{align}
where angle brackets denote quantum-mechanical expectation values. Hence to quantize \eqref{GenDynSys} exactly, it is necessary that we find a generator of quantum dynamics consistent with \eqref{<GenDynSys>}. However, \eqref{<GenDynSys>} is not sufficient. Additional constraints on the quantum generator are required for it to be physically valid, and these are discussed briefly below and in detail in Sec.~\ref{LindbladianDefn}. Note also that $\bar{f}(\xhat,\yhat)$ and $\bar{g}(\xhat,\yhat)$ are operator polynomials of degree $m$ since they are defined in terms of $f(x,y)$ and $g(x,y)$. Our main result here is a systematic procedure for quantizing any polynomial system in the form of \eqref{GenDynSys} even when \eqref{HamiltonianCondition} is not satisfied.\footnote{Our method works for any sign of $\partial f/\partial x + \partial g/\partial y$. However, due to the omnipresence of damping forces in practice, the literature has focused mostly on dissipative systems, which are defined by $\partial f/\partial x + \partial g/\partial y<0$. Aside from designing realistic models, dissipative effects also play an essential role in stabilizing dynamical systems.}

The lack of a general quantization method in the literature by no means reflects a lack of effort \cite{Dek81,KMT97,UYG02,Bol04,YX05,Tar08,Wei12,Raz17}. Existing proposals for quantization may be broadly classified by whether a method uses variational principles or not. In essence, the variational approach attempts to restore the applicability of canonical quantization. Its roots can be traced back to a paper due to Bateman in 1931, in which he derived phenomenologically the position evolution of a damped harmonic oscillator from the \el\ equation \cite{Bat31}. The caveat Bateman noticed was that an extra fictitious oscillator had to be introduced. Though this means the underlying Lagrangian is for two oscillators, one can still derive a Hamiltonian from it, which in turn can be used for quantization. The variational approach has since become the most intensely studied quantization method for non-Hamiltonian systems (such as in Refs.~\cite{BFG11,Phi12,Gal13,SCVC15,Tak18a,Tak18b,PNC18,SAAJENR18,DF20,BT21,MH24,JVHHCA23} to give some relatively recent examples). A more thorough but still inexhaustive literature review of the variational paradigm is contained in Appendix~\ref{LiteratureReview}. The aim of this appendix is to help the reader better contextualize our main result.

In this work, we abandon the variational approach altogether. Instead, we adopt an open-systems approach by mapping \eqref{GenDynSys} directly to a quantum master equation \cite{BP02,RH12,Man20}. A master equation operates in the \sch\ picture of quantum dynamics, and have as its generator of time evolution, a linear superoperator $\Lcal$. It is common to abbreviate master equations as $\rho'=\Lcal\rho$ with $\rho$ being the state of the system (a density operator). For Hamiltonian systems, $\Lcal$ is simply determined by $\Lcal\rho=-i\,[\Hhat,\rho]$ where $\Hhat$ is the Hamiltonian operator. In general however, $\Lcal$ permits non-Hamiltonian processes. The question as to how $\Lcal$ should be parameterized in the more general case received an answer in the 1970s. It was shown that if the dynamics of $\rho$ is to be physically valid, defined formally by complete positivity and trace invariance, then $\Lcal$ must have a particular parameterization in terms of non-Hermitian operators and real-valued scalars known as the Lindblad form (or the Gorini-Kossakowski-Sudarshan--Lindblad form) \cite{GKS76,Lin76}.\footnote{See Sec.~\ref{LindbladianDefn} for a more precise statement of the Lindblad form.} We will call an $\Lcal$ in the Lindblad form a Lindbladian for short.

In principle, one can now quantize non-Hamiltonian systems by searching for an effective Lindbladian consistent with \eqref{<GenDynSys>}. This constitutes an inverse problem. However, such problems are generally difficult to solve. Essentially the same problem arises for Lagrangians in variational mechanics \cite{Dou41,Eng75,HU81,Hen82,Mus09,ST14}. In the case of Lindbladians, the problem is compounded by the noncommutative algebra of the operators that parameterize it, and by having $\Lcal$ depend nonlinearly on them. These issues make finding Lindbladians highly nontrivial in practice.

A map which takes \eqref{GenDynSys} directly to $\Lcal$ is extremely valuable as it dispenses with a first-principles approach to quantization.\footnote{Historically, quantum master equations grew out of the variational literature and were developed from first principles (see Appendix~\ref{LiteratureReview}). In this approach one starts with a Hamiltonian that prescribes how the system interacts with a bath (another system with an infinite number of degrees of freedom, but whose dynamics are not of interest) \cite{Sen60,Sen61}. Then by invoking the rotating-wave and Born--Markov approximations, a physically valid equation of motion for the system's state is derived. These steps now constitute the so-called ``microscopic'' derivation of master equations in open-systems theory \cite{BP02,RH12}.} In the vast majority of cases, the actual microscopic physics giving rise to a non-Hamiltonian process is either unattainable, or simply unnecessary.\footnote{The word microscopic here has a different meaning to its usage in the previous footnote, where we mentioned the ``microscopic'' derivation of master equations. Here, we are referring to the real inter-particle interactions that occur between the system and its surrounding. In contrast, the term microscopic in the previous footnote merely emphasizes the possibility of deriving a Markovian master equation from a system-bath Hamiltonian. Whether the system-bath Hamiltonian corresponds to anything real is generally not of concern. Thus, the ``microscopic'' derivation of master equations as we explained in the previous footnote can be understood as a formal procedure \cite{BP02,RH12}, and writing down a system-bath Hamiltonian such that it leads to a desired master equation is just as phenomenological as writing down the master equation directly.} This makes an effective quantization method especially suitable for non-Hamiltonian systems. The importance of this point is difficult to overstate as it has motivated much of the quantization literature on dissipative systems. To provide some context, and to drive this point home, we cite a few examples from the literature here. We find from Riewe \cite{Rie96}:
\begin{quote}
	\sl{The most realistic approach is to include the microscopic details of the dissipation directly in the Lagrangian or Hamiltonian [...] However, it is not intended to be a general method of introducing friction into classical Lagrangian mechanics. It can be complex in practice and does not allow the functional form of the frictional force to be chosen arbitrarily.}
\end{quote}
From Katz and Gossiaux, in motivating the so-called Schr\"{o}dinger--Langevin equation (abbreviated as SLE in the following quote) \cite{KG16}:
\begin{quote}
	 \sl{Thanks to its straightforward formulation---in principle only two `classical' parameters need to be known: the friction coefficient $A$ and the bath temperature $T_{\rm bath}$---and its numerical simplicity, the SLE can be considered as a solid candidate for effective description of open quantum systems hardly accessible to quantum master equations. Indeed, in a number of complex applications, defining the bath/interaction Hamiltonian and calculating the Lindblad operators without too many approximations is rather complicated, and some effective approaches---possibly of the Langevin type---are unavoidable.}
\end{quote}
And finally we note from Blacker and Tilbrook \cite{BT21}:
\begin{quote}
	 \sl{Another class of approaches involve coupling an undamped oscillator to a loss mechanism, including through a spin-boson model or a Lindbladian master equation formalism. While these approaches have proved successful, they require a model of how the dissipation occurs. However, such an approach is not often practical, for example when characterizing superconducting quantum circuits, which rely on experimentally determined estimations of the damping conditions.}
\end{quote}
An effective quantization method, especially one based on master equations, is also well suited to a number of fields in quantum science and technology such as optomechanics \cite{AKM14,BM15}, quantum optics \cite{Car93,Car02,Car10,DH14}, circuit quantum electrodynamics \cite{VD17,BGGW21,Rip22,DBD+16}, and chemical physics \cite{Nit06,MK11}. In addition, master equations play an important role in the theory of continuous quantum measurements and control \cite{DHJMT00,WM10,Jac14,CW11a,CW11b}. These disciplines already employ master equations, but in most cases, their applications have been restricted to processes with rotational symmetry, such as linear damping.

It should also be mentioned that our quantization method arose from previous works where we considered the \vdp\ oscillator \cite{CKN20} (and a variant of it \cite{SMN+23}). The classical \vdp\ model is a paradigm for relaxation oscillations that first appeared in the context of electrical circuits \cite{vdP20,vdP26}.\footnote{Relaxation oscillations are characterized by the occurrence of two widely separated timescales within a single cycle \cite{Str15}. See also Refs.~\cite{Car60,Isr04,GL12} for a historical account.} It is now a textbook example for nonlinear oscillations \cite{Str15,TS02,GPS02,JS07}, and not surprisingly, others have also sought to quantize the \vdp\ oscillator. One approach uses Bateman's idea \cite{SCVC15}, but fails to produce bounded phase-space trajectories. Other authors have instead turned to a Lindbladian, but only an approximate quantization was attempted \cite{ACL21}. To our knowledge, Ref.~\cite{CKN20} has been the only physically valid and exact quantization of the \vdp\ oscillator. We will come back to this point about exactness and the issues that make an exact quantization difficult in Secs.~\ref{P3Sys} and \ref{Comparison}.

Without further delay, we now outline the contents of our paper and how to navigate through it. A precise formulation of quantization as an inverse problem for Lindbladians is given in Sec.~\ref{ProbDefn}. No prior knowledge about master equations and the Lindblad form is assumed. Readers familiar with these may skip Sec.~\ref{LindbladianDefn}. This is all that is required to read our constructive proof that a Lindbladian $\Lcal$ always exists for any polynomial system, i.e.~a system in the form of \eqref{GenDynSys} for $f(x,y),g(x,y) \in \mathbb{P}_m$ (see Sec.~\ref{GenSoln} and Appendix~\ref{ConsProof}). We call our solution to the Lindbladian inverse problem cascade quantization. In practice, $\mathbb{P}_3$, i.e.~third-degree polynomials, already cover several systems of interest in the literature. Therefore we first provide an efficient means to quantize degree-three systems in Sec.~\ref{P3Sys} (see Fig.~\ref{QTable}) and highlight some desirable features of our quantization method. To illustrate its usage, we show how common bifurcations appear in quantum theory in Sec.~\ref{Bifurcations}. Examples include the saddle-node, transcritical, pitchfork, Hopf, and infinite-period bifurcations. Then in Sec.~\ref{ExcitableSys}, we consider stochastic systems where we quantize the \fn\ model driven by classical white noise. The stochastic quantum dynamics is seen to exhibit a resonance phenomenon akin to coherence resonance, but outside the so-called excitable regime of the quantum \fn\ model \cite{LGONSG04}. The \fn\ model is a toy model of a neuron---a prime example of an excitable system. Previous attempts to take such a system into the quantum realm has relied on a circuit realization for the neuron \cite{GRCE+19,GRSS20}. In contrast, we do not require such a blueprint for how the model can be realized in order to quantize it. We then consider the well-known \lie\ class of systems that lead to limit cycles in Sec.~\ref{LienardSys}. Cascade quantization enables us to parameterize an entire family of quantum limit cycles based on \lie's theorem. We then compare cascade quantization to the variational approach in Sec.~\ref{Comparison}. This section highlights how Lindbladians overcome the difficulties facing Lagrangians in quantizing non-Hamiltonian systems. Peripheral, but new results related to variational quantization have also been included in Appendix~\ref{VarAppNLSys} to make Sec.~\ref{Comparison} complete and self-contained. To further highlight the power of cascade quantization, we provide a critique of other non-variational approaches in Sec.~\ref{NonVariationalMethods}. Sections~\ref{Comparison} and \ref{NonVariationalMethods}, along with Appendix~\ref{LiteratureReview}, should help clarify the context of our main result. This finally brings us to the quantization of arbitrary polynomial systems. In Sec.~\ref{GenSoln} we sketch a strategy for solving the Lindbladian inverse problem for systems in $\mathbb{P}_m$ with any $m$. In essence, we show how $\mathbb{P}_{m+1}$ can be quantized provided that we know how to quantize $\mathbb{P}_m$. It is here that our choice of terminology---cascade quantization---starts to make sense. As the detailed derivation of cascade quantization is extremely tedious, we have included it in Appendix~\ref{ConsProof} so as not to disrupt the flow of our paper. Then in Sec.~\ref{SimpleCases} we illustrate how cascade quantization works explicitly for $\mathbb{P}_3$ before finally concluding in Sec.~\ref{Conclusion}, where we have suggested further applications and generalizations of cascade quantization.

\section{Inverse problem for Lindbladians}
\label{ProbDefn}

\subsection{Complex coordinates}
\label{Reparam}

We stated in \eqref{<GenDynSys>} what it means to quantize \eqref{GenDynSys} exactly in terms of the real variables $(x,y)$. In particular, we mapped these coordinates to the Hermitian operators $(\xhat, \yhat)$ which satisfy $[\xhat,\yhat]=i\,\hat{1}$. However, physicists often like to view quantum processes as quantal exchanges. It then becomes more natural to use the annihilation and creation operators $(\ann, \adg)$, which satisfy $[\ann,\adg]=\hat{1}$, and which are related to $(\xhat, \yhat)$ by 
\begin{align}
	\ann = \frac{1}{\rt{2}} \, (\xhat + i \yhat)   \; ,   \quad    \adg = \frac{1}{\rt{2}} \, (\xhat - i \yhat)   \; .
\end{align}
For this reason, it helps to rewrite \eqref{GenDynSys} in terms of the complex variables 
\begin{align}
	\alpha  \equiv  \frac{1}{\rt{2}} \, (x + i\,y)    \; ,  \quad    \alphastar  \equiv  \frac{1}{\rt{2}} \, (x - i\,y)   \; .
\end{align}
This gives
%
\begin{align}
\label{CSys(a,a*)}
	\alpha' = {}& h(\alpha,\alphastar)   \\
	    \equiv {}& \frac{f\big(x(\alpha,\alphastar),y(\alpha,\alphastar)\big)}{\rt{2}} + i \, \frac{g\big(x(\alpha,\alphastar\big),y\big(\alpha,\alphastar)\big)}{\rt{2}}  \, ,  \nn
\end{align}
where $x=(\alpha+\alphastar)/\rt{2}$ and $y=-i(\alpha-\alphastar)/\rt{2}$. We then define, without loss of generality, the quantum analog to \eqref{CSys(a,a*)} in terms of the normally-ordered expectation value (see also Sec.~\ref{OpOrderTimeDepSys} with regards to operator ordering),
\begin{align}
\label{QSys(a,a*)}
	\an{\ann}' = \langle \, : \! h(\ann,\adg) \! : \, \rangle  \; .
\end{align}
The colons on each side of $h(\ann,\adg)$ in \eqref{QSys(a,a*)} effect normal ordering. For a simple product of annihilation and creation operators, normal ordering symbolically rearranges all creation operators to stand on the left of all annihilation operators. Note that $h(\ann,\adg)$ is derived by simply letting $\alpha\longrightarrow\ann$ and $\alphastar\longrightarrow\adg$ in $h(\alpha,\alphastar)$. For example, if $h(\alpha,\alphastar)=\alpha^2\alphastar$, then $h(\ann,\adg)=\ann^2\adg$. Assuming $f(x,y), g(x,y) \in \mathbb{P}_m$, implies that $h(\alpha,\alphastar) \in \mathbb{P}_m$ (treating $\alpha$ and $\alphastar$ as independent variables). For a general polynomial, the process of normal ordering is defined term wise. For example, if $m=3$, 
\begin{align}
	h(\alpha,\alphastar) = {}& c_{0,0} + c_{1,0} \, \alpha + c_{1,1} \, \alphastar   \\
	                                      & + c_{2,0} \, \alpha^2 + c_{2,1} \, \alpha \, \alphastar + c_{2,2} \, \alphastar{}^2   \nn \\
	                                      & + c_{3,0} \, \alpha^3 +  c_{3,1} \, \alpha^2 \alphastar + c_{3,2} \, \alpha \, \alphastar{}^2 + c_{3,3} \, \alphastar{}^3   \; .  \nn
\end{align}
Normal ordering of $h(\ann,\adg)$ is then defined by\footnote{Normal ordering is an ambiguous operation unless it is specified explicitly. In particular, it is not a linear operation as can be seen by considering $\ann\adg$. By definition, $\normord{\ann\adg}\;=\adg\ann$. But since $\ann\adg=\hat{1}+\adg\ann$, we also have $\normord{\ann\adg}\;=\;\normord{\hat{1}+\adg\ann}$\,. If normal ordering is linear, and defining $\normord{\hat{1}}\;=\hat{1}$, we then get $\normord{\ann\adg}\;=\hat{1}+\adg\ann \ne \adg\ann$. Equation \eqref{NormOrdDefn} is therefore necessary for \eqref{QSys(a,a*)} to be well defined. We can also define $\normord{h(\ann,\adg)}$ by first writing every $\alphastar$ to the left of $\alpha$ in $h(\alpha,\alphastar)$, and then letting $\alpha\longrightarrow\ann$ and $\alphastar\longrightarrow\adg$.}
\begin{align}
\label{NormOrdDefn}
	\normord{h(\ann,\adg)} \; = {}& c_{0,0} + c_{1,0}  \normord{\ann}  + \; c_{1,1}  \normord{\adg} \nn \\
	                                                & + c_{2,0}  \normord{\ann^2} + \; c_{2,1}  \normord{\ann \, \adg} + \; c_{2,2} \, \normord{\adg{}^2}   \nn \\
	                                                & + c_{3,0}  \normord{\ann^3} +  \; c_{3,1}  \normord{\ann^2 \adg} \nn  \\
	                                                & + c_{3,2}  \normord{\ann \, \adg{}^2} + \; c_{3,3}  \normord{\adg{}^3}   \\
	                                          = {}& c_{0,0} + c_{1,0} \, \ann + c_{1,1} \, \adg  \nn \\
	                                                & + c_{2,0} \, \ann^2 + c_{2,1} \, \adg \ann + c_{2,2} \, \adg{}^2   \nn \\
	                                                & + c_{3,0} \, \ann^3 +  c_{3,1} \, \adg \ann^2 + c_{3,2} \, \adg{}^2 \, \ann + c_{3,3} \, \adg{}^3   \; .  \nn
\end{align}

Quantization then entails finding a physically valid generator of time evolution consistent with \eqref{QSys(a,a*)}. Since \eqref{QSys(a,a*)} is picture independent, and the density operator $\rho(t)$ contains all the information one can extract about the system, it makes sense to define what it means for quantum dynamics to be physical in the \sch\ picture. This leads us to the Lindbladian.

\subsection{Formal statement}
\label{LindbladianDefn}

Let us denote the generator of time evolution by a linear superoperator $\Lcal$. The state of the system then evolves according to $\rho'(t)=\Lcal\rho(t)$. We have assumed $\Lcal$ to be time independent but it can be made time dependent if necessary (e.g.~if the system is driven, or if some parameters acquire time dependence). It is actually more intuitive to define a physically valid quantum evolution by thinking directly in terms $\rho(t)$, as opposed to $\rho'(t)$. Hence we consider the map generated by $\Lcal$,
\begin{align}
\label{N=exp(Lt)}
	\Ncal(t) \equiv e^{\Lcal \, t}  \; , 
\end{align}
for which $\rho(t) = \Ncal(t) \rho(0)$. In mathematics, the set $\{\Ncal(t) \,|\, t \ge 0 \}$ is called a one-parameter semigroup, because it satisfies the semigroup property \cite{BP02}
\begin{align}
\label{SemigroupDefn}
	\Ncal(t) \, \Ncal(s) = {}& \Ncal(t+s)  \;,  \quad  \forall \; t, s \ge 0   \; .
\end{align}
We now define a valid quantization of \eqref{QSys(a,a*)} to be any linear superoperator $\Lcal$ such that, for any $t \ge 0$,
\begin{itemize}
	\item[(i)] $\an{\ann}'_t = \Tr\big[ \ann \, \Lcal \, \rho(t) \big] = \an{\,\normord{h(\ann,\adg)}\,}_t$ \,,
	\item[(ii)] $\Ncal(t) \, \rho(0) \in \mathbb{V}(\mathbb{H})$\,,  $\,\forall \, \rho(0) \in \mathbb{V}(\mathbb{H})$ \,,
	\item[(iii)] $\Ncal(t)$ is completely positive \,,
\end{itemize}
where $\mathbb{V}(\mathbb{H})$ denotes the set of density operators on the system Hilbert space $\mathbb{H}$ that are Hermitian, positive, and trace one (i.e.~the set of valid states for the system). Condition (i) simply expresses \eqref{QSys(a,a*)} in the \sch\ picture. Condition (ii) demands that $\Lcal$ is such that $\Ncal(t)$ maps a valid initial state at $t=0$ to a valid state at $t>0$. It turns out that (ii) is insufficient for $\Ncal(t)$ to produce valid states if the quantized system appears as part of a larger system, as will be the case here if we are to account for non-Hamiltonian dynamics. Condition (iii) then ensures that $\Ncal(t)$ maps valid states to valid states for the larger system as well. For this larger system we define whatever is not part of the system to be the ancilla, and denote its Hilbert space as $\mathbb{H}_{\rm A}$. The map $\Ncal(t)$ is then completely positive if and only if $\Ncal(t) \otimes {\bf 1}_{\rm A}$ maps positive density operators to positive density operators on $\mathbb{H}\otimes\mathbb{H}_{\rm A}$, where ${\bf 1}_{\rm A}$ is the identity superoperator for the ancilla \cite{BP02}. Note that the conditions of positivity and Hermiticity in (ii) are implied by (iii), but not trace invariance \cite{BP02}. It is therefore conventional to call a map satisfying both (ii) and (iii) completely positive and trace preserving.

While the quantization problem defined by conditions (i)--(iii) is precise, it only constrains the form of $\Lcal$ implicitly. To make the problem more approachable, we leverage a powerful theorem from open-systems theory due to Lindblad \cite{Lin76}, Gorini, Kossakowski, and Sudarshan \cite{GKS76}. It states that \eqref{N=exp(Lt)} forms a semigroup of completely-positive trace-preserving maps if and only if $\Lcal$ has the form,
\begin{align}
\label{Lindbladian}
	\Lcal = - i \, [\Hhat, \supopdot\,] + \sum_{s=1}^N  \, \eta_s \, \Dcal[\hat{L}_s]   \; ,  
\end{align}
for some Hamiltonian operator $\Hhat$, $N$ arbitrary operators $\hat{L}_1,\hat{L}_2,\ldots,\hat{L}_N$, and an associated set of non-negative constants $\eta_1, \eta_2, \ldots, \eta_N$. We have defined in \eqref{Lindbladian} the dissipator (another superoperator parameterized by $\hat{L}$),
\begin{align}
\label{Dissipator}
	\Dcal[\hat{L}\,] \equiv \hat{L} \, \supopdot \, \hat{L}\dg - \frac{1}{2} \, \hat{L}\dg \, \hat{L}  \, \supopdot - \frac{1}{2} \, \supopdot \, \hat{L}\dg \, \hat{L}  \; .  
\end{align}
Note that we are now using a dot to denote where an input operator appears when it is acted on by a superoperator. For example, if $\hat{A}$ and $\hat{B}$ are any two operators, and if $\Scal = \hat{A}\,\supopdot\,\hat{A}\dg$, then $\Scal\hat{B}=\hat{A}\hat{B}\hat{A}\dg$. As advertised in Sec.~\ref{Introduction}, we call a generator in the form of \eqref{Lindbladian} a Lindbladian. We will also refer to $\hat{L}_1,\hat{L}_2,\ldots,\hat{L}_N$ and $\eta_1,\eta_2,\ldots,\eta_N$ respectively as Lindblad operators and coefficients. Note that \eqref{Lindbladian} includes, as its first term, the \sch\ equation. Dissipators therefore represent non-Hamiltonian processes. They tend to introduce noise into a quantum system, the effect of which broadens the system's quasiprobability distribution in quantum phase space. Dissipators can also be difficult to interpret, except for sufficiently simple systems, such as those described by $\Dcal[\ann]$ and $\Dcal[\adg]$ (the damped harmonic oscillator and linear amplifier respectively \cite{Aga12}). However, if the Lindblad operator is nonlinear in $\ann$ and $\adg$, even a seemingly simple dissipator can be nontrivial to interpret. One such example is $\Dcal[\adg{}^2]$. Although superficially it can be understood as generalizing one-photon gain to two photons, the noise introduced by $\Dcal[\adg{}^2]$ was shown to be genuinely quantum \cite{Lam67,CHN+20,CMNK23}.

We have now defined an inverse problem for Lindbladians by \eqref{QSys(a,a*)} and \eqref{Lindbladian}. It is not obvious whether a valid quantization exists for a classical system defined by $h(\alpha,\alphastar) \in \mathbb{P}_m$ for any $m>0$. Furthermore, knowing the existence of a valid quantization does not imply knowledge of its construction. Our problem is analogous to the inverse problem for Lagrangians in classical mechanics which asks if there exists a Lagrangian whose \el\ equation returns,
\begin{align}
\label{GenSecOrdEqn}
	x'' = q(x,x')  \; ,
\end{align}
where $q(x,x')$ is given. The answer to this quesion is given by the Helmholtz conditions, which provide the necessary and sufficient criteria for such a Lagrangian to exist \cite{Dou41,Eng75,Mus09,Lop99}. However, to construct the appropriate Lagrangian from $q(x,x')$ is another question, one that is at least equally important if not more (see e.g.~Refs.~\cite{Yan72,HL02,NL07,MRS08,CN10,PM23} and the papers therein). In Sec.~\ref{GenSoln} we prove that the inverse problem for Lindbladians has a solution for all $h(\alpha,\alphastar) \in \mathbb{P}_m$, and every $m>0$. We do so constructively, using a strategy that we call cascade quantization (derived in detail in Appendix~\ref{ConsProof}).

\section{Solution to the inverse problem for Lindbladians}
\label{P3Sys}

\subsection{Efficient cascade quantization}
\label{EfficientQuantization}

\begin{figure*}
\centering
\includegraphics[width=0.91\textwidth]{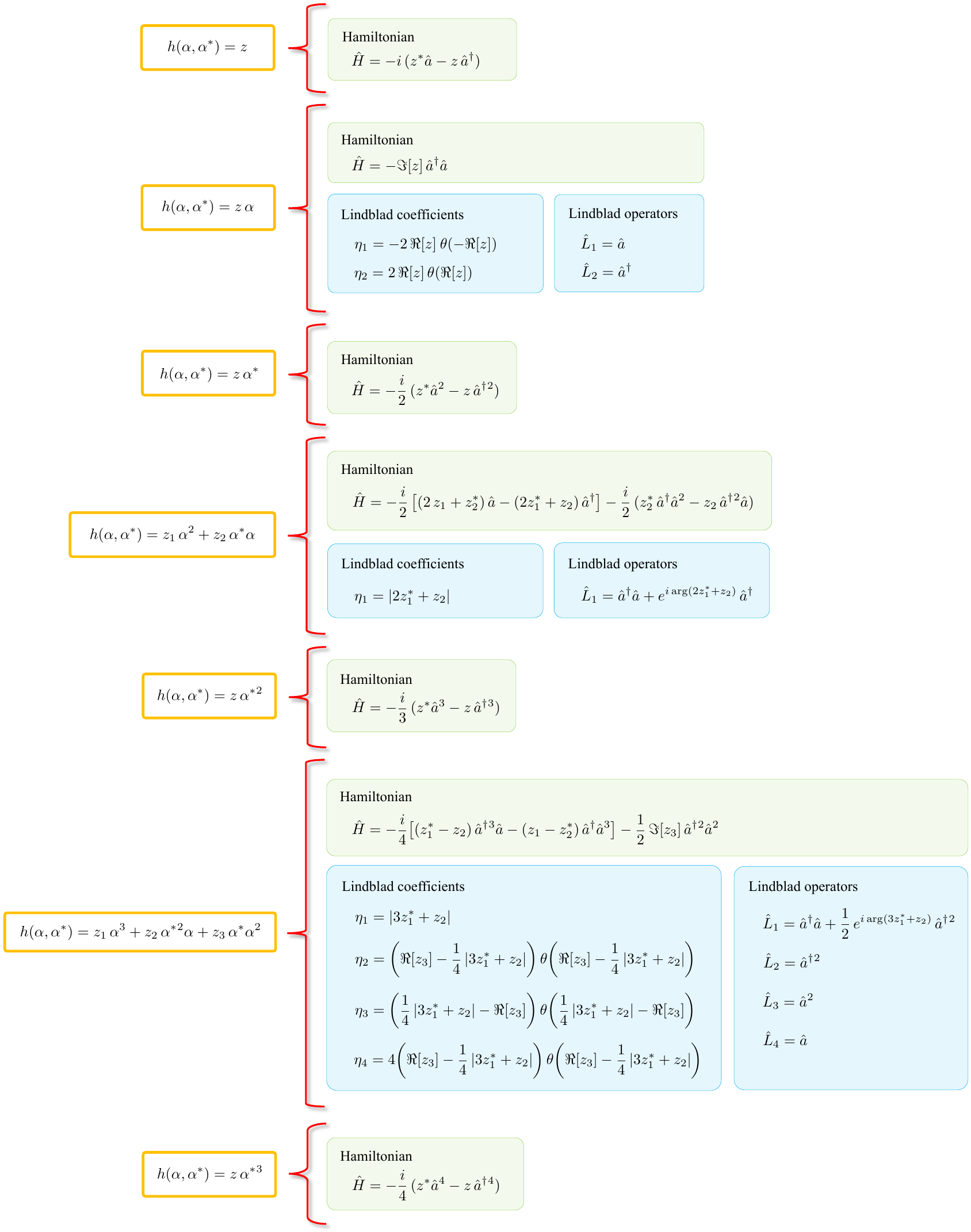}
 \caption{\label{QTable} Efficient cascade quantization for $\alpha'=h(\alpha,\alphastar) \in \mathbb{P}_3$ (polynomials of degree three). The full solution for $\mathbb{P}_m$ (polynomials of degree $m$) with arbitrary $m$ is deferred to Sec.~\ref{GenSoln}, with details provided in Appendix~\ref{ConsProof}. Arbitrary complex coefficients are denoted by the letter $z$ (differentiated with subscripts when more than one such coefficient appears). We have also used the step function defined by $\theta(x)=0$ for $x\le0$, and $\theta(x)=1$ for $x>0$. For each $h(\alpha,\alpha^*)$, the corresponding Hamiltonian and Lindblad terms are indicated respectively in the green and blue boxes on the right. By linearity of the Lindbladian, the contributions for each monomial of $h(\alpha,\alpha^*) \in \mathbb{P}_m$ are simply summed together to obtain the full Lindbladian.}
\end{figure*}

We now give an explicit solution to the inverse problem for Lindbladians that is simple to use. The solution is presented in the form of a table in Fig.~\ref{QTable}. 
It quantizes all systems defined by cubic polynomials, which includes several important models. In Fig.~\ref{QTable} we have written $h(\alpha,\alphastar)$, which defines the classical system to be quantized, inside yellow boxes on the left. The corresponding Lindbladian (indicated by a red brace), is then given in terms of its Hamiltonian, and if necessary, the associated Lindblad coefficients and operators according to \eqref{Lindbladian}. Here, and for the remainder of our paper, the real and imaginary parts of a complex number $z$ are written as $\Re[z]$ and $\Im[z]$ respectively, while its polar angle is denoted by $\arg(z)$. That is,
\begin{align}
\label{ReImParts}
	z = \Re[z] + i \, \Im[z] = |z| \, e^{i \arg(z)}  \; .
\end{align}
We have also used the Heaviside step function defined by
\begin{align}
\label{HeavisideDefn}
	\theta(x) = \begin{cases} 0 , \; x \le 0   \\
	                                         1 , \; x > 0 \end{cases} . 
\end{align}
To use our quantization table, one just needs to sum the various parts of the Lindbladian appropriately. That is, if $\Lcal_{\rm A}$ quantizes $\alpha'=h_{\rm A}(\alpha,\alphastar)$, and if $\Lcal_{\rm B}$ quantizes $\alpha'=h_{\rm B}(\alpha,\alphastar)$, then $\Lcal=\Lcal_{\rm A}+\Lcal_{\rm B}$ quantizes $\alpha' = h_{\rm A}(\alpha,\alphastar)+h_{\rm B}(\alpha,\alphastar)$. This is simply because the Lindbladian is a linear superoperator. For example, to quantize $\alpha'=\lambda_1\,\alpha+\lambda_2\,\alpha^2$ for any $\lambda_1,\lambda_2\in\mathbb{C}$, we simply look at our table for $h(\alpha,\alphastar)=z\,\alpha$ (with $z = \lambda_1$), and $h(\alpha,\alphastar)=z_1\,\alpha^2+z_2\,\alphastar\alpha$ (with $z_1 = \lambda_2$, and $z_2 = 0$). Then by inspection, the appropriate Lindbladian is 
\begin{align}
\label{QTableExL}
	\Lcal = {}& - i \, [\Hhat, \supopdot \, ]  + 2 \, |\lambda_2| \, \Dcal\big[ \adg\ann+ e^{i \arg(2\lambda_2^*)}\adg \big]    \\
	              & - 2 \, \Re[\lambda_1] \, \theta(-\Re[\lambda_1]) \, \Dcal\big[ \ann \big] + 2 \, \Re[\lambda_1] \, \theta(\Re[\lambda_1]) \, \Dcal\big[ \adg \big] \; ,  \nn
\end{align}
where we have noted that $|\lambda_2^*|=|\lambda_2|$, and 
\begin{align}
	\Hhat = -\Im[\lambda_1] \, \adg\ann -i \, (\lambda_2 \, \ann - \lambda_2^* \, \adg)  \; .
\end{align}
Note the step function ensures that only one of the dissipators in the second line of \eqref{QTableExL} will survive for a given $z_1$.

We have provided the quantization table without proof as one may simply verify that our prescribed $\Lcal$ satisfies condition (i) for the given $h(\alpha,\alphastar)$. What is unclear is why we have broken up the quantization of a cubic polynomial as shown in Fig.~\ref{QTable}. The complete answer requires Secs.~\ref{GenSoln} and \ref{SimpleCases}, so here we provide only a brief explanation: The solution to the Lindbladian inverse problem, like that of the Lagrangian inverse problem, is not unique.\footnote{It is well known in classical mechanics that for a single degree of freedom $x$, the \el\ equation $\partial L/\partial x = (\partial L/\partial x')'$ does not change if $L(x,x',t)\longrightarrow \lambda L(x,x',t)$ for $\lambda$ a constant, or, $L(x,x',t)\longrightarrow L(x,x',t)+dF/dt$ for $F$ an arbitrary function of $x$ and $t$, but not $x'$.} Therefore our table in Fig.~\ref{QTable} is not the only way to quantize degree-three systems. But this also does not mean that one should regard all valid quantizations of a classical system as being equal. We mentioned underneath \eqref{Dissipator} that dissipators in the Lindbladian introduce noise. Such noise in quantum systems tend to wash out their quantumness by the process of decoherence \cite{Sch07,Sch19}. So while dissipators are a necessary element of quantization, it is preferable to keep them to a minimum. The quantization table we have presented here tries to do this---introduce as few dissipators as possible for degree-three systems. We arrived at the table in Fig.~\ref{QTable} by fine-tuning the $m=3$ case in Sec.~\ref{SimpleCases}. This is the sense in which Fig.~\ref{QTable} is an efficient version of cascade quantization.

\subsection{Generalizations and advantages}
\label{Gen+Adv}

The most significant aspect of our quantization method is its exactness. This is so even when we compare our method to other proposals taking an open-systems approach. The literature typically makes one or more approximations to make quantization possible. One such example is to consider the semiclassical limit \cite{ACL21} (or assumptions to the same effect \cite{BGR21}), whereby $\an{h(\ann,\adg)}$ is replaced by $h(\an{\ann},\an{\adg})$. Another example is taking a perturbative limit, such as assuming a vanishing nonlinearity \cite{BGR21,LS13,WNB14}, or vanishing quantumness \cite{HdMPGGZ14}. Our method does not resort to any such approximations, taking only the classical model $(x',y')=\big(f(x,y),g(x,y)\big)$ as input. Below we highlight some difficulties and generalizations that can be handled by our quantization method.

\subsubsection{Arbitrary operator orderings and time-dependent systems}
\label{OpOrderTimeDepSys}

Although we have defined the quantum analog of $h(\alpha,\alphastar)$ by a normally-ordered $h(\ann,\adg)$, this does not restrict the applicability of our method in any way. Given an $h(\ann,\adg)$ for which $\ann$ and $\adg$ are ordered arbitrarily, say using anti-normal or symmetric ordering, we can always use $[\ann,\adg]=\hat{1}$ to rewrite $h(\ann,\adg)$ in normal order. Thus, if we can quantize $h(\alpha,\alphastar)$ corresponding to normal order, i.e.~find an $\Lcal$ such that $\an{\ann}'=\an{\,\normord{h(\ann,\adg)}\,}$, then we are able to quantize $h(\alpha,\alphastar)$ corresponding to any order of $\ann$ and $\adg$. Hence, no loss of generality has incurred by imposing normal ordering in \eqref{QSys(a,a*)}.

We have so far assumed the Lindbladian to be time independent. This means that only classical systems which are time independent can be quantized. However, the theorem leading to \eqref{Lindbladian} can actually be generalized to allow for time-dependent systems. In this case \eqref{N=exp(Lt)} is generalized to the time-ordered exponential,
\begin{align}
	\Ncal(t,t_0) = {\rm T}_{\triangleleft} \bigg\{ \exp\!\bigg[ \int_{t_0}^t d\tau \, \Lcal(\tau) \bigg] \bigg\}  \; ,
\end{align}
where ${\rm T}_{\triangleleft}$ is defined to be a linear operation such that for any two time-dependent superoperators ${\cal A}$ and ${\cal B}$,
\begin{align}
	{\rm T}_{\triangleleft}\big\{ {\cal A}(t_1) \, {\cal B}(t_2) \big\} 
	= \begin{cases} 
			{\cal A}(t_1) \, {\cal B}(t_2) \,, \; t_1 > t_2 \;, \\
	   		{\cal B}(t_2) \, {\cal A}(t_1) \,, \; t_2 > t_1  \; .
	  \end{cases}
\end{align}
The semigroup property in \eqref{SemigroupDefn} is then replaced by the divisibility condition, defined for any $t>t_1>t_0$, by
\begin{align}
\label{DivisibilityDefn}
	\Ncal(t,t_1) \, \Ncal(t_1,t_0) = \Ncal(t,t_0)    \; .
\end{align}
One may then show that an $\Lcal(t)$ is the generator of a divisible set of completely-positive and trace-preserving maps if and only if \cite{RH12,Bre12},
\begin{align}
\label{TimeDepLindbladian}
	\Lcal(t) = - i \, [\Hhat(t), \supopdot\,] + \sum_{s=1}^N  \, \eta_s(t) \, \Dcal[\hat{L}_s(t)]   \; .  
\end{align}
This has the same structure as \eqref{Lindbladian} and the same conditions are placed on the parameterizing operators and coefficients as before, but now for all times. In particular, every Lindblad coefficient must now be non-negative for all times, i.e.~$\eta_s(t) \ge 0$ for all $t \ge t_0$ and $s=1,2,\ldots,N$.

The generalization to time-dependent systems allows us to probe how quantum nonlinear systems respond to external forcing. Most often, the external force is given by a simple sinusoidal function. However, there is nothing that forbids one to consider an external drive that is stochastic. This means that our quantization method also applies to nonlinear models driven by noise. We will explore such a case in Sec.~\ref{ExcitableSys} by quantizing a classical nonlinear system with white noise. However, in doing so, the master equation for the density operator becomes a stochastic differential equation, and the question of how one should interpret such an equation arises (\ito\ versus \str) \cite{GZ04,Eva13,SS19}. Since we have derived our quantization method by assuming normal calculus, the resulting evolution with white noise must be interpreted as a \str\ equation. Or stated differently, our quantization rules are derived assuming that $\Lcal\,dt$ is of order $dt$, which can only be consistent with a \str\ interpretation.

\subsubsection{First-order systems}
\label{1stOrderSys}

A significant drawback of the variational approach to quantization is that it relies on Hamiltonians derived from Lagrangians (see Appendix~\ref{LiteratureReview}). This is problematic because a Lagrangian, through the \el\ equation, is designed to produce \eqref{GenSecOrdEqn}, not \eqref{GenDynSys}. Lindbladians on the other hand, do not have this defect.

At the core of this problem is the inequivalence between a system defined by a second-order differential equation, and a system defined by a pair of first-order differential equations: Given $x''=q(x,x')$, it is always possible to find an $f(x,y)$ and $g(x,y)$ so as to express the second-order equation as a pair of first-order ones. We denote this property as
\begin{align}
\label{2ndOrder=>1stOrder}
	x'' = q(x,x')    \; \implies \;   (x',y') = \big( f(x,y),g(x,y) \big)   \; .
\end{align}
All one has to do is let $f(x,y)=y$ and $g(x,y)=q(x,y)$. However, the converse is not true: 
\begin{align}
\label{2ndOrder<=/1stOrder}
	x'' = q(x,x')    \; \, \quad  \nLongleftarrow  \; \;   (x',y') = \big( f(x,y),g(x,y) \big)    \; .
\end{align}
That is to say, not every first-order system of equations can be written in closed form as $x''=q(x,x')$, or as $y''=p(y,y')$ for some $p(y,y')$. Furthermore, a given second-order equation may correspond to more than one set of first-order equations. Namely, it is possible to find two systems, say $(x',y')=\big(f(x,y),g(x,y)\big)$ and $(x',y')=\big(r(x,y),s(x,y)\big)$, both satisfying $x''=q(x,x')$, but for which 
\begin{align}
\label{Ineq1stOrderSets}
	\big( r(x,y),s(x,y) \big) \ne \big( f(x,y),g(x,y) \big)  \; .
\end{align}
We can illustrate this point with a simple example. Consider the system 
\begin{align}
\label{ExSys1}
	x' = f(x,y) = y \;,  \quad   y' = g(x,y) = - x  \; .
\end{align}
This is simply a linear oscillator rotating in phase space in the clockwise direction with constant angular speed. Similarly, a linear oscillator rotating in the counterclockwise direction is given by
\begin{align}
\label{ExSys2}
	x' = r(x,y) = -y  \;,  \quad  y' = s(x,y) = x  \; . 
\end{align}
It is trivial to see that both \eqref{ExSys1} and \eqref{ExSys2} satisfies $x''=-x$, but clearly $f(x,y) \ne r(x,y)$ and $g(x,y) \ne s(x,y)$.

We have thus shown that a dynamical system can only be defined unambiguously by a set of first-order differential equations. This gives Lindbladians an advantage over Lagrangians for quantization. In Sec.~\ref{Comparison} we give two concrete examples from the literature that are problematic for Lagrangians, but not for the Lindbladian. These are the \vdp\ oscillator \cite{SCVC15}, and the so-called unusual Li\'{e}nard-type oscillator \cite{CSL05,CRSL12}. The essential point we wish to make here is that the Lagrangian approach to quantization is flawed at the outset.\footnote{One possible reason for using a second-order differential equation to define a system is Newton's second law. But in nonlinear models, $x$ and $y$ often represent non-mechanical properties and have no direct relation to Newton's second law. For example, nonlinear oscillator models are commonly realised using electrical circuits. In this case $x$ and $y$ obey the familiar circuit laws for voltages and currents. Another example is chemical reactions. In this case one derives the equations of motion for $x$ and $y$ using the principles of chemical kinetics such as the law of mass action. The resulting model then describes how the concentration of reactant molecules evolve. Even more generally, nonlinear models are sometimes introduced phenomenologically without appealing to any physical principles.}

\subsubsection{Rotationally-asymmetric systems}
\label{RotationalSymmetry}

Given the Lindblad form to be extremely well known, one might wonder why the inverse problem for Lindbladians had not been solved earlier. Although \eqref{Lindbladian} significantly reduces the search space of physical generators of time evolution, finding $\Hhat$, $\hat{L}_1,\hat{L}_2,\ldots,\hat{L}_N$, and $\eta_1,\eta_2,\ldots,\eta_N$ is still difficult. This task is made even more nontrivial by the fact that $\Dcal[\hat{L}]$ depends nonlinearly on $\hat{L}$, and the fact that $\hat{L}$ acts on the system state from both the left and right.

There is, however, a special class of systems whose corresponding Lindbladian may be guessed with relative ease. These are systems with rotational symmetry. By this we mean a system $\alpha'=h(\alpha,\alphastar)$ for which $h(\alpha,\alphastar)$ satisfies \cite{GMK12}
\begin{align}
\label{RotSymProc}
	h(\alpha \, e^{-i\theta}, \alphastar \, e^{i\theta}) = h(\alpha,\alphastar) \, e^{-i\theta}     \; , 
\end{align}
where $\theta \in \mathbb{R}$ is a constant. Such processes can be quantized by dissipators whose Lindblad operator is a function $w(\ann,\adg)$ such that
\begin{align}
\label{QRotSymProc}
	\Dcal\big[ w(\ann e^{-i\theta},\adg e^{i\theta}) \big] = \Dcal[w(\ann,\adg)]    \; . 
\end{align}
Examples of $w(\ann,\adg)$ that satisfy \eqref{QRotSymProc} are $\ann{}^n$ and $\adg{}^n$ for all $n>0$. Previous literature using dissipators to quantize non-Hamiltonian terms in $h(\alpha,\alphastar)$ has mostly considered rotationally-symmetric processes due to their simplicity.

It helps to see an example of quantization that highlights the difference between systems with rotational symmetry and systems that are rotationally asymmetric. This also puts cascade quantization in the context of ongoing research on quantum nonlinear dynamics. We begin with the simplest rotationally-symmetric processes, given by linear dissipation and amplification with rate $\kappa>0$. These are described by $h(\alpha,\alphastar)=\mp\,\kappa \,\alpha$, where the negative sign applies for dissipation, and positive sign for amplification. Rotational symmetry means the system is damped or amplified at equal rates in all directions in phase space. We find, along $x$ and $y$,
\begin{align}
\label{LinProc1}
	x' = \mp \, \kappa \, x   \; ,     \quad     y' = \mp \, \kappa \, y  \; .
\end{align}
Linear dissipation and amplification can be quantized respectively by $2\kappa\,\Dcal[\ann]$ and $2\kappa\,\Dcal[\adg]$. This example provides another illustration of the inequivalence in defining a system by first- and second-order differential equations. We note that \eqref{LinProc1} gives the familiar second-order equation for a linearly-damped system 
\begin{align}
\label{LinDamp}
	x''  = \mp \kappa \, x'    \; .
\end{align}
This same second-order equation can also be derived from the rotationally-asymmetric process $h(\alpha,\alphastar)=(-i\mp\kappa)(\alpha-\alphastar)/2$, which is the same as,
\begin{align}
\label{LinProc2}
	x' = y  \; ,    \quad     y' =  \mp \kappa \, y   \; .
\end{align}
The difference between \eqref{LinProc1} and \eqref{LinProc2} has been noted in the context of quantum Brownian motion \cite{KMT97,Gao98a}, which is a linearly-damped system driven by white noise. We will meet noise-driven systems in Sec.~\ref{ExcitableSys}.

To give at least one example of a rotationally-symmetric process that is nonlinear, we consider here $h(\alpha,\alphastar)=-\,\gamma\,\alphastar\alpha^2$ with $\gamma>0$. This generalizes \eqref{LinProc1} with negative coefficients to radially-dependent dissipation, which can also be seen from 
\begin{align}
\label{RotSymNonlinLoss}
	x' = - \frac{\gamma}{2} \, (x^2 + y^2) \, x  \; ,    \quad    y' = - \frac{\gamma}{2} \, (x^2 + y^2) \, y   \; .
\end{align}
It is straightforward to see that such nonlinear dissipation is quantized by $\gamma\,\Dcal[\ann^2]$. We can in fact combine \eqref{LinProc1} and \eqref{RotSymNonlinLoss} with a simple harmonic oscillator to obtain a rotationally-symmetric system with a limit cycle, known as the \ls\ oscillator \cite{Kur03}:
\begin{align}
\label{a'SLOsc}
	\alpha' = -i \, \alpha + \kappa \, \alpha - \gamma \, \alphastar \alpha^2  \; .
\end{align}
Note that we have assumed the free oscillator to have unit angular frequency, and the strength of linear gain to be $\kappa>0$. Equation \eqref{a'SLOsc} is equivalent to 
\begin{align}
\label{x'SLOsc}
	x' = {}& y + \kappa \, x - \frac{\gamma}{2} \, (x^2 + y^2) \, x   \; ,    \\
\label{y'SLOsc}
	y' = {}& - x + \kappa \, y - \frac{\gamma}{2} \, (x^2 + y^2) \, y    \; .
\end{align}
The rotational symmetry of the \ls\ oscillator is manifested in its circularly-shaped limit cycle.\footnote{One might be tempted to think that a circular limit cycle implies the existence of rotational symmetry in the system, as in the case of the \ls\ oscillator. However, this is not the case. An example of a system with a circular limit cycle that is not rotationally symmetric is $x'=y$ and $y'=-x + \lambda \, y - \lambda \, (x^2 + y^2) \, y$ for $\lambda$ a real positive constant \cite{Str15}. We will in fact quantize an example of a circular but rotationally-asymmetric limit cycle in Sec.~\ref{InfPerBif}. We will then see how such quantum limit cycles are represented in quantum phase space by Wigner functions.} As an illustration of \eqref{2ndOrder<=/1stOrder}, we note that \eqref{x'SLOsc} and \eqref{y'SLOsc} do not decouple to give a single second-order differential equation in $x$ or $y$. Since the harmonic oscillator is quantized by a Hamiltonian, the above knowledge allows us to quantize the \ls\ oscillator exactly by a Lindbladian. Due to its rotational symmetry, the quantum \ls\ oscillator has been the basis of several results motivated by classical nonlinear dynamics \cite{LS13,WNB14,WNB15,LCW14,MH15,LANB16,IK17,SHM+18,AKLB18,BKBB20,KN21,BKB21,KN22,BB22,BB23}. New theoretical work on quantum synchronization has also considered rotationally symmetric systems with multiple attractors \cite{JLA20,KBS25}, while a new model of the quantum \ls\ oscillator based on spins has been proposed \cite{KN24}. On the experimental front, the original quantum \ls\ oscillator based on \eqref{a'SLOsc} was realized just very recently \cite{LXY+25}, also demonstrating the effects of squeezing \cite{SHM+18}, and linear dissipation \cite{MKH20} on quantum synchronization.

We now compare \eqref{a'SLOsc} to a limit cycle without rotational symmetry. Perhaps the most famous such limit-cycle system is the \vdp\ oscillator \cite{Str15,vdP26}, given by 
\begin{align}
\label{a'vdPOsc}
	\alpha' = - i \, \alpha + \frac{\mu}{2} \, (\alpha - \alphastar) - \frac{\mu}{4} \, (\alpha^3 - \alphastar{}^3 + \alphastar \alpha^2 - \alphastar{}^2 \alpha)  \; ,
\end{align}
where $\mu>0$ is called the nonlinearity parameter, and we have assumed a unit angular frequency when the nonlinearity is absent. The \vdp\ oscillator actually attains the form of a \ls\ model in the limit of weak nonlinearity, i.e.~when $\mu\longrightarrow0^+$. For this reason, \eqref{a'SLOsc} is said to produce quasilinear (or quasiharmonic) oscillations. On the other hand, \eqref{a'vdPOsc} is capable of relaxation oscillations, which only occur when $\mu$ is nonvanishing \cite{Str15}. Also unlike \eqref{a'SLOsc}, a second-order equation in $x$ can be derived from \eqref{a'vdPOsc}. We first express \eqref{a'vdPOsc} in terms of its real and imaginary parts, 
\begin{align}
\label{vdPOsc}
	x' = y   \;,    \quad   y' = - x - \mu \, (x^2 - 1) \, y  \; ,
\end{align}
Taking the time derivative of $x'$ and using $y'$ we obtain
\begin{align}
\label{x''vdPOsc}
	x'' = - x - \mu \, (x^2 - 1) \, x'  \; .
\end{align}
The second term on the right-hand side of \eqref{x''vdPOsc} can be understood as a position-dependent friction coefficient which effects both amplification (when $x<1$), and dissipation (when $x>1$). Unlike \eqref{a'SLOsc}, where each term can be quantized by a Hamiltonian or dissipator, \eqref{a'vdPOsc} no longer admits such a simple correspondence. In general, a given $h(\alpha,\alphastar)$, even if it has no Hamiltonian terms, requires a combination of dissipators and Hamiltonians to quantize. The \vdp\ oscillator was first quantized exactly in Ref.~\cite{CKN20} by translating \eqref{vdPOsc} to the Weyl-ordered averages:
\begin{align}
\label{QvdPx}
	\an{\xhat}' = {}& \an{\yhat}   \;,   \\
\label{QvdPy}
	\an{\yhat}' = {}&  - \an{\xhat} - \mu \, \an{\yhat} + \frac{\mu}{3} \, \an{\xhat^2 \, \yhat + \xhat \, \yhat \, \xhat + \yhat \, \xhat^2}   \; .
\end{align}
The quantization of the \vdp\ oscillator makes strongly-nonlinear effects accessible to quantum theory \cite{CKN20,SMN+23}. We will return to the \vdp\ oscillator again in Sec.~\ref{Comparison} when we compare cascade quantization to the Lagrangian approach.

\section{Bifurcations}
\label{Bifurcations}

Just as classical bifurcations occur in vastly different fields, one can also find bifurcations in a variety of quantum systems. For example, they are found in models motivated by atomic physics (BECs) \cite{STFL06}, solid-state physics (Jahn--Teller and Bose--Hubbard models) \cite{HDMKM04,MDMKM10,IKV+17,YI19}, quantum optics (Jaynes--Cummings and Dicke models) \cite{AM06,NFP+06}, nanomechanical systems \cite{MMKM11}, superconducting circuits \cite{MNDM14,PBB17}, and recently, in condensed matter physics (the Kitaev chain) \cite{RPR23}. But unlike the classical theory of nonlinear dynamics, there has been no systematic work on quantum bifurcations in the sense of treating their normal forms due to the lack of a quantization method. Here we take a first step towards this end.

An essential point to keep in mind here is that, unlike classical systems, quantum systems do not follow well-defined paths in phase space due to the inherent noise they possess. Thus, to identify bifurcations in quantum systems we rely on the phase-space quasiprobability representation of quantum states. Amongst the most commonly discussed quasiprobability distributions, the Wigner function is the most favored as it is well-behaved and acts as a witness to nonclassicality when it becomes negative. Given a quantum state $\rho$, the Wigner function is defined by
\begin{align}
\label{W(x,y)Defn}
	W(x,y) = \frac{1}{2\pi} \int^\infty_{-\infty} ds \; \bra{x+s/2} \rho \ket{x-s/2} \, e^{-i s y}  \; .
\end{align}
Important features of a classical system such as attractors are then manifested in the quantized system by the peaks of its steady-state Wigner function.\footnote{The steady-state Wigner function being given by \eqref{W(x,y)Defn} but with $\rho$ replaced by a steady-state density operator $\rho_{\rm ss}$ (defined by $\Lcal \rho_{\rm ss}=0$). Alternatively, one may also define it by the long-time limit of \eqref{W(x,y)Defn} if $\rho$ is a time-dependent state.} This is intuitive since an attractor draws all initial conditions in its basin towards it. By the same token, repellers can be expected to have either zero or relatively low Wigner density at steady state. Low- and high-velocity regions of phase space can also be identified, respectively, with high and low values of the Wigner function. This is because a classical ensemble of initial conditions tend to clump more when travelling through phase-space regions of lower velocity. Bifurcations identified using such a correspondence between a deterministic system and the steady-state distributions of their noisy counterparts are referred to as P-bifurcations in stochastic systems (P denoting phenomenological) \cite{Nam90,AB92,ANSH96,LN99,ZVAKK10}. The same correspondence is used to describe classical noise-induced transitions \cite{HL06} (bifurcations induced by multiplicative noise in stochastic differential equations), and very recently, its quantum analog \cite{CMNK23}.

The phenomenological approach can only explain certain qualitative features of quantum bifurcations. First, information about the quantized system such as the bifurcation point (the parameter value at which the bifurcation occurs) cannot be directly inferred from the classical system. Like classical noise, quantum noise prevents the quantized system from having a well-defined bifurcation point. A precise definition of quantum bifurcations would need to be formulated in order to pinpoint when the quantized system bifurcates, but that will take us too far from the theme of quantization. Second, and more interestingly, quantum analogs of even very simple bifurcations (in particular the saddle-node and transcritical bifurcations as we shall see) contain negative values in their Wigner functions. However, the phenomenological method for identifying the occurrence of a bifurcation does not take into account such negative regions of the steady-state Wigner function.

\subsection{Saddle-node bifurcation}

\begin{figure*}
\centering
\includegraphics[width=\textwidth]{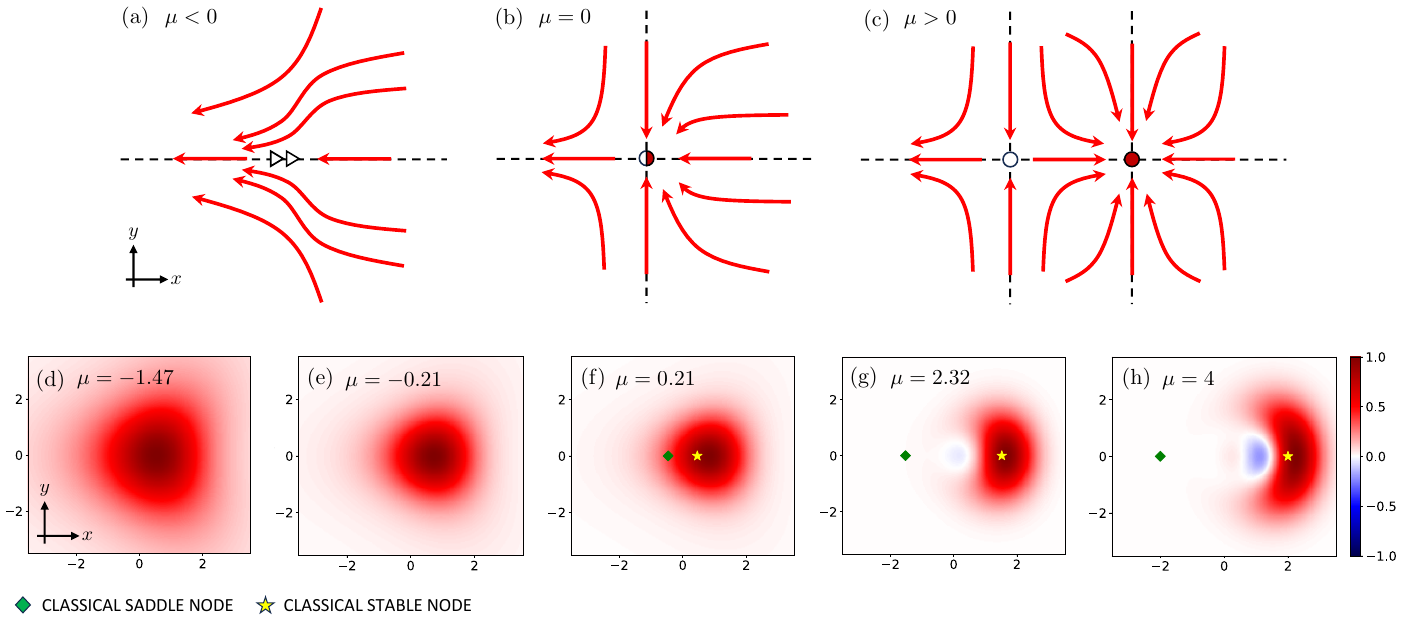}
 \caption{\label{SadNodeBif} Saddle-node bifurcation. (a)--(c) Phase portraits of \eqref{SadNodeNF} as $\mu$ varies from a negative to positive value. The bifurcation occurs at $\mu=0$. In (a)--(c) horizontal dashed lines denote $y$ nullclines [all $(x,y)$ for which $y'=0$], while vertical dashed lines indicate $x$ nullclines [all $(x,y)$ for which $x'=0$]. The origin of phase space is placed at the middle of the $y$ nullcline. We have used the symbol $\triangleright\triangleright$ to denote the bottleneck/ghost region around the origin in (a). The closer $\mu$ is to zero, the greater the slowdown is in the bottleneck. In (d)--(h) we plot the steady-state Wigner function obtained from \eqref{SadNodeBifL} and \eqref{SadNodeBifH} for different values of $\mu$. These and other plots of steady-state Wigner functions in this paper were obtained using the QuTiP package in Python \cite{JNN12,JNN13,LGM+24}. We have scaled the Wigner function so that its value is in $[-1,1]$. It is clear from inspecting (d) to (h) that a saddle-node bifurcation has occurred in the quantum system. However, the exact quantum bifurcation point cannot be inferred from \eqref{SadNodeNF}, and a more nuanced study is required. Despite the simplicity of \eqref{SadNodeNF}, the Wigner negativity seen in (g) and (h) means its quantum analog in \eqref{SadNodeBifL} and \eqref{SadNodeBifH} exhibits fundamentally different behavior, which cannot be understood even from stochastic generalizations of \eqref{SadNodeNF}.}
\end{figure*}

The normal form of a saddle-node bifurcation is given by \cite{Str15}
\begin{align}
\label{SadNodeNF}
    x' = \mu - x^2  \; ,     \quad   y' = - y  \; ,
\end{align}
where $\mu\in\mathbb{R}$ is the bifurcation parameter. The flow governed by \eqref{SadNodeNF} is sketched for different values of $\mu$ in Fig.~\ref{SadNodeBif}(a)--(c). Equation \eqref{SadNodeNF} has no fixed points for $\mu<0$ and all phase-space points flow towards the left. However, as $\mu \longrightarrow 0^-$ a bottleneck region (also called a ghost region \cite{Str15}) starts to develop around the origin which we have symbolized by a $\triangleright\triangleright$ pointing against the flow. Then at precisely $\mu=0$ the flow comes to a halt, and two fixed points are created at the origin, as shown by the half-filled node in Fig.~\ref{SadNodeBif}(b). The two fixed points subsequently move apart for $\mu>0$: One being a saddle node, at $(-\sqrt{\mu},0)$, while the other is a stable node at $(\sqrt{\mu},0)$. Note in Fig.~\ref{SadNodeBif}, and the remainder of this section, we use dashed lines to denote invariant sets (a subset of phase space which the system does not leave once inside it). For the case of Fig.~\ref{SadNodeBif}(a)--(c), the nullclines of \eqref{SadNodeNF} are invariant lines.
\begin{figure*}[t]
\centering
\includegraphics[width=\textwidth]{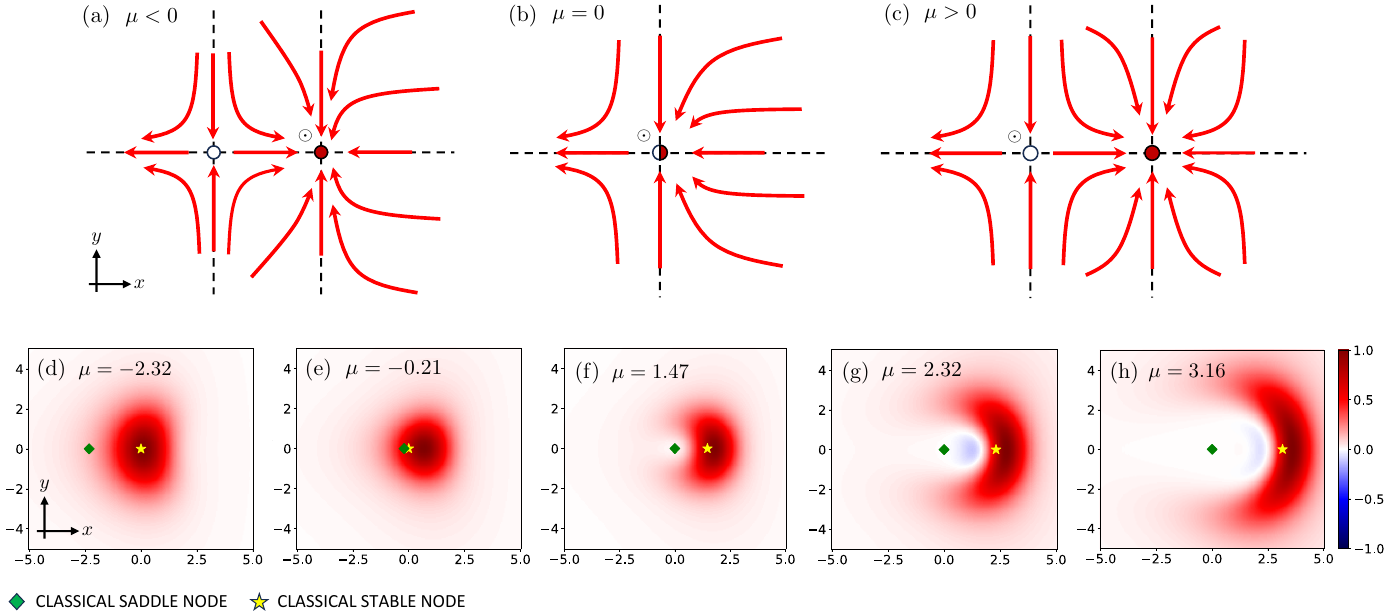}
 \caption{\label{TransCritBif} Transcritical bifurcation. (a)--(c) Phase portraits of \eqref{TransCritNF} as $\mu$ varies from a negative to positive value. The bifurcation occurs at $\mu=0$. In (a)--(c) horizontal dashed lines denote $y$ nullclines [all $(x,y)$ for which $y'=0$], while vertical dashed lines indicate $x$ nullclines [all $(x,y)$ for which $x'=0$]. We have marked the fixed point at the origin by an $\odot$ symbol. In (d)--(h) we plot the steady-state Wigner function defined by \eqref{TransCritBifL} and \eqref{TransCritBifH} corresponding to different values of $\mu$. We have scaled the Wigner function so that its value is in $[-1,1]$. While it is clear that a quantum transcritical bifurcation has occurred, the exact moment at which it happens is not derivable from \eqref{TransCritNF}. Particularly interesting are the regions in (g) and (h) where the Wigner function becomes negative, highlighting again that even extremely simple dynamical systems can have nonclassical features when quantized.}
\end{figure*}

The quantization table of Fig.~\ref{QTable} can now be employed to study the analog of \eqref{SadNodeNF} in quantum mechanics. Expressing \eqref{SadNodeNF} in terms of complex coordinates we get  
\begin{align}
\label{SadNodeNF(a,a*)}
	\alpha' = \frac{\mu}{\sqrt{2}} - \frac{1}{2} \, (\alpha-\alphastar) - \frac{1}{2\sqrt{2}} \, (\alpha^2 + 2 \, \alphastar \alpha + \alphastar{}^2)  \; .
\end{align}
Then by inspection of Fig.~\ref{QTable}, we find \eqref{SadNodeNF(a,a*)} to be quantized by
\begin{align}
\label{SadNodeBifL}
    \Lcal = - i \, [\Hhat, \supopdot\,] + \Dcal[\ann] + \sqrt{2} \, \Dcal[\adg \ann - \adg]  \; ,
\end{align}
where
\begin{align}
\label{SadNodeBifH}
      \Hhat = {}& - i \, \frac{(\mu-1)}{\sqrt{2}} \, (\ann - \adg) - \frac{i}{4} \, (\ann^2 - \adg{}^2)   \nn \\
                     & + \frac{i}{6\sqrt{2}} \, (\ann^3 - \adg{}^3) + \frac{i}{2\sqrt{2}} \, (\adg \ann^2 - \adg{}^2 \ann)   \; .
\end{align}
Steady-state Wigner functions generated from different values of $\mu$ in \eqref{SadNodeBifL} and \eqref{SadNodeBifH} are shown in Fig.~\ref{SadNodeBif}(d)--(h). We have, for ease of comparison, and for the remainder of this section, scaled all Wigner functions to lie in the interval $[-1,1]$. Since for $\mu<0$, all trajectories tend towards infinity in the classical theory [Fig.~\ref{SadNodeBif}(a)], one might not expect the existence of a normalizable steady state in the quantum case. Interestingly, the quantum system does have a normalizable steady state as can be seen in Fig.~\ref{SadNodeBif}(d) and (e), for which the bottleneck region in Fig.~\ref{SadNodeBif}(a) is manifested by a Wigner function peaked near the origin. We also find this peak to be more concentrated in going from Fig.~\ref{SadNodeBif}(d) to (e). This is consistent with the classical intuition that as $\mu$ gets closer and closer to zero, the slow-down near the origin becomes more and more dramatic. The quantum saddle-node bifurcation is then revealed in the steady-state Wigner functions corresponding to $\mu>0$, shown in Fig.~\ref{SadNodeBif}(f)--(h). We have also plotted the fixed points of \eqref{SadNodeNF} as a green diamond (saddle node) and a yellow star (stable node) in Fig.~\ref{SadNodeBif}(f)--(h). The quantum analog to Fig.~\ref{SadNodeBif}(c) can now be seen in the Wigner function in terms of its mode (i.e.~peak) moving towards the right, and a region of zero moving towards the left, as $\mu$ increases. However, as we said above, not all features can be qualitatively captured from classical intuition. We find a region of Wigner negativity starting to develop around the origin in Fig.~\ref{SadNodeBif}(g). And as $\mu$ is increased, this negativity becomes stronger, as shown in Fig.~\ref{SadNodeBif}(h). Note that such effects cannot be captured by any classical model, not even stochastic extensions to \eqref{SadNodeNF}. Thus, we see that even an extremely simple dynamical system like \eqref{SadNodeNF} can exhibit genuine nonclassicality when quantized.

\subsection{Transcritical bifuration}

The normal form of a transcritical bifurcation is 
\begin{align}
\label{TransCritNF}
	x' = \mu \, x - x^2 \; ,   \quad   y' = - y   \; .
\end{align}
Flows corresonding to different values of $\mu$ in \eqref{TransCritNF} are shown in Fig.~\ref{TransCritBif}(a)--(c). When $\mu < 0$, there are two fixed points in \eqref{TransCritNF}: A stable node at the origin $(0,0)$, and a saddle node at $(\mu,0)$. As $\mu$ increases the saddle node moves towards the right, until at $\mu=0$ the two fixed points collide at the origin and exchange stabilities. For $\mu>0$, we then find $(0,0)$ to be a saddle, while the fixed at point at $(\mu,0)$ becomes stable. This exchange of stabilities is known as the transcritical bifurcation. Note there is always a fixed point at the origin in Fig.~\ref{TransCritBif}(a)--(c), which is marked by an $\odot$ next to it.
\begin{figure*}[t]
\centering
\includegraphics[width=\textwidth]{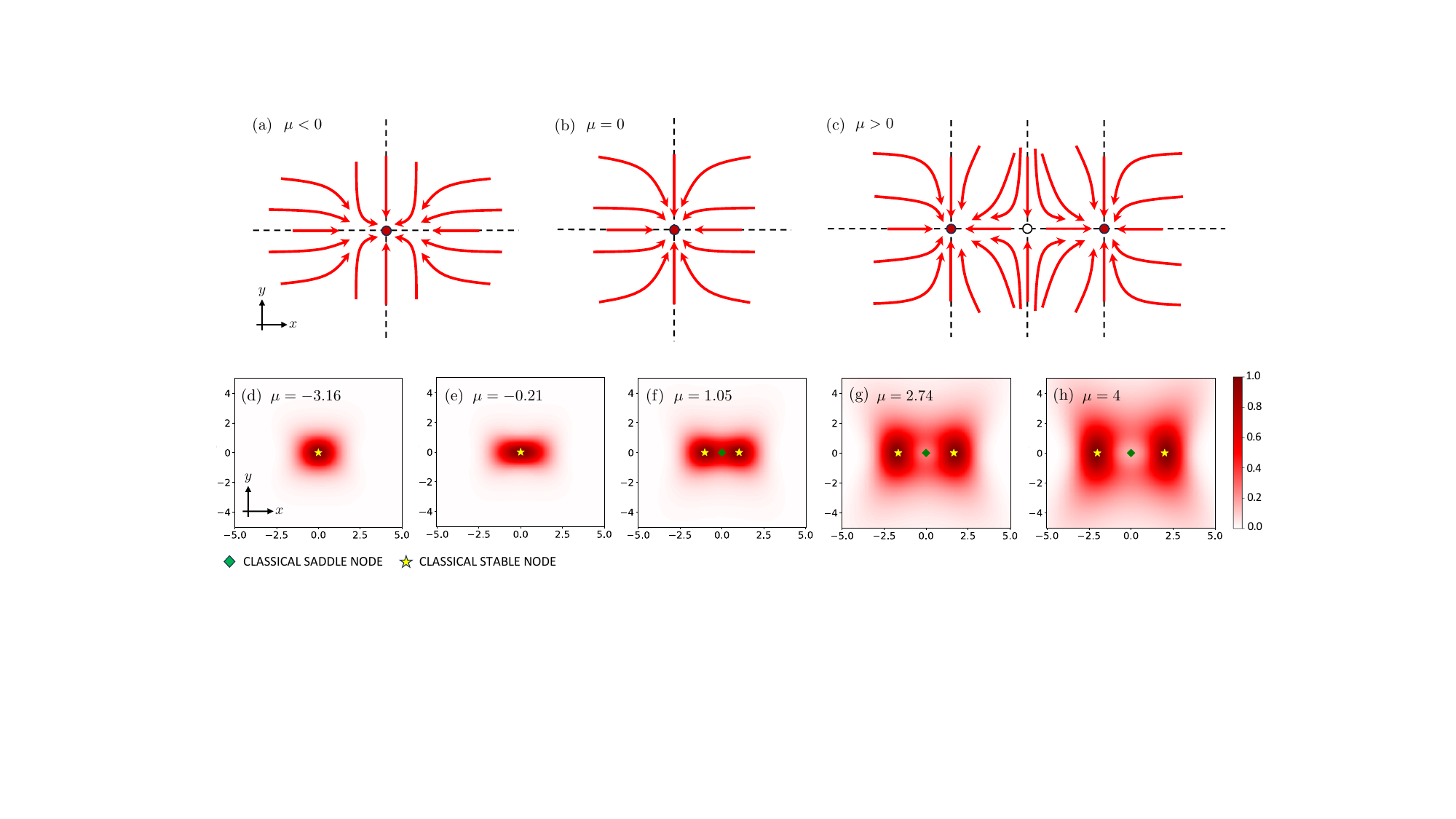}
 \caption{\label{PitchForkBif} Pitchfork bifurcation. (a)--(c) Phase portraits of \eqref{PitchForkNF} as $\mu$ varies from a negative to positive value. The bifurcation occurs at $\mu=0$. In (a)--(c) horizontal dashed lines denote $y$ nullclines [all $(x,y)$ for which $y'=0$], while vertical dashed lines indicate $x$ nullclines [all $(x,y)$ for which $x'=0$]. The origin of phase space is at the center of the $y$ nullcline. In (d)--(h) we plot the steady-state Wigner function corresponding to \eqref{PitchForkBifL} and \eqref{PitchForkBifH} for different values of $\mu$. We have scaled the Wigner function so that its value is in $[-1,1]$. A clear resemblance of the classical pitchfork bifurcation can be seen in the quantized system where a single peak in the steady-state Wigner function (before the bifurcation) splits into two lobes (after the bifurcation) that move apart as $\mu$ increases.}
\end{figure*}

To consider the transcritical bifurcation in quantum theory we again rewrite \eqref{TransCritNF} in complex coordinates first, giving
\begin{align}
\label{TransCritNF(a,a*)}
	\alpha' = \frac{(\mu-1)}{2} \, \alpha + \frac{(\mu+1)}{2} \, \alphastar - \frac{1}{2\sqrt{2}} \, (\alpha^2 + 2 \, \alphastar \alpha + \alphastar{}^2)   \; . 
\end{align}
Applying the quantization table in Fig.~\ref{QTable} to \eqref{TransCritNF(a,a*)} then gives us the appropriate Lindbladian 
\begin{align}
\label{TransCritBifL}
    \Lcal = {}& - i \, [\Hhat, \supopdot\,] + \sqrt{2} \, \Dcal[\adg \ann - \adg]   \\
                  & + (1-\mu) \, \theta(1-\mu) \, \Dcal[\ann] + (\mu-1) \, \theta(\mu-1) \, \Dcal[\adg]    \; ,     \nn
\end{align}
where 
\begin{align}
\label{TransCritBifH}
      \Hhat = {}& \frac{i}{\sqrt{2}} \, ( \ann - \adg) - i \, \frac{ (\mu+1)}{4} \, (\ann^2 - \adg{}^2)   \nn \\ 
                      & + \frac{i}{6\sqrt{2}} \, (\ann^3 - \adg{}^3) + \frac{i}{2} \, (\adg \ann^2 - \adg{}^2 \ann)   \; ,
\end{align}
and $\theta(x)$ is the Heaviside step function defined in \eqref{HeavisideDefn}. We then obtain from \eqref{TransCritBifL} and \eqref{TransCritBifH} steady-state Wigner functions for different values of $\mu$, shown in Fig.~\ref{TransCritBif}(d)--(h). As before, the Wigner function is peaked near the classical stable node. We also find a small region of zeros where the classical saddle node is in Fig.~\ref{TransCritBif}(f), with some negativity developing in the Wigner function in Fig.~\ref{TransCritBif}(g) and (h). Thus, as with the quantum saddle-node bifurcation, the quantum transcritical bifurcation cannot be simulated by any classical stochastic system.

\subsection{Pitchfork bifurcation (supercritical)}

The normal form of the supercritical pitchfork bifurcation is
\begin{align}
\label{PitchForkNF}
	x' = \mu \, x - x^3 \; ,    \quad   y' = - y   \; .
\end{align}
Sketches of the flow for \eqref{PitchForkNF} with different values of $\mu$ are shown in Fig.~\ref{PitchForkBif}(a)--(c). Equation \eqref{PitchForkNF} has only one stable node at the origin for $\mu<0$. It remains stable as we increase $\mu$, up till and including $\mu=0$. The origin then becomes a saddle for $\mu>0$ and two new stable nodes are created at $(-\rt{\mu},0)$ and $(\rt{\mu},0)$. The two stable nodes move away from the origin as $\mu$ increases.

To quantize \eqref{PitchForkNF} we first convert it to complex coordinates, 
\begin{align}
\label{PitchForkNF(a,a*)}
	\alpha' = {}& \frac{(\mu-1)}{2} \, \alpha + \frac{(\mu+1)}{2} \, \alphastar  \nn  \\
	                 & - \frac{1}{4} \, \big( \alpha^3 + 3 \, \alphastar{}^2 \alpha + 3 \, \alphastar \alpha^2 + \alphastar{}^3 \big)   \; .
\end{align}
Applying the quantization table in Fig.~\ref{QTable} to \eqref{PitchForkNF(a,a*)} then gives
\begin{align}
\label{PitchForkBifL}
    \Lcal = {}& - i \, [\Hhat, \supopdot\,] + \frac{9}{8} \, \Dcal[\ann^2] + \frac{3}{2} \, \Dcal[\adg \ann - \adg{}^2/2]   \\
                  & + (1-\mu) \, \theta(1-\mu) \, \Dcal[\ann] + (\mu-1) \, \theta(\mu-1) \, \Dcal[\adg]    \; ,     \nn
\end{align}
where 
\begin{align}
\label{PitchForkBifH}
      \Hhat = {}& - i \, \frac{ (\mu+1)}{4} \, (\ann^2 - \adg{}^2) + \frac{i}{8} \, (\adg{} \ann^3 - \adg{}^3 \ann)  \nn  \\
                     & + \frac{i}{16} \, (\ann^4 - \adg{}^4)   \; .
\end{align}
The steady-state Wigner functions generated by \eqref{PitchForkBifL} and \eqref{PitchForkBifH} for different values of $\mu$ are shown in Fig.~\ref{PitchForkBif}(d)--(h). The quantum pitchfork bifurcation is qualitatively similar to the classical pitchfork bifurcation, with the peaks and troughs of the Wigner function occurring where the classical stable and saddle nodes are.

The quantum pitchfork bifurcation has cropped up in different guises elsewhere. For example, in quantum synchronization, where squeezing was shown to enhance the synchronization of a \ls\ oscillator to an external drive \cite{SHM+18}. It has also appeared in the stabilization of \sch\ cat states, which are useful for bosonic quantum error correction and fault-tolerant quantum computing \cite{CNAA+22}.

\subsection{Hopf bifurcation (supercritical)}
\label{HopfBifSup}

\begin{figure*}[t]
\centering
\includegraphics[width=\textwidth]{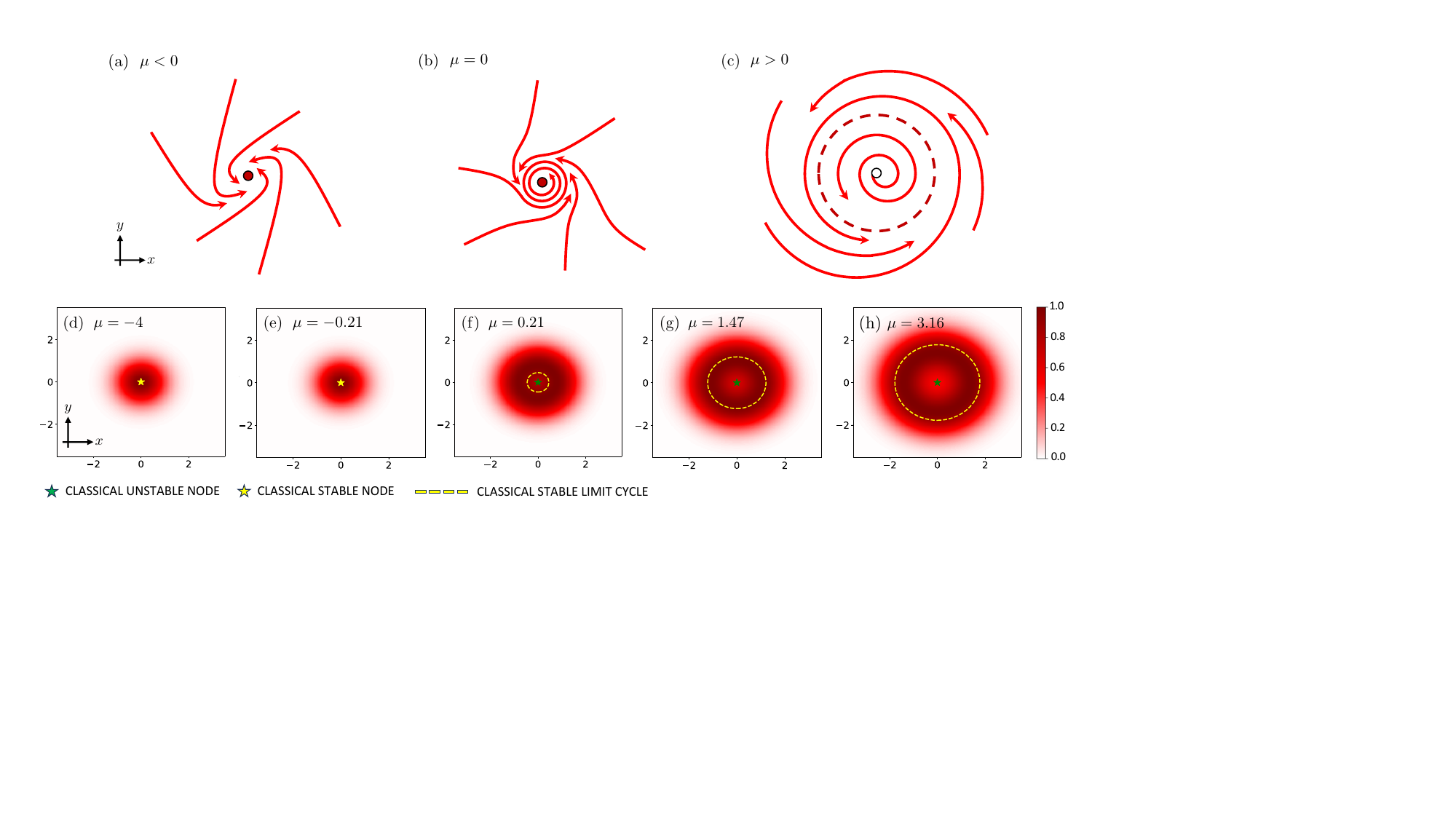}
 \caption{\label{HopfBif} Hopf bifurcation. (a)--(c) Phase portraits of \eqref{HopfNFx} and \eqref{HopfNFy} as $\mu$ varies from a negative to positive value. The bifurcation occurs at $\mu=0$. The origin of phase space is at the center, always occupied by a fixed point. Given the rotational symmetry of \eqref{HopfNFx} and \eqref{HopfNFy}, it is actually best to express them in polar coordinates. Doing so gives $r'=\mu\,r-r^3$ and $\phi'=1$ for the radial and phase variables, which are related to the Cartesian coordinates by $x=r \cos\phi$ and $y=r\sin\phi$. The radial equation is thus nothing more than a one-dimensional pitchfork bifurcation with $r$ restricted to be positive. It is then straightforward to see that only one stable fixed point exists at $r=0$ for $\mu\le0$, while there are two fixed points for $\mu>0$, one being unstable at the origin, and the other stable at $r=\rt{\mu}$. In (d)--(h) we plot the steady-state Wigner functions generated from \eqref{HopfBifL} with different values of $\mu$. We have scaled the Wigner function so that its value is in $[-1,1]$. The limit cycle is created in passing from (e) to (f), when the Wigner-function peak starts to dip and a crater starts to form around the origin.} 
\end{figure*}

The supercritical Hopf bifurcation occurs when a stable fixed point loses its stability, and in the process a stable limit cycle is created. The literature also refers to this as the Andronov--Hopf, or Poincar\'{e}--Andronov--Hopf bifurcation. It is the most common way in which stable limit cycles arise. It has a normal form given by
\begin{align}
\label{HopfNFx}
	x' = {}& \! -y + \mu \, x - (x^2+y^2) \, x   \; ,    \\
\label{HopfNFy}
	y' = {}& x + \mu \, y - (x^2+y^2) \, y   \; ,
\end{align}
where we have again assumed the system to have a unit natural frequency. We illustrate the phase-space flow of \eqref{HopfNFx} and \eqref{HopfNFy} in Fig.~\ref{HopfBif}(a)--(c). If $\mu\le0$, there is only one stable node at the origin. When $\mu>0$, the origin loses its stability and a circular limit cycle with radius $\rt{\mu}$ is born. Note that \eqref{HopfNFx} and \eqref{HopfNFy} have a similar form as the  \ls\ oscillator defined in \eqref{x'SLOsc} and \eqref{y'SLOsc}, except with all its coefficients positive [whereas $\mu$ may be negative in \eqref{HopfNFx} and \eqref{HopfNFy}], and with $\gamma/2$ set to one.\footnote{The direction of phase-space rotation is also different---clockwise in \eqref{x'SLOsc}, \eqref{y'SLOsc}, and counterclockwise in \eqref{HopfNFx}, \eqref{HopfNFy}---but this is a trivial modification.} Thus \eqref{HopfNFx} and \eqref{HopfNFy} corresponds to the \ls\ oscillator for $\mu>0$, when the system has a limit cycle. The only important parameter in this case is the ratio of linear gain to nonlinear damping, which determines the radius of the limit cycle. In terms of \eqref{x'SLOsc} and \eqref{y'SLOsc} the radius of the limit cycle is simply $\rt{2\kappa/\gamma}$, which can be produced just as well from \eqref{HopfNFx} and \eqref{HopfNFy} with the appropriate choice of $\mu$.

To quantize \eqref{HopfNFx} and \eqref{HopfNFy} we again transform them into a single equation in terms of complex variables,
\begin{align}
\label{HopfNF(a,a*)}
	\alpha' =  i \, \alpha + \mu \, \alpha - 2 \, |\alpha|^2 \alpha  \; .
\end{align}
Then from Fig.~\ref{QTable} we get
\begin{align}
\label{HopfBifL}
    \Lcal = {}& i \, [\, \adg\ann, \supopdot ] + 2 \, \Dcal[\ann^2]   \nn \\ 
                  & + 2  \mu \, \theta(\mu) \, \Dcal[\adg] -2 \mu \, \theta(-\mu) \, \Dcal[\ann]   \; ,
\end{align}
where we have used $\Hhat=-\adg\ann$ in the Lindbladian.~\footnote{Although the Hamiltonian is unbounded from below, the Lindbladian can be physically realized in a suitable rotating frame.} Equation \eqref{HopfBifL} for $\mu>0$ has been erroneously referred to as the quantum \vdp\ oscillator in recent physics literature, e.g.~in Refs.~\cite{LS13,WNB14}. Strictly speaking, the \vdp\ oscillator refers exclusively to the dynamical system in \eqref{vdPOsc}.\footnote{Equation \eqref{HopfNF(a,a*)} describes \emph{any} limit cycle just after the onset of a supercritical Hopf bifurcation. Since many nonlinear systems other than the \vdp\ oscillator have limit cycles created from a Hopf bifurcation, one cannot uniquely associate the \vdp\ oscillator with \eqref{HopfNF(a,a*)}. For this reason, and appropriately so, the nonlinear dynamics literature refers to \eqref{HopfNF(a,a*)} as the \ls\ oscillator, which is also the terminology used here. In fact, historically, and as we already hinted in Sec.~\ref{RotationalSymmetry} under \eqref{a'vdPOsc}, the \vdp\ oscillator is a prototype for relaxation oscillations. These oscillations exist far from the Hopf bifurcation point, and as such, they are non-circular and non-uniformly traversed orbits in phase space. In fact, older publications in physics (as far back as the 1960s to our knowledge) actually considered the difference between the \ls\ and \vdp\ oscillators by referring to the former as a rotating-wave \vdp\ oscillator \cite{LL67,Lou73,LCX06}. See Ref.~\cite{Ris96} for a pedagogical account of how the rotating-wave approximation is used to reduce the \vdp\ oscillator to the \ls\ form. Essentially the same method goes by the name of Krylov--Bogoliubov time averaging in nonlinear dynamics \cite{BJPS09}.} Also worth noting is the subcritical flavor of the Hopf bifurcation. Although it is encountered much less frequently, a quantized version of it has appeared in the context of quantum synchronization \cite{JLA20} (see also Ref.~\cite{AM06}).
\begin{figure*}[t]
\centering
\includegraphics[width=\textwidth]{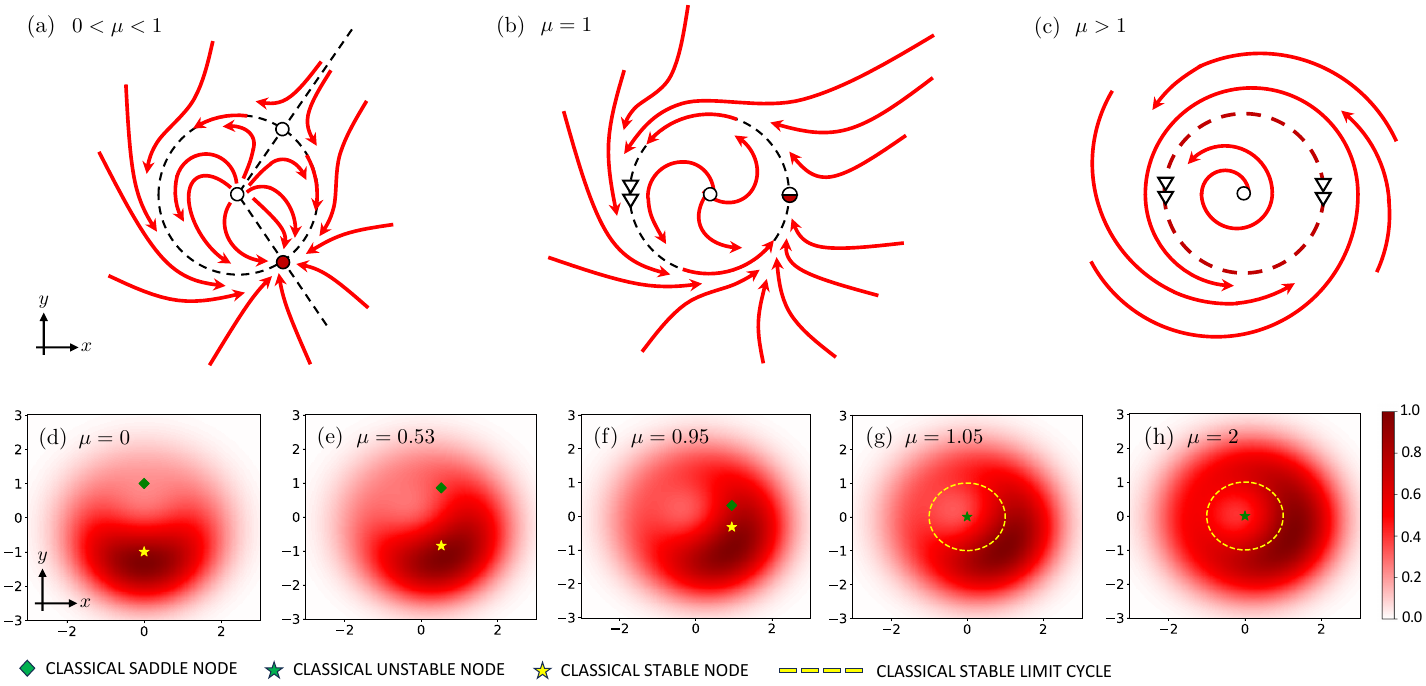}
\caption{\label{SNICBif} Infinite-period bifurcation. (a)--(c) Phase portraits of \eqref{SNICBifx} and \eqref{SNICBify} as $\mu$ varies from a positive number less than one to a value greater than one. The bifurcation occurs at $\mu=1$. Note that $\mu$ can be negative in \eqref{SNICBifx} and \eqref{SNICBify}, so more generally, the important regimes are $|\mu|<1$, $|\mu|=1$ (bifurcation point), and $|\mu|>1$. The origin of phase space is at the center, always occupied by an unstable node. As with the Hopf bifurcation, this system is most suitably expressed in polar coordinates, given by $r' = r - r^3$, and $\phi' = \mu - r \, \cos\phi$, where $x=r\cos\phi$ and $y=r\sin\phi$. In the literature the phase equation is often given as $\phi'=\mu-\cos\phi$, or $\phi'=\mu-\sin\phi$, without the extra factor of $r$ in front of the trigonometric function. The additional $r$ in our phase equation is introduced to simplify the quantization problem, but it is actually inconsequential for the steady-state behavior of the system. For all $\mu$, there is always an unstable node at $r=0$, and a stable invariant circle at $r=1$ (shown as a dashed circle). In addition, we find for $0 \le |\mu| \le 1$, a saddle node at $r=1$ and $\phi= \cos^{-1}\mu$, and a stable node at $r=1$ and $\phi=2\pi-\cos^{-1}\mu$. Note there are also two invariant lines extending outwards from the origin defined by the angles of the saddle and stable nodes [shown as straight dashed lines in (a)]. The fixed points then collide at $|\mu| = 1$. When $|\mu| > 1$, the fixed points disappear and the invariant circle becomes a limit cycle. We have used $\triangleright\triangleright$ to indicate where the flow is fastest or slowest on the limit cycle, as defined by the value of $\phi'$ (by having $\triangleright\triangleright$ point in the same or opposite direction as the flow). In (d)--(h) we plot the steady-state Wigner functions obtained from \eqref{SNICBifL} and \eqref{SNICBifH} for different values of $\mu$. The Wigner function is scaled so that its value falls within $[-1,1]$.  In (d) and (e) we find the Wigner-function peak to be displaced in accordance to the motion of the classical stable node as $\mu$ increases. The transition to a non-uniform limit cycle is then manifested by a diffusing Wigner-function peak along the phase direction. This can be seen incrementally when going from (g) to (h).}
\end{figure*}
Figure~\ref{HopfBif}(d)--(h) shows the steady-state Wigner functions obtained from \eqref{HopfBifL} when $\mu$ is varied. A quantum Hopf bifurcation can be seen to occur when passing from Fig.~\ref{HopfBif}(e)
 to (f). For $\mu<0$, the Wigner function is peaked at the origin, resembling the corresponding classical stable node there. The quantum Hopf bifurcation then occurs when the Wigner function starts to dip at the origin and simultaneously develops a ring around it. This craterlike structure in Fig.~\ref{HopfBif}(f)--(h) resembles the classical limit cycle with an unstable node at the origin. In fact, a similar craterlike structure can also be found in the steady-state probability density function for classical systems with a stochastic Hopf bifurcation \cite{ANSH96,PAV89,Hop96,DELR18}. Though not of direct relevance to bifurcations, the transient dynamics leading to the limit-cycle state in a quantum \ls\ oscillator has been explored just recently \cite{LDUN24}.

Not all aspects of the limit cycle produced by a quantum Hopf bifurcation can be explained with classical intuition. One aspect, attributed to quantum fluctuations, is that it gives birth to a quantum limit cycle whose radius is larger than the classical value when $\mu\lesssim1$ \cite{MKH20}. Besides fundamental physics, it may also be possible to exploit the quantum Hopf bifurcation for sensing weak signals \cite{DC19}.

\subsection{Infinite-period bifurcation}
\label{InfPerBif}

While the Hopf bifurcation is the most common path to a stable limit cycle, it is not the only way. Another path to a stable limit cycle is provided by an infinite-period bifurcation, also known as a saddle-node bifurcation on an invariant circle. The infinite-period bifurcation gives birth to a rotationally asymmetric limit cycle, and may be described by
\begin{align}
\label{SNICBifx}
	x' = {}& \! -\mu \, y + x + x \, y - (x^2+y^2) \, x   \; ,   \\
\label{SNICBify}
	y' = {}& \mu \, x + y - x^2 - (x^2+y^2) \, y   \; .
\end{align}
Sketches of the phase portrait for different values of $\mu$ are shown in Fig.~\ref{SNICBif}(a)--(c). The system always has an unstable origin surrounded by an invariant circle. As in previous figures, invariant subsets are denoted by dashed lines. If $0\le\mu\le1$, we find two more fixed points which lie on the invariant circle, one being a saddle node at $(\mu,\rt{1-\mu^2})$, and the other a stable node at $(\mu,-\rt{1-\mu^2})$. As $\mu$ is increased, the saddle and stable nodes move towards each other along the invariant circle, until at $\mu=1$ they collide, as shown by the half-filled node in Fig.~\ref{SNICBif}(b). The invariant circle is thus a homoclinic orbit at $\mu=1$, which is traversed at a non-uniform speed. We have again used $\triangleright\triangleright$ to indicate where the flow is notably fast (by having $\triangleright\triangleright$ point in the same direction as the flow lines). For $\mu>1$ the invariant circle then turns into a limit cycle, shown by a dashed circle in dark red in Fig.~\ref{SNICBif}(c). Similar to the usual saddle-node bifurcation from earlier, the half-filled node in Fig.~\ref{SNICBif}(b) develops into a bottleneck/ghost region (shown by pointing $\triangleright\triangleright$ in the opposite direction as the flow lines). For $\mu>1$, the limit cycle has a period given by
\begin{align}
	T = \int^{2\pi}_0 d\phi \; (\mu-\cos\phi)^{-1}  =  \frac{2 \pi}{\rt{\mu^2-1}}  \; ,
\end{align}
which tends to infinity as $\mu\longrightarrow1^+$ (hence the name infinite-period bifurcation) \cite{Str15}.

To quantize \eqref{SNICBifx} and \eqref{SNICBify} we note that they are equivalent to 
\begin{align}
\label{SNICNF(a,a*)}
	\alpha' = i \, \mu \, \alpha + \alpha - \frac{i}{\rt{2}} \, ( \alpha^2 + \alphastar \alpha ) - 2 \, |\alpha|^2 \alpha  \; .
\end{align}
From Fig.~\ref{QTable} we find that \eqref{SNICNF(a,a*)} has a valid quantization given by
\begin{align}
\label{SNICBifL}
	\Lcal = - \, i \, [\Hhat, \supopdot\,] + 2 \, \Dcal[\adg] + 2 \, \Dcal[\ann^2] + \frac{1}{\rt{2}} \, \Dcal[\adg\ann + i \, \adg]  \; ,
\end{align}
where 
\begin{align}
\label{SNICBifH}
	\Hhat = - \mu \, \adg\ann - \frac{1}{\rt{2}} \, (\ann + \adg) + \frac{1}{2\rt{2}} \, (\adg \ann^2 + \adg{}^2 \ann)  \; .
\end{align}
The steady-state Wigner functions obtained from \eqref{SNICBifL} and \eqref{SNICBifH} are shown in Fig.~\ref{SNICBif}(d)--(h). The classical behavior seen in Fig.~\ref{SNICBif}(a) is clearly reflected in Fig.~\ref{SNICBif}(d) and (e), where the Wigner function can be seen to peak around the classical stable node. A quantum infinite-period bifurcation can then be inferred in going from Fig.~\ref{SNICBif}(e) to (h). Most notably, a non-uniform quantum limit cycle can be seen in Fig.~\ref{SNICBif}(h), where the Wigner-function peak tries to wrap around the origin by diffusing along the phase direction. This is in contrast to Fig.~\ref{SNICBif}(d) and (e), where the Wigner-function peak is simply displaced. A non-uniform circular limit cycle in classical mechanics thus translates into an uneven crater for the steady-state Wigner function in quantum mechanics. Henceforth, we call a steady-state Wigner function obtained from quantizing a classical non-uniform limit cycle, such as Fig.~\ref{SNICBif}(h), a non-uniform quantum limit cycle. A uniform quantum limit cycle would then refer to a steady-state Wigner function like Fig.~\ref{HopfBif}(h).

We can get an intuitive understanding of why \eqref{SNICBifx} and \eqref{SNICBify} contains a non-uniform circular limit cycle by rewriting them as 
\begin{align}
\label{AmpDepFreqSLOx}
	x' = {}& \! - \omega(x) \, y + x - (x^2+y^2) \, x     \; ,   \\
\label{AmpDepFreqSLOy}
	y' = {}& \omega(x) \, x + y - (x^2+y^2) \, y    \; ,
\end{align}
where these equations have the \ls\ form but with an $x$-dependent angular frequency, defined by $\omega(x)=\mu-x$. In complex coordinates, the $\omega(x)$ terms translate to the second-degree terms in \eqref{SNICNF(a,a*)}, which are the only terms that do not satisfy \eqref{RotSymProc} from Sec.~\ref{RotationalSymmetry}. We can thus isolate the $x$-dependent frequency as the sole source of rotational asymmetry in this system.

\section{Stochastic systems}
\label{ExcitableSys}

We now turn to the quantization of classical stochastic dynamics. That is, we apply cascade quantization to a classical nonlinear system driven by a random process $\Upsilon(t)$. Such stochastic models may capture realistic systems but also produce nonlinear effects which are interesting in their own right \cite{MM12,AAVNSG07,Dua15}. A well-known model in classical stochastic nonlinear dynamics is the \fn\ model. This model is often used to describe a neuron and its response to a noisy input \cite{Fit61,NAY62,Ede05,Fra21,SSGO07}.

\subsection{Noise-free \fn\ model}

It is instructive to first consider the deterministic, i.e.~noise-free \fn\ model \cite{Fit61,NAY62,Ede05,Fra21}. This is given by
\begin{align}
\label{FNModel}
	x' = - \varepsilon  \bigg( \frac{x^3}{3} - x_0^2 \, x \bigg) + y   \; ,     \quad   y' = -x + \mu  \; .
\end{align}
We have included in our \fn\ model an effective nonlinearity parameter $\varepsilon>0$, and a parameter $x_0$ that adjusts the shape of the limit cycle when it exists. The parameter $\mu\in\mathbb{R}$ may then be regarded as a bifurcation parameter for a fixed $\varepsilon$ and $x_0$. The system has a fixed point given by 
\begin{align}
\label{FNfxpt}
	x_\fxpt = \mu \, ,  \quad  y_\fxpt = \varepsilon\bigg( \frac{\mu^3}{3} - \mu \, x_0^2 \bigg)  \; .
\end{align}
It is straightforward to show that $(x_\fxpt,y_\fxpt)$ is stable when $|\mu|>x_0$, and as $|\mu|$ decreases, the system undergoes a supercritical Hopf bifurcation at $|\mu|=x_0$. We thus find a stable limit cycle and an unstable $(x_\fxpt,y_\fxpt)$ for $|\mu|<x_0$. 
\begin{figure*}[t]
\centering
\includegraphics[width=\textwidth]{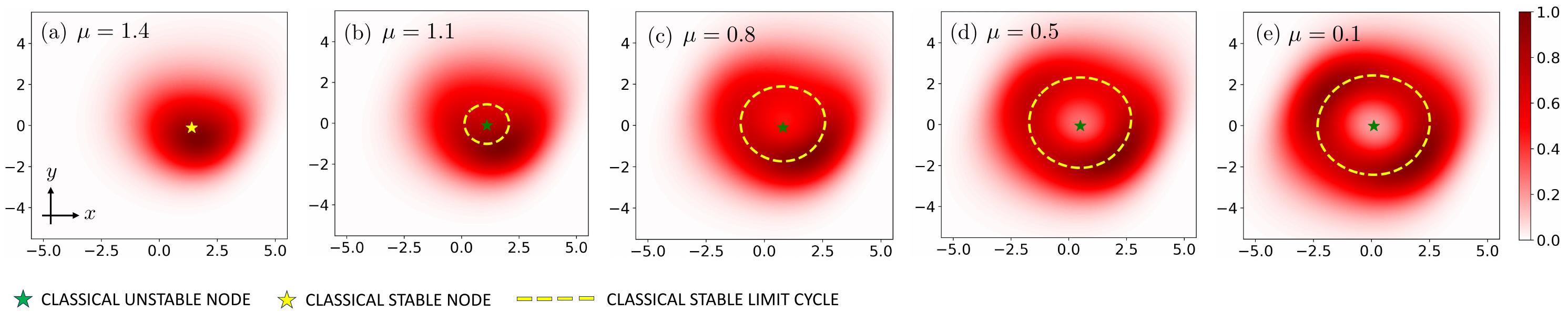}
\caption{\label{QFNModel}Steady-state Wigner functions for the quantum \fn\ model as defined by \eqref{NoiseFreeFNL} and \eqref{NoiseFreeFNH} with different values of $\mu$ (shown in the inset) at a fixed $\varepsilon=0.1$, and a fixed $x_0=1.2$.}
\end{figure*}

It is straightforward to show that \eqref{FNModel} may be written equivalently as, 
\begin{align}
	\alpha' = {}& i \, \frac{\mu}{\rt{2}} + \bigg( \frac{\varepsilon \, x_0^2}{2} - i \bigg) \, \alpha + \frac{\varepsilon \, x_0^2}{2} \, \alphastar  \nn \\
	                 & - \frac{\varepsilon}{12} \, (\alpha^3 + 3 \, \alphastar{}^2 \alpha + 3 \, \alphastar \alpha^2 + \alphastar{}^3)  \; .
\end{align}
Looking up Fig.~\ref{QTable} we find the Lindbladian corresponding to $\kappa=0$ to be
\begin{align}
\label{NoiseFreeFNL}
	\Lcal_0 = {}& - i \, [\Hhat,\supopdot\,] + \varepsilon \, x_0^2 \, \Dcal[\adg] + \frac{3\,\varepsilon}{8} \, \Dcal[\ann^2]  \nn \\
	            & + \frac{\varepsilon}{2} \, \Dcal\big[ \adg \ann - \adg{}^2/2 \big]   \; ,
\end{align}
where
\begin{align}
\label{NoiseFreeFNH}
	\Hhat = {}& \adg \ann - \frac{\mu}{\rt{2}} \, (\ann + \adg) - i \, \frac{\varepsilon \, x_0^2}{4} \, ( \ann^2 - \adg{}^2 )  \nn \\
	           & - i \frac{\varepsilon}{24} \, (\adg{}^3 \ann - \adg \ann^3) + i \frac{\varepsilon}{48} \, (\ann^4 - \adg{}^4)   \; .
\end{align}
Equations \eqref{NoiseFreeFNL} and \eqref{NoiseFreeFNH} now define a quantum \fn\ model. Its steady-state Wigner function alongside the attractors of \eqref{FNModel} are plotted in Fig.~\ref{QFNModel} for different values of $\mu$ with $\varepsilon$ and $x_0$ held constant. It can be seen from Fig.~\ref{QFNModel}(a) that, when $\mu>x_0$, the steady-state Wigner function is unimodal, concentrated around the classical stable fixed point given by \eqref{FNfxpt}. If instead $\mu<x_0$, a classical Hopf bifurcation will have occurred, which is the regime illustrated in Fig.~\ref{QFNModel}(b)--(e). Beginning with Fig.~\ref{QFNModel}(b), we find the steady-state Wigner function just starting to diffuse around the classical limit cycle when $\mu$ is only slightly smaller than $x_0$, i.e., the classical Hopf bifurcation has only just occurred. Then as we move progressively from Fig.~\ref{QFNModel}(c) to (e), the Wigner function starts to dip around the classical unstable fixed point. This is consistent with the classical fixed point becoming more unstable the smaller $\mu$ is than $x_0$. This property may be verified by calculating the real part of the eigenvalue of the Jacobian matrix for \eqref{FNModel}. The Wigner function can also be seen to diffuse more as we transition from Fig.~\ref{QFNModel}(c) to (e). This eventually gives rise to a second peak as seen in Fig.~\ref{QFNModel}(e). Again, though we have not shown this, the bimodal nature of the steady-state Wigner function does correspond to a classical limit cycle with two opposing sides that move relatively slowly. This can be shown from \eqref{FNModel}, e.g.~by considering its nullclines. We will in fact do this later in Sec.~\ref{LienardSys} for a class of limit-cycle systems (see Fig.~\ref{Lienard}).

\subsection{Noise-driven \fn\ model}

To illustrate cascade quantization on a stochastic model, we add a Gaussian white-noise process $\Upsilon(t)$ to \eqref{FNModel}. To keep things simple and for ease of comparison, we follow Refs.~\cite{PK97,LSG00} of only adding noise to one of the phase-space variables:
\begin{align}
\label{NoisyFNModel}
	X' = - \varepsilon \bigg(  \frac{X^3}{3} - x_0^2 \, X  \bigg) + Y    \; ,  \quad   Y' = -X + \mu + \kappa \, \Upsilon  \; ,
\end{align}
where the stochastic term is assumed to have a strength of $\kappa$. Being white, $\Upsilon(t)$ is defined by
\begin{align}
	{\rm E}\big[ \Upsilon(s) \Upsilon(t) \big] = \delta(s-t)  \; .
\end{align}
Note that we are following the convention of denoting a stochastic process by an uppercase letter, and its realizations by the corresponding lowercase letter. The ensemble average of a function of $n$ real-valued random variables, say $w(\bm{Z}) \equiv w(Z_1,Z_2,\ldots,Z_n)$, is then defined by
\begin{align}
    {\rm E}[w(\bm{Z})] \equiv \int_{\mathbb{R}^n}  d\bm{z} \; w(\bm{z}) \, \wp(\bm{z})  \; ,
\end{align} 
where we have assumed $Z_1,Z_2,\ldots,Z_n$ to have the joint probability density $\wp(\bm{z})$. We will see next that $\Upsilon(t)$ has to be interpreted as a \str\ process for it to be consistent with cascade quantization [a point which we foreshadowed under \eqref{TimeDepLindbladian} in Sec.~\ref{OpOrderTimeDepSys}].

\subsubsection{Noise-activated spikes and quasi-regular oscillations}

The influence of $\Upsilon(t)$ on the quantum \fn\ model can be accounted for by \eqref{NoiseFreeFNH} if we simply let $\mu \longrightarrow \mu+\kappa\,\Upsilon(t)$. It will then be useful to separate out the stochastic component of the noise-added Lindbladian by defining
\begin{align}
\label{LUpsilon}
	\Lcal_\Upsilon(t) = \Lcal_0 + i \, \frac{\kappa}{\rt{2}} \, \Upsilon(t) \, [ \, \ann+\adg, \supopdot \, ]    \; .
\end{align}
Equation \eqref{LUpsilon} now quantizes the noisy \fn\ model \eqref{NoisyFNModel}, but now $\Lcal_\Upsilon(t)$ is a stochastic superoperator that depends on the history of $\Upsilon(t)$, i.e.~on the particular realization of $\Upsilon(t)$. Note that in writing \eqref{LUpsilon}, we have assumed the rules of normal calculus.\footnote{Or in terms of stochastic differential equations, we are assuming $\Upsilon(t)\,dt$ to be of order $dt$ in \eqref{NoisyFNModel} and \eqref{LUpsilon}.} Consequently, this results in a \str\ stochastic differential equation for the density operator when $\Lcal_\Upsilon(t)$ is applied to a quantum state. It is common in stochastic differential equations to define $dW(t) = \Upsilon(t) dt$, and write products containing $dW$ using a $\circ$ for \str\ equations \cite{Eva13,SS19}. We thus write, on using \eqref{LUpsilon},
\begin{align}
\label{StrCohResStochME}
	d\rho_\Upsilon = \Lcal_0 \, \rho_\Upsilon \, dt + i \, \frac{\kappa}{\rt{2}} \, [ \ann+\adg, \rho_\Upsilon ] \circ dW   \; .
\end{align}
Equation \eqref{StrCohResStochME} now describes how the noise-driven quantum \fn\ system evolves for each realization of $\Upsilon(t)$. Its solution $\rho_\Upsilon$ is now stochastic, conditioned on a realization of the white-noise process $\Upsilon(t)$, as we have indicated by its subscript.
\begin{figure*}[t]
\centering
\includegraphics[width=\textwidth]{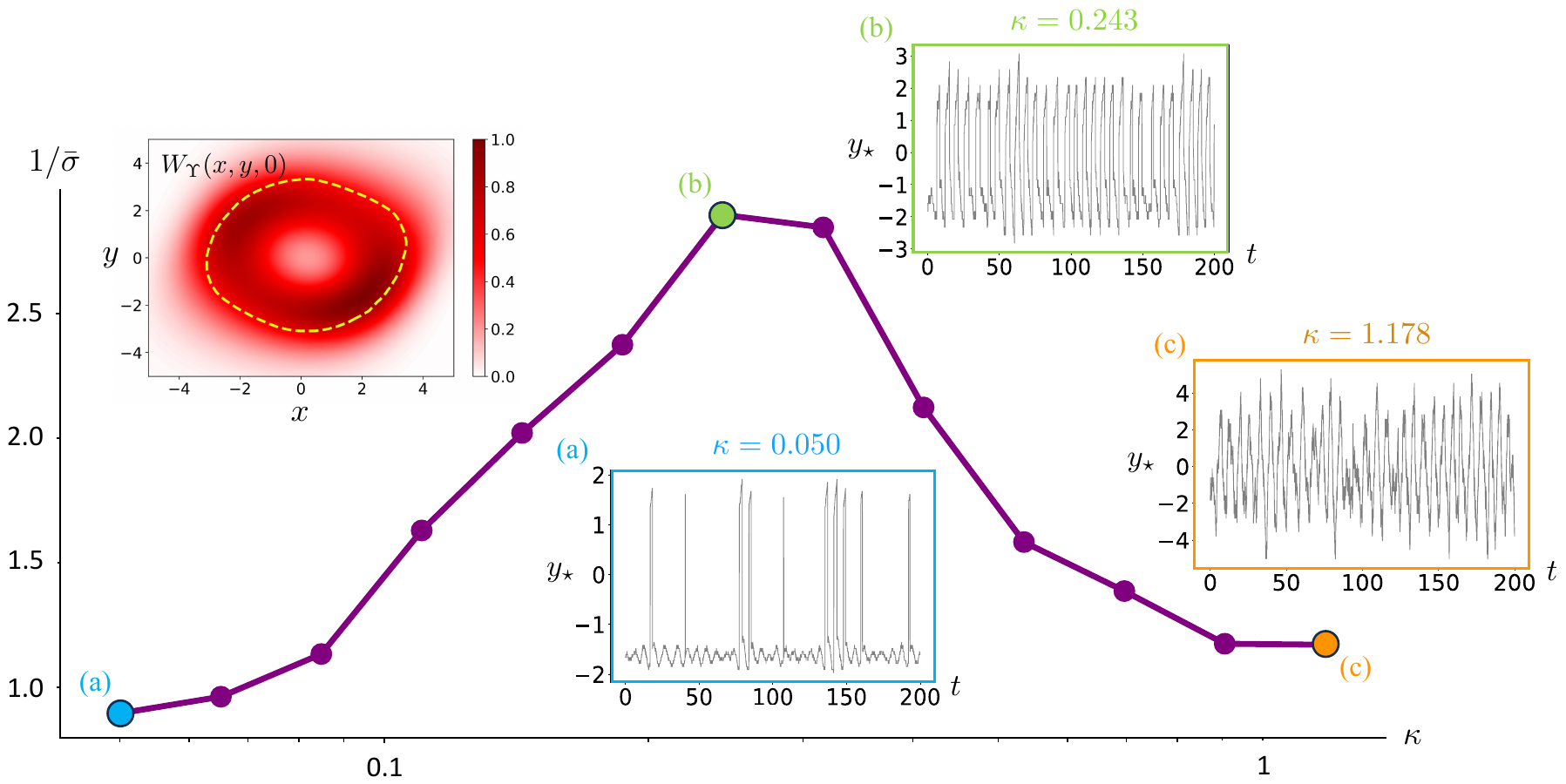}
\caption{\label{CoeffVar}Inverse normalized standard deviation for $Y_\star(t)$ as a function of $\kappa$ (log scale). The initial Wigner function $W_\Upsilon(x,y,0)$ corresponds to a classical limit cycle (yellow dashed line), and is derived from the steady state of $\Lcal_0$ with $\mu=0.2$, $\varepsilon=0.1$, and $x_0=1.2$. We have biased the bottom right peak of $W_\Upsilon(x,y,0)$ to be slightly higher than its top left peak. Plots of $y_\star(t)$ [a realization of $Y_\star(t)$] are shown in (a), (b), and (c) for $\kappa=0.050$, 0.243, and 1.178 respectively. These also correspond to the data points for $1/\bar{\sigma}$ with the same labels and color code. Note however that each data point for $1/\bar{\sigma}$ is obtained from a much longer spike train than that shown in (a)--(c). The ability of noise to elicit quasi-regular oscillations in classical stochastic systems is well documented in the literature \cite{LGONSG04,SSGO07}. The phenomenon is then said to be a resonance effect when there exists an optimal value of the noise intensity for which the stochastic oscillations become most regular \cite{LGONSG04,SSGO07}.}
\end{figure*}

To illustrate the effect of noise, we simulate \eqref{StrCohResStochME} for different values of $\kappa$. The results may be visualized by following the peak of the Wigner function for $\rho_\Upsilon$. This approach is still very much in the spirit of P-bifurcations, except the Wigner function is no longer at steady state. The noise-driven Wigner function is a direct generalization of the deterministic version with the quantum state replaced by $\rho_\Upsilon$:
\begin{align}
\label{StochWigner}
	W_\Upsilon(x,y,t) = \frac{1}{2\pi} \int^\infty_{-\infty} ds \, \bra{x+s/2} \rho_\Upsilon(t) \ket{x-s/2} e^{-i s y}  \, .
\end{align}
We then track the mode of $W_\Upsilon(x,y,t)$ for each $t$, defined by 
\begin{align}
	\big(X_\star(t),Y_\star(t)\big) = \underset{(x,y)}{\text{argmax}} \; W_\Upsilon(x,y,t)  \; .
\end{align}
Note that $X_\star(t)$ and $Y_\star(t)$ are stochastic processes because the state $\rho_\Upsilon$ is stochastic, not because $W_\Upsilon(x,y,t)$ has stochastic arguments [$x$ and $y$ are simply independent variables that label the Wigner function, which can be seen from \eqref{StochWigner}]. The resulting dynamics of $Y_\star(t)$ at low, moderate, and large values of $\kappa$ are shown respectively in Fig.~\ref{CoeffVar}(a), (b), and (c) (with the remainder of Fig.~\ref{CoeffVar} to be explained below). Our initial state is a noise-free steady state obtained under $\Lcal_0$ (see figure caption for parameter values). Its Wigner function $W_\Upsilon(x,y,0)$ is shown in Fig.~\ref{CoeffVar} (top left), which corresponds to a classical limit cycle (yellow dashed line). Note $W_\Upsilon(x,y,0)$ is such that its bottom right peak is slightly higher than the top left peak. The noise-driven quantum system then shows a spike train in $Y_\star(t)$. As can be seen in Fig.~\ref{CoeffVar}(a), when the noise intensity is relatively low, there are fewer spikes, and the spikes are fired at seemingly random times. As we increase the noise intensity (i.e.~$\kappa$) to that seen in Fig.~\ref{CoeffVar}(b), the pulsations appear to be quasi-regular. Increasing the noise intensity further then tends to ruin the regularity of the spike train for which an example is shown in Fig.~\ref{CoeffVar}(c).

To be more precise, a measure of regularity for the spike train is required. For this we use the normalized standard deviation of the interspike interval (also called the coefficient of variation \cite{LGONSG04}): Consider a sample of $N+1$ spikes in a time interval $[0,t]$, occurring at the times $s_1,s_2,\ldots,s_{N+1}$. An interspike interval is defined as the time between two successive spikes. For a sequence of $N+1$ spikes, there are thus $N$ interspike intervals, given by
\begin{align}
	z_k = s_{k+1} - s_{k} \; ,  \quad k=1,2,\ldots,N  \;.
\end{align}
For a sufficiently large sample, i.e.~large $t$, and hence large $N$, we may estimate the standard deviation of the interspike interval $\sigma$ as
\begin{align}
	\sigma = \left[ \frac{1}{N} \sum_{k=1}^N \, (z_k-\nu)^2 \right]^{\frac{1}{2}} \; ,  \quad   \nu = \frac{1}{N} \sum_{k=1}^N z_k  \; .
\end{align}
The normalized standard deviation of the interspike interval is then simply 
\begin{align}
\label{DefnCOV}
	\bar{\sigma} = \frac{\sigma}{\nu}  \; .
\end{align}
If a spike train is exactly periodic then $\bar{\sigma}=0$. If it is instead a Poisson process, then $\bar{\sigma}=1$. We have computed $1/\bar{\sigma}$ for $Y_\star(t)$ at various values of $\kappa$. The results are shown in Fig.~\ref{CoeffVar} as purple dots [joined for ease of visualization, and colored differently for (a), (b), and (c)]. The data indicates a resonance effect of $\Upsilon(t)$ on $Y_\star(t)$. This refers to the existence of an optimal value of $\kappa$ at which $Y_\star(t)$ oscillates most regularly. The same can also be seen in $X_\star(t)$ (the results for which are not shown). Starting at point (a) ($\kappa=0.050$), the spike train can be seen to be more and more periodic as $\kappa$ increases, until we reach (b) ($\kappa=0.243$) when the period is optimized. Adding more noise thereafter only reduces the regularity of the spike train, which can be seen to level off around point (c) ($\kappa=1.178$).

The manner in which $Y_\star(t)$ changes with noise is reminiscent of coherence resonance in excitable systems \cite{LGONSG04}. The effect is well known for classical systems \cite{PK97,LSG00,SH89,GDNH93,DNH94,LNK98}, but has appeared only relatively recently in quantum systems \cite{YXCF18,KN21}. A notable difference between our results in Fig.~\ref{CoeffVar} and coherence resonance is that we have not operated the noise-free quantum \fn\ system in the excitable regime (defined classically by having a stable fixed point in the system while being close to the supercritical Hopf bifurcation). If we define quantum excitability based on the classical theory, then the excitable regime of the noise-free quantum \fn\ model should look something like Fig.~\ref{QFNModel}(a). We have in fact calculated $\bar{\sigma}(\kappa)$ using such a state for $W_\Upsilon(x,y,0)$. The resonance effect of $\Upsilon(t)$ on $Y_\star(t)$ is retained, i.e.~we get a similar $\bar{\sigma}^{-1}(\kappa)$ as in Fig.~\ref{CoeffVar}, but the mode of $W_\Upsilon(x,y,t)$ is also a lot noisier for such a $W_\Upsilon(x,y,0)$, so the spikes are no longer apparent. A more careful study is required to understand the noise-driven quantum \fn\ model in this regime. This is beyond the scope of this paper as our goal here is simply to illustrate the range of systems amenable to cascade quantization. There is however a separate note to this end that is worth mentioning: That is---excitable systems come in different flavors, depending on the bifurcation that gives rise to their excitability. In the case of the \fn\ model, it is a supercritical Hopf bifurcation that facilitates its excitability. Another commonly used mechanism is the infinite-period bifurcation, an example of which was considered in Sec.~\ref{InfPerBif} (and recall that this is the same as a saddle-node bifurcation on an invariant circle). In this case, \eqref{SNICBifx} and \eqref{SNICBify} become excitable for $|\mu|\lesssim1$, i.e.~near the onset of the infinite-period bifurcation. In the classical literature, these two bifurcations---the Hopf and the infinite period---lead to substantially different interspike intervals and the two modes of excitability are classified as type I (saddle-node on invariant circle) and type II (Hopf) \cite{LGONSG04}. Therefore as a byproduct, the Lindbladian in \eqref{SNICBifL} [and \eqref{SNICBifH}] can also be said to quantize a type-I excitable system for an appropriate value of $\mu$. This classification also puts the system in Ref.~\cite{KN21} as type I excitable, but in contrast to \eqref{SNICBifL}, Ref.~\cite{KN21} considers a bistable system instead of a monostable one. It may thus be interesting to also investigate how quantum versions of the different excitable systems respond to noise.

\subsubsection{Average effect of noise}

We can also get an understanding of how $\Upsilon(t)$ affects the system by considering its effect on average. Suppose we are interested in how a function $s(\ann,\adg)$ might change. Then in principle we could use \eqref{StrCohResStochME} to obtain many runs of $\an{s(\ann,\adg)}_\Upsilon=\Tr[s(\ann,\adg)\rho_\Upsilon]$ and calculate their ensemble average ${\rm E}[\an{s(\ann,\adg)}_\Upsilon]$. This approach works, but is rather indirect. It does not provide a simple way to capture the average effect of $\Upsilon$ on the system. Here we derive an equation that propagates the quantum state of the noise-driven system while averaging over $\Upsilon$. That is, we seek the evolution of a state $\rho$ such that, for any $s(\ann,\adg)$ and any time, 
\begin{align}
\label{<s(a,adg)>}
	\Tr\big\{ s(\ann,\adg) \, \rho \big\} = {\rm E}\big[ \an{s(\ann,\adg)}_\Upsilon \big]   \; .
\end{align}
Such a state can be found by noting that
\begin{align}
	{\rm E}\big[ \an{s(\ann,\adg)}_\Upsilon \big] \equiv {}& \int^\infty_{-\infty} d\upsilon \, \wp_{\Upsilon}(\upsilon,t) \, \Tr\big[ s(\ann,\adg) \, \rho_\Upsilon(t) \big]   \nn \\
\label{E[<s(a,adg)>Ups]}
											           = {}& \Tr\bigg[ s(\ann,\adg) \, \int^\infty_{-\infty} d\upsilon \, \wp_{\Upsilon}(\upsilon,t) \, \rho_\Upsilon(t) \bigg]   \; ,
\end{align}
where $\wp_\Upsilon(\upsilon,t)$ is such that $\wp_\Upsilon(\upsilon,t)d\upsilon$ gives the probability that $\Upsilon(t)\in[\upsilon,\upsilon+d\upsilon]$. Comparing \eqref{<s(a,adg)>} to \eqref{E[<s(a,adg)>Ups]}, we see that
\begin{align}
\label{E[rhoUpsilon]}
	\rho(t) = {\rm E}[\rho_\Upsilon(t)] = \int^\infty_{-\infty} d\upsilon \, \rho_\Upsilon(t) \, \wp_{\Upsilon}(\upsilon,t)   \; ,
\end{align}
Equation \eqref{E[rhoUpsilon]} implies that an equation of motion for $\rho(t)$ can be obtained by averaging \eqref{StrCohResStochME}. One way to calculate the ensemble average of \eqref{StrCohResStochME} is to convert it to its \ito\ equivalent first \cite{SS19}, given by 
\begin{align}
\label{ItoCohResStochME}
	d\rho_\Upsilon = {}& \Lcal_0 \, \rho_\Upsilon \, dt + \frac{\kappa^2}{2} \, \Dcal[\ann+\adg] \rho_\Upsilon \, dt    \nn \\ 
	                              & + i \, \frac{\kappa}{\rt{2}} \, [ \ann+\adg, \rho_\Upsilon ] \, dW   \; .
\end{align}
As \eqref{ItoCohResStochME} is now an \ito\ stochastic differential equation, it obeys \ito\ calculus for which $dW$ has zero mean and $(dW)^2=dt$. In particular, $dW$ is independent of $\rho_\Upsilon$ at all times, so the average of \eqref{ItoCohResStochME} can be computed easily using
\begin{align}
	{\rm E}[\rho_\Upsilon \, dW] = {\rm E}[\rho_\Upsilon] \, {\rm E}[dW] = 0   \; . 
\end{align}
We thus have, using \eqref{E[rhoUpsilon]} and \eqref{ItoCohResStochME},
\begin{align}
\label{EnsAvgSME}
	\rho' = \Lcal_0 \, \rho + \frac{\kappa^2}{2} \, \Dcal[\ann+\adg] \rho  \; .
\end{align}
Equation \eqref{EnsAvgSME} is again a master equation in Lindblad form. It shows that averaging over $\Upsilon$ modifies $\Lcal_0$ according to
\begin{align}
\label{EnsAvgL}
	\Lcal_0   \; \longrightarrow \;   \Lcal_\kappa = \Lcal_0 + \frac{\kappa^2}{2} \, \Dcal[\ann+\adg]   \; .
\end{align}
Let us therefore consider the effect of $(\kappa^2/2) \, \Dcal[\ann+\adg]=\kappa^2\,\Dcal[\xhat]$ on the system. Using only $\kappa^2\,\Dcal[\xhat]$ as the generator of time evolution, we find the time-dependent means of $\xhat$ and $\yhat$ to be
\begin{align}
\label{FNMean}
	\an{\xhat}_t = \an{\xhat}_0  \; ,   \quad  \an{\yhat}_t = \an{\yhat}_0   \; .
\end{align}
That is, the noise has no effect on the system's mean motion. The noise does, however, introduce diffusion around the system's mean trajectory. It is straightforward to show that $\kappa^2\Dcal[\xhat]$ gives
\begin{align}
\label{FNVariance}
	\an{(\Delta\xhat)^2}_t = \an{(\Delta\xhat)^2}_0   \; ,    \quad   \an{(\Delta\yhat)^2}_t = \an{(\Delta\yhat)^2}_0 + \kappa^2 \, t   \; ,
\end{align}
where we have defined $\Delta\hat{s}=\hat{s}-\an{\hat{s}}$ for any $\hat{s}$. Hence we find the variance along $x$ to be constant, and independent of the noise. However, the variance along $y$ has acquired a linearly-increasing contribution due to $\Upsilon(t)$.

\section{\lie\ systems}
\label{LienardSys}

\subsection{\lie's theorem}

The study of periodic motion, especially limit cycles, is a major branch of nonlinear dynamics. Such cyclic behavior can be found in a variety of natural and man-made systems, such as the weather, the beating heart, or chemical reactions. Limit-cycle oscillators are also indispensable for the study of synchronization, which examines how multiple such oscillators behave when coupled \cite{PRK01,BJPS09}. A fundamental question in the theory of nonlinear oscillators is how stable limit cycles arise. More precisely, given $(x',y')=\big(f(x,y),g(x,y)\big)$, how should one choose $f(x,y)$ and $g(x,y)$ so that a limit cycle can be obtained? \lie\ provided an answer to this question with the following theorem \cite{Str15}. Consider the system defined by
\begin{align}
\label{LienardEqns}
	x' = y   \; ,  \quad    y' = - \, u(x) - v(x) \, y   \; ,
\end{align}
where $u(x)$ and $v(x)$ are continuously differentiable functions for all $x$. The system \eqref{LienardEqns} then has a unique stable limit cycle around the origin if
\begin{itemize}
	\item[(i)] $u(x) = -u(-x)$\,,
	\item[(ii)] $u(x) > 0 \,, \;  \forall \; x>0$\,,
	\item[(iii)] $v(x)=v(-x)$\,,
\end{itemize}
and if the antiderivative of $v(x)$, i.e.
\begin{align} 
	V(x) = \int^x_0 ds \; v(s)  \; ,
\end{align} 
is such that\footnote{The symbol $\exists\,!$ (used in mathematics and logic), is the quantifier for unique existence. In words, (iv)--(vi) says $V(x)$ is such that it has exactly one positive root $x_0$; is negative for $0<x<x_0$; is positive and monotonically increasing for $x>x_0$.}
\begin{itemize}
	\item[(iv)] $\exists\;\! ! \, x_0>0:$ $V(x_0)=0$\,,
	\item[(v)] $V(x)<0$\,, $\,\forall \; x \in (0,x_0)$\,,
	\item[(vi)] $V(x_2) \ge V(x_1) > 0\,,  \;  \forall \; x_2 > x_1 > x_0$\,.
\end{itemize}

\subsection{A family of quantum limit cycles}

\lie's theorem, coupled with our quantization technique, now permits us to parameterize a family of quantum limit cycles---namely those obeying \eqref{LienardEqns} and for which $u(x)$ and $v(x)$ are polynomial functions satisfying conditions (i)--(vi). Although our quantization method applies to arbitrary polynomials, we shall continue to focus on systems of degree three as we have been doing. One can then check that 
\begin{align}
\label{LienardPoly}
	u(x) = \gamma_3 \, x^3 + \gamma_1 \, x   \; ,  \quad     v(x) = \gamma_2 \, x^2 - \gamma_0     \; ,
\end{align}
where $\gamma_0, \gamma_1, \gamma_2$, and $\gamma_3$ are all positive real numbers, satisfy (i)--(vi). Substituting \eqref{LienardPoly} into \eqref{LienardEqns} and expressing the resulting model in terms of complex variables we get
\begin{align}
\label{3rdOrdLieSys(a,a*)}
	\alpha' = {}& \frac{1}{2} \, \big[ \gamma_0 - i \, (\gamma_1 + 1) \big] \alpha - \frac{1}{2} \, \big[ \gamma_0 + i \, (\gamma_1 - 1) \big] \alphastar      \nn  \\
	                 & - \frac{1}{4} \, (\gamma_2 + i \, \gamma_3) \,  \alpha^3 + \frac{1}{4} \, (\gamma_2 - i \, 3 \gamma_3) \, \alphastar{}^2 \alpha      \nn  \\
	                 & - \frac{1}{4} \, (\gamma_2 + i \, 3 \gamma_3) \, \alphastar \alpha^2 + \frac{1}{4} \, (\gamma_2 - i \, \gamma_3) \,  \alphastar{}^3    \; .
\end{align}
Using our table in Fig.~\ref{QTable}, the following Lindbladian therefore captures a family of quantum limit cycles,
\begin{align}
\label{LienardL}
	\Lcal = {}& - i \, [\Hhat,\supopdot\,] + \gamma_0 \, \Dcal[\adg] + \frac{3\gamma_2}{8} \, \Dcal[\ann^2]    \nn \\
	              & + \frac{\gamma_2}{2} \, \Dcal[\adg\ann-\adg{}^2/2]  \; ,
\end{align}
where
\begin{align}
\label{LienardH}
	\Hhat = {}& \adg \ann + \frac{3\gamma_3}{8} \, \adg{}^2 \, \ann^2   \nn  \\
	                & + \frac{i}{4} \, \big[ \gamma_0 - i (\gamma_1-1) \big] \ann^2 - \frac{i}{4} \, \big[ \gamma_0 + i (\gamma_1-1) \big] \adg{}^2   \nn  \\
	                & - \frac{i}{8} \, (\gamma_2 + i \, 2 \gamma_3) \, \adg \ann^3 + \frac{i}{8} \, (\gamma_2 - i \, 2 \gamma_3) \, \adg{}^3 \ann   \nn  \\
	                & - \frac{i}{16} \, (\gamma_2 + i \, \gamma_3) \, \ann^4 + \frac{i}{16} \, (\gamma_2 - i \, \gamma_3) \, \adg{}^4  \; .
\end{align}

\begin{figure}
\centering
\includegraphics[width=0.49\textwidth]{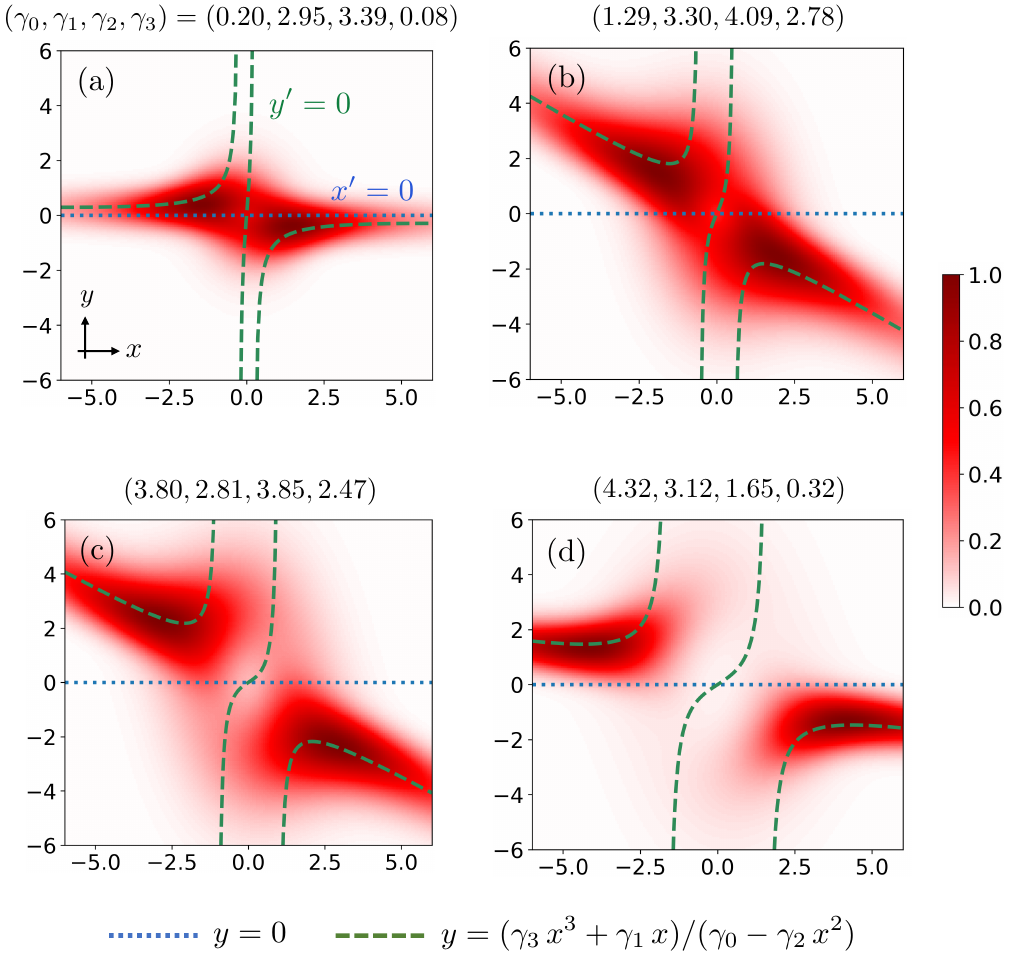}
 \caption{\label{Lienard} Steady-state Wigner functions of \eqref{LienardL} and \eqref{LienardH} for different values of $(\gamma_0,\gamma_1,\gamma_2,\gamma_3)$ (shown on top of each subplot). Since \eqref{3rdOrdLieSys(a,a*)} lacks rotational symmetry [see \eqref{RotSymProc} from Sec.~\ref{RotationalSymmetry}], we expect \eqref{LienardL} and \eqref{LienardH} to generate non-uniform quantum limit cycles, as can be seen in the peaks of the steady-state Wigner functions. 
 To better understand the Wigner functions we have also plotted the $x$ and $y$ nullclines corresponding to \eqref{LienardEqns} and \eqref{LienardPoly}. They are generally useful for explaining the shape and non-uniformity of the classical limit cycle \cite{Str15}, and in our case, particularly the $y$ nullcline (see Fig.~4(b) of Ref.~\cite{CKN20}).}
\end{figure}

An obvious and interesting question is whether there is a quantum Lienard's theorem. If we define quantum limit cycles phenomenologically as we have been doing (i.e.,~by matching where the steady-state Wigner function is peaked to the speed at which the corresponding classical limit cycle is traversed), then a numerical search within the parameter space $\gamma_j\in(0,5]$ ($j = 0,1,2,3$) has not revealed any counterexample for limit cycles described by degree-three polynomials. That is, when \lie's theorem applies to the classical system, we find a limit cycle in the quantized version. Our numerical search identifies a quantum limit cycle by finding the peak position of the steady-state Wigner function and demanding that it has a nonzero distance away from the phase-space origin. Some examples of quantum limit cycles produced by \eqref{LienardL} and \eqref{LienardH} are shown in Fig.~\ref{Lienard}, along with the nullclines of \eqref{LienardEqns} and \eqref{LienardPoly} [the set of $(x,y)$ such that $x'=0$ ($x$ nullcline), and $y'=0$ ($y$ nullcline)]. 

There is yet a more general application of cascade quantization to limit-cycle systems that one could consider. Of course, there is the trivial generalization of letting $u(x)$ and $v(x)$ in \eqref{LienardEqns} be polynomials of degree higher than those in \eqref{LienardPoly}. But a more interesting application is to consider nonlinear oscillators with multiple limit cycles. A generalization of \lie's theorem, sometimes referred to as the \lie--Levinson--Smith theorem \cite{LS42}, provides the conditions for one or more limit cycles to exist in a system of the form\footnote{We refer the reader to the literature (see e.g.~Refs.~\cite{LS42,Mic96}) for more details as we will not be quantizing such a system here.} 
\begin{align}
\label{LLSEqns}
    x' = y \; ,  \quad  y' = - u(x) - v(x,y) \, y   \; .
\end{align}
Equation \eqref{LLSEqns} now generalizes the damping coefficient in \eqref{LienardEqns} to allow for both $x$ and $y$ dependence. In principle cascade quantization can then be used to find a quantum oscillator with multiple limit cycles when $u(x)$ and $v(x,y)$ are polynomials.\footnote{If we restrict to rotationally-symmetric systems then a quantum oscillator with two limit cycles has been studied in Ref.~\cite{KBS25}.} Classical polynomial systems have also been studied in the literature, where issues such as the design of an oscillator with a desired number of limit cycles have been investigated (see e.g.~Ref.~\cite{SGR20}, and other works cited therein). Similar questions could then be addressed for an analogous quantum oscillator using cascade quantization.

\section{Cascade quantization versus variational techniques}
\label{Comparison}

\subsection{Van der Pol oscillator}
\label{VdPComparison}

\subsubsection{Variational quantization}

It is most appropriate to begin comparing our quantization method against the literature by referring to the \vdp\ oscillator. It was defined in \eqref{vdPOsc}, which we recall here,
\begin{align}
\label{DefvdP}
	x' = y \; ,  \quad  y'= - x  - \mu \, ( x^2 - 1 ) \, y   \; ,
\end{align}
where $\mu>0$, is called a nonlinearity parameter. We introduced the \vdp\ oscillator in Sec.~\ref{RotationalSymmetry} as an example of a limit-cycle system without rotational symmetry. An attempt to quantize the \vdp\ oscillator was made in Ref.~\cite{SCVC15}, in which Shah and coworkers generalized Bateman's dual-oscillator paradigm to include nonlinear damping.\footnote{The results in Ref.~\cite{SCVC15} are entirely classical, for which the authors reported the following reasons: (i) Nonlinear and non-Hamiltonian systems are difficult to quantize, and (ii) a working Hamiltonian will most likely be required as a first step towards quantization.} That is, they introduced an ancillary oscillator to solve the Langrangian inverse problem defined by the second-order equation
\begin{align}
\label{Sec2vdP}
	x'' + \mu \, (x^2 - 1)  \, x' + \, x = 0   \; .
\end{align}
For reasons that will soon be clear, we dispense with the term \vdp\ oscillator when referring to the variables $x$ and $y$. Instead, we use the term primary system. And to distinguish the ancillary and primary systems, we shall use, respectively, $q$ and $p$ for the position and momentum of the ancillary oscillator. And like Bateman's theory in Appendix~\ref{BatemanOrigin}, the ancillary oscillator here also satisfies an equation analogous to \eqref{ClassicalAHO}, 
\begin{align}
\label{ShahAncOsc}
 	q'' - \mu \, (x^2-1) \, q' + q = 0   \; .
\end{align}
Hence, the ancillary oscillator is amplified ($x^2-1<0$) or damped ($x^2-1>0$) when the primary oscillator is correspondingly damped or amplified.\footnote{We note that \eqref{Sec2vdP} and \eqref{ShahAncOsc} are in fact special cases of the results in Ref.~\cite{SCVC15}, though this is not important to our critique here.} Most importantly, one can now derive the following Hamiltonian from the solution to the Langrangian inverse problem,
\begin{align}
\label{ShahH}
	H =  y \, p + x \, q + \mu\; ( x^2 - 1 ) \, q \, p  \; .       
\end{align}
This leads to the Hamilton equations 
\begin{align}
\label{xOsc}
	x' = {}& \frac{\partial H}{\partial y} = p  \; ,  \\
\label{yOsc}
	y' = {}& - \frac{\partial H}{\partial x} = - q - 2 \, \mu \, x \, q \, p  \; , \\
\label{qOsc}	
	q' = {}& \frac{\partial H}{\partial p} = y + \mu \, ( x^2 - 1 ) \, q \; ,  \\
\label{pOsc}
	p' =  {}& - \frac{\partial H}{\partial q} = - x - \mu \, ( x^2 - 1 ) \, p \; .
\end{align}
Inspecting \eqref{xOsc}--\eqref{pOsc}, we see that \eqref{DefvdP} is actually reproduced by $x$ (the position of the primary system), and $p$ (the momentum of the ancillary system). This is why we have avoided associating the \vdp\ oscillator to $x$ and $y$ as mentioned underneath \eqref{Sec2vdP}. The Hamiltonian in \eqref{ShahH} has thus embedded the \vdp\ oscillator in a four-dimensional phase space. Of course, there is nothing that forbids one from defining the \vdp\ oscillator using degrees of freedom from different systems. The fundamental problem that prevents \eqref{ShahH} from being useful for quantization is that it generates unstable dynamics, for both the primary and ancillary oscillators. The time-dependent solutions for $y$ and $q$ grow rapidly without bound.\footnote{See (14) and the commentary under (17) in Ref.~\cite{SCVC15} for the actual time dependence. We have also simulated \eqref{xOsc}--\eqref{pOsc} directly to check the unboundedness in $y$ and $q$.} The divergence of the two-oscillator dynamics thus ruins any chance of seeing a limit cycle in a quantum \vdp\ oscillator founded on \eqref{ShahH}. In Appendix~\ref{AppendixVdP} we show that symbolically, \eqref{ShahH} can still lead to some sensible results in the \hei\ picture if one carries out the quantization.

\subsubsection{Cascade quantization}

What we would like is a quantum \vdp\ oscillator based on \eqref{DefvdP}, but free of any instabilities. As we mentioned in Sec.~\ref{RotationalSymmetry}, this problem was solved in Ref.~\cite{CKN20} by the use of a Lindbladian, albeit with a different operator ordering and scaling. This permits the paradigmatic relaxation oscillations and associated limit cycle to be seen in quantum phase space, via the Wigner function at steady state \cite{CKN20}. Such steady-state Wigner functions are qualitatively similar to the probability density of a classical, but stochastic, \vdp\ oscillator \cite{YWMYA13}. However, the quantum \vdp\ oscillator derives its probabilistic nature from quantum mechanics. The classical stochastic \vdp\ oscillator on the other hand, derives its probabilistic nature simply from an external input that is random.

Here we quantize the \vdp\ oscillator as an application of our general method. The \vdp\ oscillator \eqref{DefvdP} is in fact a \lie\ system in the form of \eqref{LienardEqns} and \eqref{LienardPoly}, with the following choice of coefficients,
\begin{align}
\label{vdPgammas}
	\gamma_0 = \gamma_2 = \mu  \; ,  \quad  \gamma_1 = 1  \; ,  \quad  \gamma_3 = 0  \; .
\end{align}
Hence we can simply substitute \eqref{vdPgammas} in \eqref{LienardL} and \eqref{LienardH} to obtain a valid quantization, given by
\begin{align}
\label{LvdP}
	\Lcal = - i \, [\Hhat, \supopdot ] + \mu \, \Dcal[\adg] + \frac{3\mu}{8} \, \Dcal[\ann^2] + \frac{\mu}{2} \, \Dcal[\adg\ann-\adg{}^2/2]  \, ,    
\end{align}
where 
\begin{align}
\label{HvdP}
	\Hhat = {}& \adg \ann + i \, \frac{\mu}{4} (\ann^2 - \adg{}^2) + i \, \frac{\mu}{8} \, (\adg{}^3\ann - \adg \ann^3)   \nn \\
	                & - i \, \frac{\mu}{16} \, (\ann^4 - \adg{}^4)  \; .
\end{align}
Recall that we also stated the \vdp\ oscillator in terms of complex variables in \eqref{a'vdPOsc} from Sec.~\ref{RotationalSymmetry}. As a consistency check, one can look up Fig.~\ref{QTable} for \eqref{a'vdPOsc} and arrive at \eqref{LvdP} and \eqref{HvdP}. We also mentioned in Sec.~\ref{RotationalSymmetry} that the \vdp\ oscillator was first quantized in Ref.~\cite{CKN20}. Interestingly, soon after the publication of Ref.~\cite{CKN20}, an approximate Lindbladian for the \vdp\ oscillator was given by Arosh and collaborators \cite{ACL21}. Their Lindbladian also breaks the rotational symmetry in quantum phase space (necessary for quantizing the \vdp\ oscillator), but is approximate in that it ignores second-order effects in the nonlinearity. The approximation can be seen from the steady-state Wigner functions in Ref.~\cite{ACL21} as $\mu$ is increased. The effects of the symmetry-breaking dissipators in Refs.~\cite{CKN20} and \cite{ACL21} have been studied in detail recently by Sudler and coworkers, especially in the context of quantum synchronization \cite{STB24}.

\subsection{Unusual \lie\ oscillator}

\subsubsection{Variational quantization}

Another system which has received some attention in the literature is an oscillator whose position evolution has the following \lie\ structure \cite{CSL05},
\begin{align}
\label{UnusualOsc2ndOrder}
	x'' + k \, x \, x' + \frac{k^2}{9} \; x^3 + x = 0   \; ,
\end{align}
where $k$ denotes the strength of nonlinearity, and we have again assumed the oscillator to have unit mass and unit frequency. Chandrasekar and coauthors have shown that although \eqref{UnusualOsc2ndOrder} describes a nonlinear oscillator with position-dependent damping, it contains non-isolated closed orbits. Such orbits are more commonly found in conservative systems. This can be seen by defining \eqref{UnusualOsc2ndOrder} to correspond to \cite{CSL05}
\begin{align}
\label{xUnusualOscNonH}
	x' ={}& f(x,y) = y     \; ,    \\
\label{yUnusualOscNonH}
	y' = {}& g(x,y) = - x - k \, x \, y - \frac{k^2}{9} \, x^3     \; .
\end{align}
By \eqref{HamiltonianCondition}, this system is non-conservative, since 
\begin{align}
	\frac{\partial f}{\partial x} + \frac{\partial g}{\partial y} = - k \, x \; .
\end{align}
But yet it has only one fixed point at the origin, which turns out to be a center.\footnote{Nevertheless, there is no contradiction between \eqref{HamiltonianCondition} and \eqref{xUnusualOscNonH}, \eqref{yUnusualOscNonH}, because the origin is a center only locally. That is, not all phase-space points follow non-isolated closed paths around the origin. Some in fact tend to infinity. In short, non-conservative systems, as defined by \eqref{HamiltonianCondition} in Sec.~\ref{Introduction}, need not have attractors.} Furthermore, the non-isolated orbits around the origin have an amplitude-independent frequency. For these reasons, \eqref{UnusualOsc2ndOrder} as been referred to as an unusual \lie-type oscillator \cite{CSL05}.

The quantization of \eqref{UnusualOsc2ndOrder} was then considered using a variational approach \cite{CRSL12}. A Lagrangian for \eqref{UnusualOsc2ndOrder} can be found, and from it, the following Hamiltonian was derived \cite{CSL05},
\begin{align}
\label{UnusualOscHam}
	H = - \frac{9}{k^2} \bigg( 1 - \frac{2k}{3} y \bigg)^{\!\frac{1}{2}} + \frac{1}{2}\,x^2 - \frac{3}{k}\,y - \frac{k}{3} \, x^2 \, y    \; .
\end{align}
Note that \eqref{UnusualOscHam} does not follow the Bateman approach in that it is for a single oscillator, but also not time dependent. However, it is restricted to $y<3/2k$. Keeping this in mind, the Hamilton equations of motion are given by
\begin{align}
\label{xUnusualOsc}
	x' = {}& \frac{\partial H}{\partial y} = \frac{3}{k} \, \bigg( 1 - \frac{2k}{3} \, y \bigg)^{\!-\frac{1}{2}} - \frac{3}{k} - \frac{k}{3} \, x^2    \; ,    \\
\label{yUnusualOsc}
	y' = {}& - \frac{\partial H}{\partial x} = - x + \frac{2k}{3} \, x \, y    \; .
\end{align}
These can be seen to give \eqref{UnusualOsc2ndOrder} after some algebra. The details are included in Appendix~\ref{unLienHeiEOM}. Variational quantization then proceeds by turning \eqref{UnusualOscHam} into an operator. However, as the emphasis of Ref.~\cite{CRSL12} had been on the exact solvability of the associated \sch\ equation, a rather specialized operator ordering for $\xhat$ and $\yhat$ was chosen \cite{CRSL12,vRo83,vRoM85}. Hence for completeness, we have considered the quantum dynamics based on \eqref{UnusualOscHam} under the more conventional Weyl (totally symmetric) ordering in Appendix~\ref{unLienHeiEOM}.

\subsubsection{Cascade quantization}

Ultimately, non-Hamiltonian systems do not yield to variational techniques. Although \eqref{UnusualOscHam} is a time-independent Hamiltonian for a single oscillator, it is restricted to $y<3/2k$. Furthermore, the quantized model in Appendix~\ref{unLienHeiEOM} is an analog of \eqref{xUnusualOsc} and \eqref{yUnusualOsc}, not of \eqref{xUnusualOscNonH} and \eqref{yUnusualOscNonH}. However, the unusual \lie\ oscillator, i.e.~\eqref{xUnusualOscNonH} and \eqref{yUnusualOscNonH}, can be quantized by a Lindbladian. It is straightfoward to show that 
\begin{align}
\label{h(a,a*)Unusual}
	\alpha' = {}& - i  \, (\alpha - \alphastar) - i \, \frac{\lambda}{\sqrt{2}} \, (\alpha + \alphastar)    
	                 - \frac{k}{2\sqrt{2}} \, (\alpha^2 - \alphastar{}^2)    \nn \\
	              & - i \, \frac{k^2}{9\sqrt{2}} \, (\alpha^3 + 3 \, \alphastar \alpha^2 + 3 \, \alphastar{}^2 \alpha + \alphastar{}^3)  \; . 
\end{align}
Reading off the quantization table in Fig.~\ref{QTable}, we find that \eqref{h(a,a*)Unusual} may be quantized by the Lindbladian
\begin{align}
\label{LienardTypeLindbladian}
	\Lcal = - i \, [ \Hhat, \supopdot \,] + \frac{k}{\sqrt{2}} \, \Dcal[\adg\ann-\adg{}]   \; ,
\end{align}
where 
\begin{align}
\label{LienardTypeHamiltonian}
	\Hhat = {}& \bigg( \frac{1+\sqrt{2}}{\sqrt{2}} \bigg) \adg \ann + i \, \frac{k}{2\sqrt{2}} \, (\ann - \adg)   \nn \\
	                & - \bigg( \frac{1-\sqrt{2}}{2\sqrt{2}} \bigg) (\ann^2 + \adg{}^2)  - i \, \frac{k}{6\sqrt{2}} \, (\ann^3 - \adg{}^3)   \nn \\  
	                & + \frac{k^2}{9\sqrt{2}} \, (\adg \ann^3 - \adg{}^3 \ann) + \frac{k^2}{6\sqrt{2}} \, \adg{}^2 \ann^2  \nn \\
	                & + \frac{k^2}{36\sqrt{2}} \, (\ann^4 + \adg{}^4)  \; .
\end{align}
We have therefore quantized the unusual \lie-type oscillator exactly. Furthermore, the generator of time evolution given by \eqref{LienardTypeLindbladian} and \eqref{LienardTypeHamiltonian} applies to all regions of phase space, in contrast to \eqref{UnusualOscHam}.

\section{Cascade quantization versus other non-variational techniques}
\label{NonVariationalMethods}

Relative to variational approaches, there are a lot fewer non-variational techniques, especially ones targeting general systems. Our approach in this work has been to treat the classical model as an open system. This is natural, and not surprisingly, others have also tried to quantize nonlinear systems by following an open-systems approach, though none have prescribed an exact and general way of deriving the quantum generator of time evolution \cite{ACL21,BGR21,LS13,WNB14,HdMPGGZ14} (recall also Sec.~\ref{Gen+Adv}). However, one recent approach applies the framework of operational dynamic modeling \cite{BCLIR12} to search for a Lindbladian consistent with \cite{VZC+18},
\begin{align}
\label{SpecQDynSys}
	\an{\xhat}' = \an{G(\yhat)}   \; ,   \quad    \an{\yhat}' = \an{F(\xhat)}   \; .
\end{align}
The problem defined by \eqref{SpecQDynSys} is similar to ours, except that $G$ and $F$ are functions of one variable.\footnote{A similar technique has been used to quantize a system with a specific $\yhat$ dependence in $\an{\yhat}'$ \cite{BCCMR16}.} In contrast, our approach does not require anything beyond the results of Gorini, Kossakowski, Sudarshan \cite{GKS76}, and Lindblad \cite{Lin76}. Cascade quantization simply outputs an explicit Lindbladian for a wide class of systems using purely algebraic means.

Not all non-variational methods follow an open-systems approach and here we discuss one such method due to Tarasov \cite{Tar01}. The idea of Tarasov is to map a classical generator of time evolution to its quantum analog as a generalized Weyl transform which permits differential operators \cite{Tar01}. For a general two-dimensional system given by $(x',y')=\big(f(x,y),g(x,y)\big)$, we find its generator to be
\begin{align}
	\mathscr{K} = f(x,y) \, \frac{\partial}{\partial x} + g(x,y) \, \frac{\partial}{\partial y}   \; .
\end{align}
That is, the time evolution of an arbitrary function $s(x,y)$, is described by $s'=\mathscr{K}s$. Tarasov's goal is then to map $\mathscr{K}$ to a superoperator $\Kcal$, which governs the dynamics of an arbitrary operator $\hat{s}$, i.e.~$\hat{s}'=\Kcal\,\hat{s}$ \cite{Tar01}. While the method appears to be very general, it does not respect fundamental principles. For example, we can show that Tarasov's procedure fails to ensure $[\xhat,\yhat]=\hat{1}$ at all times. To do this, it is sufficient to consider a simple system. Here we take the example of a damped harmonic oscillator\footnote{This is the same as (30) of Ref.~\cite{Tar01} for $n=1$, with $(q_1,p_1)=(x,y)$, $\beta_{111}=0$, and $\alpha_{11}=\gamma$. The quantized generator is therefore given by (34) of Ref.~\cite{Tar01}.}
\begin{align}
\label{TarDHO}
	x' = y  \; ,  \quad    y' = - x - \gamma \, y   \; , 
\end{align}
where $\gamma$ is the friction coefficient. Its generator of time evolution is thus given by
\begin{align}
	\mathscr{K} = y \, \frac{\partial}{\partial x} - (x + \gamma \, y) \, \frac{\partial}{\partial y}   \; .
\end{align}
Following Tarasov's prescription, the quantized version of $\mathscr{K}$ can be found to be (setting $\hbar=1$ in Ref.~\cite{Tar01}),
\begin{align}
\label{KsupopDHO}
	\Kcal = \frac{i}{2} \, [\yhat^2 + \xhat^2, \supopdot\,] + i \, \frac{\gamma}{2} \big( \yhat \, [\xhat, \supopdot \,] + [\xhat, \supopdot \,] \, \yhat \big)  \; .
\end{align}
Consider now the dynamics of $[\xhat,\yhat]$,
\begin{align}
	[\xhat,\yhat]' = {}& \xhat' \yhat + \xhat \, \yhat' - \yhat' \xhat - \yhat \, \xhat'  \\
\label{K[x,y]}
	                      = {}& [\xhat',\yhat] + [\xhat,\yhat'] = [\Kcal \xhat,\yhat] + [\xhat,\Kcal \yhat]   \;.
\end{align}
It is simple to see that \eqref{KsupopDHO} gives
\begin{align}
\label{(x',y')=(Kx,Ky)}
	\xhat' = \Kcal \xhat = \yhat  \; ,  \quad      \yhat' = \Kcal \yhat = - \xhat - \gamma\,\yhat  \; .
\end{align}
Substituting \eqref{(x',y')=(Kx,Ky)} into \eqref{K[x,y]}, we find the evolution of $[\xhat,\yhat]$ as determined by \eqref{KsupopDHO} to be
\begin{align}
	[\xhat,\yhat]' = - \gamma \, [\xhat,\yhat]  \; .
\end{align}
This gives an exponentially decaying canonical commutator (and thus violates the \hei\ uncertainty principle). We see that although $\Kcal$ produces operator equations similar to \eqref{TarDHO}, which might have appeared promising, it is problematic at a more fundamental level. That $\Kcal$ is unable to preserve the canonical commutator is actually not surprising from an open-systems perspective because noise operators are missing from \eqref{(x',y')=(Kx,Ky)}. Such noise operators arise from coupling the system to a bath, and are necessary for maintaining the correct canonical commutation relation \cite{Car02,GZ04} (see also Appendix~\ref{OpenSystems}).

Equation \eqref{KsupopDHO} can also be seen to be unphysical if we interpret it as the generator of an adjoint master equation \cite{BP02}. In this interpretation, $\Kcal\dg$ becomes the generator for the density operator $\rho$.\footnote{As with operators, given any superoperator $\Kcal$, we define $\Kcal\dg$ by $\langle\hat{A},\Kcal\hat{B}\rangle=\langle\Kcal\dg\hat{A},\hat{B}\rangle$ for any two operators $\hat{A}$ and $\hat{B}$, and inner product $\langle\hat{A},\hat{B}\rangle$. To obtain \eqref{KdagDHO} and \eqref{TarHamDHO} we have used the Hilbert-Schmidt inner product, given by $\langle\hat{A},\hat{B}\rangle=\Tr[\hat{A}\dg\hat{B}]$.} We then find that $\Kcal\dg$ can be written in terms of $\ann$ and $\adg$ as
\begin{align}
\label{KdagDHO}
	\Kcal\dg = - i \, [\Hhat, \supopdot \,] + \frac{\gamma}{2} \, \Dcal[\ann] - \frac{\gamma}{2} \, \Dcal[\adg]  \; ,
\end{align}
where 
\begin{align}
\label{TarHamDHO}
	\Hhat = \adg \ann - i \, \frac{\gamma}{4} \,(\ann^2-\adg{}^2)  \; .
\end{align}
Note that $\Kcal\dg$ is nearly of the Lindblad form, except for the occurrence of a negative Lindblad coefficient in the last term in \eqref{KdagDHO}. Hence \eqref{KdagDHO} is not a physically valid generator of time evolution. It does however, produce the correct quantum dynamics corresponding to \eqref{TarDHO} in expectation value. In particular, we note that \eqref{TarHamDHO} contains a squeezing Hamiltonian (second term). As we discussed in Sec.~\ref{RotationalSymmetry}, this is due to a lack of rotational symmetry in \eqref{TarDHO}.

\section{Cascade quantization: Formulae}
\label{GenSoln}

We stated at the very beginning that cascade quantization covers arbitrary polynomial systems, though we have only considered systems of degree three until now. As promised, this section will provide the most general formulae of cascade quantization, applicable to polynomial systems of any degree. We first explain the main idea behind our approach in Sec.~\ref{QuantStrategy}. This also sets up the notation used in Secs.~\ref{LnEven} and \ref{LnOdd}, where the Lindbladian corresponding to an arbitrary polynomial system is prescribed. The results reported in Secs.~\ref{LnEven} and \ref{LnOdd} are derived in detail in Appendix~\ref{ConsProof}, but we recommend reading Sec.~\ref{SimpleCases} first, where Secs.~\ref{QuantStrategy}--\ref{LnOdd} are illustrated explicitly.

\subsection{General strategy}
\label{QuantStrategy}

We will split the quantization of $\alpha'=h(\alpha,\alphastar)\in\mathbb{P}_m$ for any $m$ into subproblems that require us to quantize polynomials with fewer terms than $h(\alpha,\alphastar)$. To this end, we write a general polynomial of degree $m$ as
\begin{align}
	h(\alpha,\alphastar) = \sum_{n=0}^m \, h_n(\alpha,\alphastar)  \; ,
\end{align}
where $h_n(\alpha,\alphastar)$ is a homogeneous polynomial of degree $n$, i.e.~a linear combination of $\alphastar{}^k \alpha^{n-k}$ for $k=0,1,\ldots,n$. Thus, to quantize a general polynomial system, we only need to quantize an arbitrary homogeneous polynomial. To facilitate the quantization of $h_n(\alpha,\alphastar)$ we futher decompose it as
\begin{align}
\label{hn(a,a*)Sum}
	h_n(\alpha,\alphastar) = \lambda_n \, \alphastar{}^n + (1-\delta_{n,0}) \sum_{k=0}^{K(n)} \, h_{n,k}(\alpha,\alphastar) \; .
\end{align}
Note that for $n=0$ we have $h_0(\alpha,\alphastar)=\lambda_0$, while for $n \ge 1$ we have defined, 
\begin{align}
\label{hnk(a,a*)}
	h_{n,k}(\alpha,\alphastar) = \mu_{n,k} \, \alphastar{}^k \alpha^{n-k} + \nu_{n,k} \, \alphastar{}^{n-k-1} \alpha^{k+1}  \; ,
\end{align}
where $\lambda_n$, $\mu_{n,k}$, and $\nu_{n,k}$ are arbitrary complex constants. The top limit of the sum in \eqref{hn(a,a*)Sum} is defined by
\begin{align}	
\label{K(n)Defn}
	 K(n) = \begin{cases}  \frac{n}{2} - 1  \; ,  \quad    \text{$n$ even}    \\   
	                                   \frac{n-1}{2}  \; ,  \quad   \text{$n$ odd}         \end{cases}  ,
\end{align}
It is straightforward to check that all homogeneous polynomials of degree $n$ are encompassed by \eqref{hn(a,a*)Sum}--\eqref{K(n)Defn}.

We have written $h_n(\alpha,\alpha^*)$ in the form of \eqref{hn(a,a*)Sum} to limit our attention to quantizing $\lambda_n\,\alpha^*{}^n$ and $h_{n,k}(\alpha,\alpha^*)$. It is straightforward to see that for all $n$, the Hamiltonian which quantizes $\lambda_n \alphastar{}^n$ is 
\begin{align}
\label{adgnHamiltonian}
	\Hhat_n = - \frac{i}{n+1} \, (\lambda_n^* \, \ann^{n+1} - \lambda_n \, \adg{}^{n+1})  \; .
\end{align}
The problem of quantizing a general polynomial system now boils down to finding a valid quantization $\Lcal_{n,k}$ for $h_{n,k}(\alpha,\alphastar)$. Since the Lindbladian is a linear superoperator, we may write a valid quantization of $\alpha'=h(\alpha,\alphastar) \in \mathbb{P}_m$ as
\begin{align}
	\Lcal = \sum_{n=0}^m \, \Lcal_n  \; ,
\end{align}
where $\Lcal_n$ quantizes $h_n(\alpha,\alphastar)$, and has the form
\begin{align}
\label{HomPolyL}
	\Lcal_n = -i \,[ \Hhat_n, \supopdot \,] + (1-\delta_{n,0}) \sum_{k=0}^{K(n)} \, \Lcal_{n,k}   \; ,
\end{align}
assuming $\Lcal_{n,k}$ to quantize $h_{n,k}(\alpha,\alphastar)$. To construct $\Lcal_{n,k}$ we first note a general difficulty in constructing Lindbladians. Given a polynomial system specified by $h(\alpha,\alphastar)$, it is not at all straightforward to find a Lindbladian which generates exclusively $\an{\,\normord{h(\ann,\adg)}\,}$. In general it is easier to find a Lindbladian which produces $\an{\,\normord{h(\ann,\adg)}\,}$ plus some additional terms. With this in mind, let us denote by $\Lin_{n,k}$ the Lindbladian which quantizes all the necessary terms in $h_{n,k}(\alpha,\alphastar)$, but which also generates unwanted byproduct terms. That is, $\Lin_{n,k}$ quantizes a system of the form
\begin{align}
\label{<a>=In+Notin}
	\alpha' = h_{n,k}(\alpha,\alphastar) + h^{\notin}_{n,k}(\alpha,\alphastar)    \; ,
\end{align}
where $h^{\notin}_{n,k}(\alpha,\alphastar)$ are the byproduct terms not in $h_{n,k}(\alpha,\alphastar)$. These unwanted terms are always of a lower degree than $h_{n,k}(\alpha,\alphastar)$. This will be seen in Sec.~\ref{SimpleCases} for the case of degree-three systems, and proven in Appendix~\ref{ConsProof} for an arbitrary polynomial system. Assuming that we can quantize these lower-degree polynomials, we can then cancel them off by adding to $\Lin_{n,k}$ another Lindbladian $\Lnot_{n,k}$, which quantizes
\begin{align}
	\alpha' = - \, h^{\notin}_{n,k}(\alpha,\alphastar)     \; .
\end{align}
A valid quantization of $h_{n,k}(\alpha,\alphastar)$ is thus given by 
\begin{align}
\label{L=Lin+Lnot}
	\Lcal_{n,k} = \Lin_{n,k} + \Lnot_{n,k}     \; . 
\end{align}
We can now see a path towards quantizing a system in $\mathbb{P}_m$ for any $m$: First, we can already quantize $\mathbb{P}_3$ with the table in Fig.~\ref{QTable}. Suppose now we want to quantize $\mathbb{P}_4$. This means that we require $\Lin_{4,0}$ and $\Lin_{4,1}$. Since we can quantize $\mathbb{P}_3$,  $\Lnot_{4,0}$ and $\Lnot_{4,1}$ can be assumed to be at hand. Hence we can quantize $\mathbb{P}_4$ provided that we have $\Lin_{4,0}$ and $\Lin_{4,1}$. And if we can quantize $\mathbb{P}_4$, then by the same token we can quantize $\mathbb{P}_5$ if we can find $\Lin_{5,0}$, $\Lin_{5,1}$, and $\Lin_{5,2}$. As with the previous case, having quantized $\mathbb{P}_4$, we can assume to have $\Lnot_{5,0}$, $\Lnot_{5,1}$, and $\Lnot_{5,2}$. Below we provide the explicit forms of $\Lin_{n,k}$ for all values of $n$ and $k$ by considering the cases of even and odd $n$ in turn.

We refer to our method as cascade quantization---since to quantize $\mathbb{P}_m$ requires that we know how to quantize $\mathbb{P}_{m-1}$, and to quantize $\mathbb{P}_{m-1}$ we need to know how to quantize $\mathbb{P}_{m-2}$, and so forth. Therefore $\mathbb{P}_m$ is quantized successively, by tacking on (or cascading) Lindbladians from $\mathbb{P}_{m-1}$.

\subsection{$\Lin_{n,k}$ for even $n$}
\label{LnEven}

We prove in Appendix~\ref{nEvenApp} that for $n\ge2$ (assumed even),
\begin{align}
\label{LnkEven}
	\Lin_{n,k} = - i \, [\Hhat_{n,k}, \supopdot\,] + \Dcal[\hat{c}_{n,k}] + \kappa_{n,k} \, \Dcal[\hat{b}_{n,k}]   \; ,
\end{align}
where we have defined the Hamiltonian,
\begin{align}
\label{EvenHmk}
	\Hhat_{n,k} = {}& i \, \frac{(k+2) \, \mu_{n,k} - (k+1) \, \nu_{n,k}^* }{(n+2)(k+1)} \, \adg{}^{k+1} \, \ann^{n-k}  \\
	                        & - i \, \frac{(k+2) \, \mu^*_{n,k} - (k+1) \, \nu_{n,k}}{(n+2)(k+1)} \, \adg{}^{n-k} \, \ann^{k+1}    \; ,  \nn
\end{align}
and the Lindblad operators and coefficients,
\begin{gather}
	\hat{c}_{n,k} =  \adg{}^{\frac{n}{2}-k-1}  \, \ann^{k+1} + \sigma_{n,k} \, \ann^{\frac{n}{2}+1}   \; ,      \\                   
	\hat{b}_{n,k} = \adg{}^{\frac{n}{2}+1}   \; ,  \quad  \kappa_{n,k} = |\sigma_{n,k}|^2  \; .                                      
\end{gather}
We have defined for ease of writing,
\begin{align}
\label{sigmaEven}
	\sigma_{n,k} = {}& - \frac{2\,[(n-k) \, \mu_{n,k} + (k+1) \, \nu_{n,k}^*]}{(n+2)(k+1)}   \; .
\end{align}
The explicit form of $h^{\notin}_{n,k}(\alpha,\alphastar)$ is provided in Appendix~\ref{nEvenApp}. The important point is that it has degree less than $n$, so it is valid to assume that we have $\Lnot_{n,k}$.

\subsection{$\Lin_{n,k}$ for odd $n$}  
\label{LnOdd}

\subsubsection{$0\le k<K$}
\label{LnOddk<K}

It is shown in Appendix~\ref{nOddk<KApp}, that for $n\ge3$ (assumed odd), and $k=0,1,\ldots,K-1$, 
\begin{align}
\label{LnkOdd1}
	\Lin_{n,k} = {}& - i \, [\Hhat_{n,k}\,, \supopdot\,] + \Dcal[\hat{c}_{n,k}]   \nn \\
	                     & + \kappa^-_{n,k} \, \Dcal[\hat{b}^-_{n,k}] + \kappa^+_{n,k} \, \Dcal[\hat{b}^+_{n,k}]   \; ,
\end{align}
where we have defined the Hamiltonian,
\begin{align}
\label{OddHmk}
	\Hhat_{n,k} = {}& i \, \frac{( \mu_{n,k} - \nu_{n,k}^* )}{n+1}  \, \adg{}^{k+1} \, \ann^{n-k}  \nn \\
	                         & - i \, \frac{( \mu^*_{n,k} - \nu_{n,k} ) }{n+1} \, \adg{}^{n-k} \, \ann^{k+1}   \; ,
\end{align}
and the Lindblad operators and coefficients,
\begin{gather}
\label{cnkOdd}
	\hat{c}_{n,k} = \adg{}^{\frac{n-1}{2}-k} \, \ann^{k+1} + \sigma_{n,k} \, \ann^{\frac{n+1}{2}}    \; ,     \\ 
\label{bnk-Odd}
	\hat{b}^-_{n,k} = \adg{}^{\frac{n+1}{2}} \; ,   \quad   \kappa^-_{n,k} = - \frac{4}{n+1}  \, \theta(-\zeta_{n,k}) \, \zeta_{n,k}    \; ,    \\
\label{bnk+Odd}
	\hat{b}^+_{n,k} = \ann^{\frac{n+1}{2}} \; ,   \quad   \kappa^+_{n,k} = \frac{4}{n+1}  \, \theta(\zeta_{n,k}) \, \zeta_{n,k}    \; .
\end{gather} 
Recall for convenience that we defined the Heaviside step function in \eqref{HeavisideDefn} as
\begin{align}
\label{HeavisideDefnRecalled}
	\theta(x) = \begin{cases} 0 , \; x \le 0   \\
	                                         1 , \; x > 0 \end{cases} .
\end{align}
We have also defined in \eqref{cnkOdd}--\eqref{bnk+Odd} the constants
\begin{align}
\label{sigmaOdd}
	\sigma_{n,k} = {}& -\frac{2\,\big[ (n-k) \mu_{n,k} + (k+1) \nu^*_{n,k} \big]}{(n+1)(k+1)}  \; ,  \\
\label{zetaOdd}
	\zeta_{n,k} = {}& \frac{(n-3)-(n+1) |\sigma_{n,k}|^2}{4} - k  \; .
\end{align}
As with the case of even $n$, we leave the expression for $h^{\notin}_{n,k}(\alpha,\alphastar)$ to Appendix~\ref{nOddk<KApp} and just note here that it is a polynomial of lower degree than $n$. This means that we can again assume to have $\Lnot_{n,k}$.

\subsubsection{$k=K$}
\label{LnOddk=K}

Finally, for $n\ge3$ (assumed odd) and $k=K=(n-1)/2$, we have from Appendix~\ref{nOddk=KApp},
\begin{align}
\label{LbarnK+}
	\Lin_{n,K} = - \, i \, [\Hhat_{n,K}\,, \supopdot\,] + \gamma^-_{n,K} \, \Dcal[\hat{c}^-_{n,K}] + \gamma^+_{n,K} \, \Dcal[\hat{c}^+_{n,K}]   \; ,
\end{align}
where the Hamiltonian is
\begin{align}
\label{HnK}
	\Hhat_{n,K} = - \, \frac{2\, \Im[\epsilon_n]}{n+1} \, \adg{}^{\frac{n+1}{2}} \, \ann^{\frac{n+1}{2}}   \; ,
\end{align}
and the dissipators are defined by
\begin{gather}
\label{cnK+}
	\hat{c}^-_{n,K} = \ann^{\frac{n+1}{2}} \; ,   \quad   \gamma^-_{n,K} = - \, \frac{4\,\Re[\epsilon_n] }{n+1} \, \theta\big(\!-\!\Re[\epsilon_n] \big)    \; ,    \\
\label{cnK-}
	\hat{c}^+_{n,K} = \adg{}^{\frac{n+1}{2}} \; ,   \quad   \gamma^+_{n,K} = \frac{4\,\Re[\epsilon_n]}{n+1} \, \theta\big( \Re[\epsilon_n] \big)    \; .
\end{gather} 
Note in \eqref{HnK}--\eqref{cnK-} we have defined 
\begin{align}
	\epsilon_n = \mu_{n,K} + \nu_{n,K}    \; ,
\end{align}
and recall from \eqref{ReImParts} that we are using $\Re[z]$ and $\Im[z]$ to denote the real and imaginary parts of an arbitrary complex number $z$ respectively. As Appendix~\ref{nOddk=KApp} shows, $h^{\notin}_{n,K}(\alpha,\alphastar)$ has degree less than $n$, which allows us to construct $\Lnot_{n,K}$\,.

\section{Cascade quantization: Examples}
\label{SimpleCases}

We now illustrate how cascade quantization works by considering the four lowest degree polynomials, i.e.~$\mathbb{P}_m$ for $m=0,1,2,3$, considering each case successively. One may also wish to contrast the results here with the table in Fig.~\ref{QTable}.

\subsection{Degree-zero polynomial $(m=0)$}

Here we have the simplest case of a system defined by a complex-valued constant 
\begin{align}
	h(\alpha,\alphastar) = h_0(\alpha,\alphastar) = \lambda_0  \;. 
\end{align}
This corresponds to an arbitrary displacement in quantum phase space which can be generated by a single Hamiltonian:
\begin{align}
	\Hhat_0 = - i \, ( \lambda_0^* \, \ann -  \lambda_0 \, \adg )  \; .
\end{align}
The corresponding Lindbladian is simply
\begin{align}
\label{Lindbladian0}
	\Lcal_0 = - i \, [\Hhat_0, \supopdot \,]    \; .
\end{align}

\subsection{Degree-one polynomial $(m=1)$}

Since we have quantized $\lambda_0$, we only need to focus on a homogeneous polynomial of degree one. Setting $n=1$ in \eqref{hn(a,a*)Sum}--\eqref{K(n)Defn} we have,
\begin{align}
\label{DegOne<a>}
	h_1(\alpha,\alphastar) = \lambda_1 \, \alphastar  + h_{1,0}(\alpha,\alphastar)   \; ,
\end{align}
where 
\begin{align}
	h_{1,0}(\alpha,\alphastar) = \epsilon_1 \, \alpha  \; .
\end{align}
From \eqref{HomPolyL} the quantizing Lindbladian is thus given by
\begin{align}
	\Lcal_1 = - i \, [\Hhat_1, \supopdot\,] + \Lcal_{1,0}  \; ,
\end{align}
for which \eqref{adgnHamiltonian} gives
\begin{align}
\label{HDegOne}
	\Hhat_{1} = - \frac{i}{2} \, ( \lambda_1^* \, \ann^2 -  \lambda_1 \, \adg{}^2 )   \; .
\end{align}
To find $\Lcal_{1,0}$ we can now use Sec.~\ref{LnOddk=K} with $n=1$ and $K=0$. This gives 
\begin{align}
\label{Lindbladian1}
	\Lin_{1,0} = {}& - i \, [\Hhat_{1,0}, \supopdot\,]  + \gamma_{1,0}^- \, \Dcal[\hat{c}_{1,0}^-] + \gamma_{1,0}^+ \, \Dcal[\hat{c}_{1,0}^+]   \; .
\end{align}
where
\begin{align}
	\Hhat_{1,0} = - \Im[\epsilon_1] \, \adg \ann   \; , 
\end{align}
and 
\begin{gather}
\label{DegOneLindblad1}
	\hat{c}_{1,0}^- = \ann  \; ,   \quad   \gamma_{1,0}^- = -2 \, \theta\big(\!-\!\Re[\epsilon_1]\big) \, \Re[\epsilon_1]  \; ,  \\
\label{DegOneLindblad2}	
	 \hat{c}_{1,0}^+ = \adg  \; ,  \quad   \gamma_{1,0}^+ = 2 \, \theta\big(\Re[\epsilon_1]\big) \, \Re[\epsilon_1]   \; .
\end{gather}
Note in this case  $\Lnot_{1,0}$ vanishes so that 
\begin{align}
	\Lcal_{1,0} = \Lin_{1,0}  \; . 
\end{align}
Thus we see that a general degree-one polynomial given by
\begin{align}
\label{hLinear}
	h(\alpha,\alphastar) = h_0(\alpha,\alphastar) + h_1(\alpha,\alphastar)   \; ,
\end{align}
is quantized by the Lindbladian
\begin{align}
	\Lcal = {}& \Lcal_0 + \Lcal_1   \\
	         = {}& - i \, [\Hhat_0 + \Hhat_1, \supopdot\,] - i \, [\Hhat_{1,0}, \supopdot\,]   \nn \\ 
	               & + \gamma_{1,0}^- \, \Dcal[\hat{c}_{1,0}^-] + \gamma_{1,0}^+ \, \Dcal[\hat{c}_{1,0}^+]  \; .
\end{align}
The usefulness of cascade quantization only becomes obvious when quantizing systems of degree two or higher, where \eqref{L=Lin+Lnot} has a nonvanishing $\Lnot_{n,k}$.

\subsection{Degree-two polynomial $(m=2)$}
\label{h2Example}

We now encounter the simplest nonlinear system, given by a quadratic polynomial. As we have already quantized a general degree-one system, it is sufficient to quantize a homogeneous quadratic polynomial here. Setting $n=2$ in \eqref{hn(a,a*)Sum}--\eqref{K(n)Defn} gives
\begin{align}
	h_{2}(\alpha,\alphastar) = \lambda_2 \, \alphastar{}^2 + h_{2,0}(\alpha,\alphastar)  \; ,
\end{align}
with
\begin{align}
	h_{2,0}(\alpha,\alphastar) = \mu_{2,0} \, \alpha^2 + \nu_{2,0} \, \alphastar \alpha   \; . 
\end{align}
Using \eqref{HomPolyL} the Lindbladian corresponding to $h_2(\alpha,\alphastar)$ is thus
\begin{align}
	\Lcal_2 = - i \, [\Hhat_2, \supopdot\,] + \Lcal_{2,0}  \; ,
\end{align}
where \eqref{adgnHamiltonian} gives
\begin{align}
\label{DegTwoH1}
	\Hhat_2 = - \frac{i}{3} \, ( \lambda_2^* \, \ann^3 - \lambda_2 \, \adg{}^3 )  \; .
\end{align}
In contrast to the previous example of a degree-one polynomial, we now have
\begin{align}
	\Lcal_{2,0} = \Lin_{2,0} + \Lnot_{2,0}  \; .
\end{align}
The form of $\Lin_{2,0}$ is given by Sec.~\ref{LnEven} with $n=2$ and $k=0$ as
\begin{align}
	\Lin_{2,0} = - i \, [\Hhat_{2,0}, \supopdot\,] + \Dcal[\hat{c}_{2,0}] + \kappa_{2,0} \, \Dcal[\hat{b}_{2,0}]   \; ,
\end{align}
where
\begin{align}
\label{DegTwoH2}
	\Hhat_{2,0} = -\frac{i}{4} \, \big[ \, (2 \, \mu_{2,0}^*- \nu_{2,0} ) \, \adg{}^2 \ann - (2 \, \mu_{2,0} - \nu_{2,0}^* ) \, \adg \ann^2 \, \big]  \; ,  
\end{align}
and 
\begin{gather}
\label{DegTwoLindblad1}
	\hat{c}_{2,0} = \ann - \frac{1}{2} \, ( 2 \, \mu_{2,0} + \nu_{2,0}^* ) \, \ann^2   \; ,       \\
\label{DegTwoLindblad2}
	\hat{b}_{2,0} = \adg{}^2   \; ,  \quad    \kappa_{2,0} = \frac{1}{4} \, | 2 \, \mu_{2,0} + \nu_{2,0}^* |^2    \; . 
\end{gather}
It is straightforward to see that $\Lin_{2,0}$ generates 
\begin{align}
	\an{\ann}' = \an{\,\normord{h_{2,0}(\ann,\adg)}\,} + \an{\,\normord{h^{\notin}_{2,0}(\ann,\adg)}\,}
\end{align}
for which
\begin{align}
	h^{\notin}_{2,0}(\alpha,\alphastar) = \frac{1}{2} \, \big( | 2 \, \mu_{2,0} + \nu_{2,0}^*|^2 - 1 \big) \, \alpha   \; .
\end{align}
We can now construct $\Lnot_{2,0}$ explicitly using the fact that we know how to quantize any degree-one polynomial. It is simple to see that
\begin{align}
	\Lnot_{2,0} = \eta_2^- \, \Dcal[\hat{L}_2^-] + \eta_2^+ \, \Dcal[\hat{L}_2^+]  \; ,
\end{align}
where we have defined
\begin{align}
\label{L2-}
	\hat{L}_2^- = {}& \adg  \; ,   \\
	\eta_2^- = {}& - \theta\big(1-| 2 \, \mu_{2,0} + \nu_{2,0}^*|^2\big) \, ( | 2 \, \mu_{2,0} + \nu_{2,0}^*|^2 - 1 ) \; ,  
\end{align}
and
\begin{align}
\label{L2+}	 
	 \hat{L}_2^+ = {}& \ann  \; ,  \\
	 \eta_2^+ = {}& \theta\big(| 2 \, \mu_{2,0} + \nu_{2,0}^*|^2 - 1\big) \, ( | 2 \, \mu_{2,0} + \nu_{2,0}^*|^2 - 1 )  \; .
\end{align}
This is the simplest nontrivial example of cascade quantization. Combining our previous results, an arbitrary degree-two system
\begin{align}
\label{hQuadratic}
	h(\alpha,\alphastar) = {}& h_0(\alpha,\alphastar) + h_1(\alpha,\alphastar) + h_2(\alpha,\alphastar)   \; ,
\end{align}
can be quantized by
\begin{align}
	\Lcal = {}& \Lcal_0 + \Lcal_1 + \Lcal_2   \\
		 = {}& - i \, [\Hhat_0+\Hhat_1+\Hhat_2, \supopdot\,] - i \, [\Hhat_{1,0}+\Hhat_{2,0}, \supopdot\,]  \nn  \\
	              & + \gamma_{1,0}^- \, \Dcal[\hat{c}_{1,0}^-]  + \gamma_{1,0}^+ \, \Dcal[\hat{c}_{1,0}^+]   \\
	              & + \Dcal[\hat{c}_{2,0}] + \kappa_{2,0} \, \Dcal[\hat{b}_{2,0}] + \eta_2^- \, \Dcal[\hat{L}_2^-]  + \eta_2^+ \, \Dcal[\hat{L}_2^+]  \; .   \nn
\end{align}

\subsection{Degree-three polynomial $(m=3)$}
\label{h3Example}

Since we have quantized an arbitrary quadratic system, we only have to quantize a homogeneous cubic polynomial in order to quantize any degree-three system. Setting $n=3$ in \eqref{hn(a,a*)Sum}--\eqref{K(n)Defn} we get
\begin{align}
\label{DegThrHomPoly}
	h_3(\alpha,\alphastar) = \lambda_3 \, \alphastar + h_{3,0}(\alpha,\alphastar)+ h_{3,1}(\alpha,\alphastar)   \; ,
\end{align}
in which
\begin{align}
	h_{3,0}(\alpha,\alphastar) = {}& \mu_{3,0} \, \alpha^3  +  \nu_{3,0} \, \alphastar{}^2 \alpha  \; ,   \\
	h_{3,1}(\alpha,\alphastar) = {}& \epsilon_{3} \, \alphastar \, \alpha^2   \; .
\end{align}
The corresponding Lindbladian is thus
\begin{align}
	\Lcal_3 = - i \, [\Hhat_3, \supopdot\,] + \Lcal_{3,0} + \Lcal_{3,1}   \; .
\end{align}
where 
\begin{align}
\label{DegThrH1}
	\Hhat_3 = - \frac{i}{4} \, ( \lambda_3^* \, \ann^4 - \lambda_3 \, \adg{}^4 ) \; .
\end{align}

We consider the quantization of $h_{3,0}(\alpha,\alphastar)$ first. Its Lindbladian can be written as 
\begin{align}
	\Lcal_{3,0} = \Lin_{3,0} + \Lnot_{3,0}  \; .
\end{align}
To find $\Lin_{3,0}$ we can use Sec.~\ref{LnOddk<K} with $n=3$ and $k=0$. This gives 
\begin{align}
	\Lin_{3,0} = {}& - i \, [\Hhat_{3,0}\,, \supopdot\,] + \Dcal[\hat{c}_{3,0}] + \kappa^-_{3,0} \, \Dcal[\hat{b}^-_{3,0}]   \; ,
\end{align}
with the Hamiltonian
\begin{align}
\label{DegThrH3}
	\Hhat_{3,0} = \frac{i}{4} \, \big[ ( \mu_{3,0} - \nu_{3,0}^* ) \, \adg \ann^3 - ( \mu_{3,0}^* - \nu_{3,0} ) \, \adg{}^3 \ann \big]   \; ,
\end{align}
and the Lindblad operators and coefficients
\begin{gather}
\label{DegThrLindblad4}
	\hat{c}_{3,0} = \adg \ann - \frac{1}{2} \, (3 \, \mu_{3,0} + \nu_{3,0}^*) \, \ann^2       \; .   \\
\label{DegThrLindblad5}
	\hat{b}^-_{3,0} = \adg{}^2   \; ,   \quad    \kappa^-_{3,0} = \frac{1}{4} \, |3\,\mu_{3,0} + \nu_{3,0}^*|^2   \; .
\end{gather}
Note that $\kappa^+_{3,0}=0$ because $\zeta_{3,0}$ is negative definite. It can be verified directly that $\Lin_{3,0}$ generates 
\begin{align}
\label{h30Example}
	\an{\ann}' = \an{\,\normord{h_{3,0}(\ann,\adg)}\,} + \an{\,\normord{h^{\notin}_{3,0}(\ann,\adg)}\,}   \, .
\end{align}
where 
\begin{align}
	h^{\notin}_{3,0}(\alpha,\alphastar) = \frac{1}{2} \, \big( |3\, \mu_{3,0} + \nu_{3,0}^*|^2 - 1 \big) \, \alpha   \; .
\end{align}
This now determines the form of $\Lnot_{3,0}$. Having quantized an arbitrary linear system, it is simple to see that
\begin{align}
	\Lnot_{3,0} = \eta_3^- \, \Dcal[\hat{L}_3^-] + \eta_3^+ \, \Dcal[\hat{L}_3^+]   \; ,
\end{align}
with the Lindblad operators and coefficients given by
\begin{align}
\label{DegThrLindblad7}	
	 \hat{L}_3^- = {}& \adg   \; ,  \\ 
	 \eta_3^- = {}& - \theta\big( 1 - |3\,z_{3,0} + z_{3,2}^*|^2 \big) \, \big( |3\,z_{3,0} + z_{3,2}^*|^2 - 1 \big)   \; .
\end{align}
and 
\begin{align}
\label{DegThrLindblad6}
	\hat{L}_3^+ = {}& \ann  \; ,   \\
	\eta_3^+ = {}& \theta\big( |3\,z_{3,0} + z_{3,2}^*|^2 - 1 \big) \, \big( |3\,z_{3,0} + z_{3,2}^*|^2 - 1 \big)    \; .   
\end{align}

Turning now to $h_{3,1}(\alpha,\alphastar)$, we find that it may be quantized by
\begin{align}
	\Lcal_{3,1} = \Lin_{3,1} + \Lnot_{3,1}   \; ,
\end{align}
where $\Lin_{3,1}$ can be found by letting $n=3$ and $K=1$ in Sec.~\ref{LnOddk=K}. This leads to
\begin{align}
\label{Lbar31+}
	\Lin_{3,1} = - \, i \, [\Hhat_{3,1}\,, \supopdot\,] + \gamma^-_{3,1} \, \Dcal[\hat{c}^-_{3,1}] + \gamma^+_{3,1} \, \Dcal[\hat{c}^+_{3,1}]   \; ,
\end{align}
where the Hamiltonian is
\begin{align}
\label{H31}
	\Hhat_{3,1} = - \, \frac{1}{2} \, \Im[\epsilon_3] \, \adg{}^{2} \, \ann^{2}   \; ,
\end{align}
and the dissipators are defined by
\begin{align}
\label{c31+}
	\hat{c}^-_{3,1} = {}& \ann^{2} \; ,   \quad   \gamma^-_{3,1} = - \, \Re[\epsilon_3] \, \theta\big(\!-\!\Re[\epsilon_3] \big)    \; ,    \\
\label{c31-}
	\hat{c}^+_{3,1} = {}& \adg{}^{2} \; ,   \quad   \gamma^+_{3,1} = \Re[\epsilon_3] \, \theta\big( \Re[\epsilon_3] \big)    \; .
\end{align} 
It is then straightforward to show that $\Lin_{3,1}$ generates
\begin{align}
\label{h31Example}
	\an{\ann}' = \an{\,\normord{h_{3,1}(\ann,\adg)}\,} + \an{\,\normord{h^{\notin}_{3,1}(\ann,\adg)}\,}    
\end{align}
where
\begin{align}
	h^{\notin}_{3,1}(\alpha,\alphastar) = 2 \, \theta\big( \Re[\epsilon_3] \big) \, \alpha  \; ,
\end{align}
We should therefore let $\Lnot_{3,1}$ be
\begin{align}
	\Lnot_{3,1} = \eta_3 \, \Dcal[\hat{L}_3]  \; ,
\end{align}
with 
\begin{align}
\label{DegThrLindblad3}
	 \hat{L}_3 = \ann  \;,   \quad  \eta_3 = 4 \, \theta\big(\Re[\epsilon_3]\big) \, \Re[\epsilon_3]   \; .
\end{align}

Finally, a general cubic polynomial, given by 
\begin{align}
	h(\alpha,\alphastar) = {}& h_0(\alpha,\alphastar) + h_1(\alpha,\alphastar) + h_2(\alpha,\alphastar)   \nn \\
	                                      & + h_3(\alpha,\alphastar)   \; .
\end{align}
can now be quantized by combining all our results above to give the Lindbladian,
\begin{align}
\label{Lindbladian3}
	\Lcal = {}& \Lcal_0 + \Lcal_1 + \Lcal_2 + \Lcal_3   \nn \\
		 = {}& - i \, [\Hhat_0+\Hhat_1+\Hhat_2+\Hhat_3, \supopdot\,]  \nn  \\
		       & - i \, [\Hhat_{1,0}+\Hhat_{2,0}+\Hhat_{3,0}, \supopdot\,] - i \, [\Hhat_{3,1}, \supopdot\,]  \nn  \\
	              & + \gamma_{1,0}^- \, \Dcal[\hat{c}_{1,0}^-]  + \gamma_{1,0}^+ \, \Dcal[\hat{c}_{1,0}^+]   \nn  \\
	              & + \Dcal[\hat{c}_{2,0}] + \kappa_{2,0} \, \Dcal[\hat{b}_{2,0}] + \eta_2^- \, \Dcal[\hat{L}_2^-]  + \eta_2^+ \, \Dcal[\hat{L}_2^+]   \nn \\
	              & + \Dcal[\hat{c}_{3,0}] + \kappa^-_{3,0} \, \Dcal[\hat{b}^-_{3,0}]  + \eta_3^- \, \Dcal[\hat{L}_3^-] + \eta_3^+ \, \Dcal[\hat{L}_3^+]  \nn  \\
	              & + \gamma^-_{3,1} \, \Dcal[\hat{c}^-_{3,1}] + \gamma^+_{3,1} \, \Dcal[\hat{c}^+_{3,1}] + \eta_3 \, \Dcal[\hat{L}_3]   \; .
\end{align}
Note that several dissipators in \eqref{Lindbladian3} can be grouped together since they have the same Lindblad operator. The examples above illustrate the essence of cascade quantization and how it works in practice.

\section{Summary, related works, and outlook}
\label{Conclusion}

Our main result in this paper is a procedure that we call cascade quantization. It is an exact and effective means of quantizing nonlinear non-Hamiltonian systems for which the dynamical equations are arbitrary polynomials. Defined as an inverse problem for Lindbladians in Sec.~\ref{ProbDefn}, cascade quantization maps a classical system directly and explicitly to completely-positive and trace-preserving evolutions in the \sch\ picture of quantum theory. To our knowledge, cascade quantization is the first exact analytic result for quantizing a general class of systems while respecting physical requirements. An efficient and user-friendly version of cascade quantization was provided in Sec.~\ref{P3Sys}, in the form of a table (Fig.~\ref{QTable}). There, we also explained the value of cascade quantization by highlighting the issues that make quantizing a nonlinear non-Hamiltonian system difficult. We then illustrated the power of cascade quantization in Secs.~\ref{Bifurcations}--\ref{Comparison}, before delving into its inner workings in Secs.\ref{GenSoln}, \ref{SimpleCases}, and Appendix~\ref{ConsProof}.

For our first application of cascade quantization we considered the normal forms of some important bifurcations in Sec.~\ref{Bifurcations}. Despite being textbook examples, they have not been quantized until now. Particularly noteworthy are the saddle-node and transcritical bifurcations which, though dynamically simple, display Wigner negativity in their quantized forms. We then considered the \fn\ model and the effects of noise in Sec.~\ref{ExcitableSys}. Here we saw how a classical white-noise process can have a resonance effect on the Wigner mode of the quantum \fn\ system. This example illustrates how simple classical stochastic systems may be accounted for within cascade quantization. Another interesting class of nonlinear systems are \lie\ oscillators. By utilizing \lie's theorem and cascade quantization we parameterized an entire family of quantum limit cycles in Sec.~\ref{LienardSys}. This now generalizes the earlier results on the quantum \vdp\ and \ray\ oscillators in Ref.~\cite{CKN20}.

A major competitor to cascade quantization is the variational paradigm reviewed in Appendix~\ref{LiteratureReview}. To highlight the advantages of cascade quantization over variational methods, we examined two examples in Sec.~\ref{Comparison}. These are the \vdp\ and unusual \lie\ oscillators. For completeness, the comparison has also been supplemented by some additional calculations in Appendix~\ref{VarAppNLSys}. Section~\ref{Comparison} also compares cascade quantization to other non-variational methods, and explained the problems of these results. Appendix~\ref{LiteratureReview} and Sec.~\ref{Comparison} help set the context of our main result.

Despite the range of systems amenable to cascade quantization, its scope is not unlimited. One generalization that is obviously desirable is to extend cascade quantization to systems in $\mathbb{R}^{2n}$ for $n\ge2$.\footnote{Note that cascade quantization operates in phase space, which is always $2n$-dimensional.} It is clear from a geometrical perspective that increasing the phase-space dimension increases the range of permissible dynamics. For instance, the Poincar\'{e}--Bendixson theorem implies that strange attractors, which are characteristic of chaotic systems, are only possible for autonomous systems with at least three dimensions. Thus, such a generalization could potentially lead to a new paradigm of quantum chaos. Another useful generalization is to expand the table for efficient cascade quantization in Fig.~\ref{QTable} to include systems in $\mathbb{P}_m$ for $m\ge4$. The subcritical Hopf bifurcation with bounded phase-space dynamics is a rotationally-symmetric example of such a higher-degree system \cite{JLA20}. While the focus of this paper is the quantization of single-particle dynamics in $\mathbb{R}^2$, one can also generalize this to multiple interacting particles, which can be entangled. The quantization of nonlinear (and possibly dissipative) interactions is left as future work.

Other than enlarging the scope of our quantization method, it would also be interesting to investigate in more detail nonclassical effects in the quantized models. It is important to keep in mind that Wigner negativity is only a sufficient condition on nonclassicality, and that even relatively simple dissipators in uncoupled systems can lead to nonclassical effects \cite{CMNK23,MKH20}. An intriguing question, which we leave as an open problem, is to determine the family of classical dynamical systems whose quantum analogs produce non-classical steady states. Beyond purely theoretical interest, this has potential impact on the development of novel quantum technologies. For example, a leading candidate for bosonic quantum error correction relies on dissipative cat qubits, in which bosonic cat states are stabilized by nonlinear engineered two-photon dissipation \cite{CNAA+22,MLA+14,LTP+15}. The corresponding Lindbladian can be derived using our cascade quantization method. Dissipative cat qubits are advantageous due to the strong noise bias (where one type of logical error is exponentially suppressed), which can significantly reduce the physical overhead of quantum error correction \cite{LVP+20,PNP+24}. Dissipative stabilization of other quantum codes, such as the Gottesman-Kitaev-Preskill (GKP) code \cite{GKP01}, have also been proposed very recently \cite{SSL+25,NOBN+24}. In this context, our method provides a general framework for generating new quantum codes, motivated by their classical phase-space dynamics. From a physical perspective, another fascinating application of our method is to study the connection between quantum bifurcations and dissipative quantum phase transitions \cite{CFN+23,MBBC18}. It is also worth noting that the quantum models derived from our method, which describes bosonic systems, can be immediately applied to spin systems via the Holstein--Primakoff transformation \cite{HP40,Ser23}. In contrast to the infinite-dimensional Hilbert space of bosonic systems, the Hilbert space of a spin-$S$ system has $2S+1$ dimensions. The boundedness of the corresponding phase space, together with nonlinear dissipation, can give rise to interesting physical effects such as synchronization phase transitions \cite{MTK24}.

In summary, there has always been a strong curiosity in how concepts in classical nonlinear dynamics would translate to quantum mechanics. Well known examples of this are chaos \cite{Haa00,Wim14}, and stochastic resonance (a close cousin of coherence resonance) \cite{WSB04,LC94,GH96a,GH96b,AMK01,WTB+19,HMB+21,LSY24}. Another, and much more recent example is synchronization \cite{CKN20,HdMPGGZ14,LS13,WNB14,DRSF23}. The most direct path to exploring how a classical nonlinear effect would play out in the quantum world is to quantize a classical model with that effect. This is in fact how the burgeoning field of quantum synchronization began \cite{LS13,WNB14}. Today we find several publications in similar vein, such as the quantum versions of relaxation oscillations \cite{CKN20,ACL21,STB24}, chimera states of partial synchrony \cite{BOZSB15}, amplitude and oscillation death \cite{IK17,BKBB20,BKB21}, Turing instability \cite{KN22}, aging transition \cite{BB23}, and pure noise-induced transitions \cite{CMNK23}. However, as we discussed in Sec.~\ref{RotationalSymmetry}, and with the exception of Refs.~\cite{CKN20,ACL21,STB24}, all these quantum generalizations are based on the rotationally symmetric \ls\ model. Equipped with cascade quantization we can now go beyond rotationally-symmetric models with ease. Moreover, cascade quantization is effective, i.e.~it does not require its user to understand the microscopic derivation of quantum master equations. Therefore we hope, as a consequence of its effectiveness, that cascade quantization will also bridge the gap between quantum dynamicists who are not familiar with open-systems theory, and those who are.

\section*{Acknowledgement}

 The Institute for Quantum Information and Matter is an NSF Physics Frontiers Center. AC and LCK acknowledges support from the Ministry of Education, Singapore and the National Research Foundation, Singapore. CN acknowledges support by the National Research Foundation of Korea (NRF) grant funded by the Korea government (MSIT) (NRF-2022R1F1A1063053) and by the Institute of Information \& Communications Technology Planning \& Evaluation (IITP) grant funded by the Korea government(MSIT) (No. 2022-0-01029). AC would also like to thank Pawe{\l} Kurzy\'{n}ski and Benjamin Stickler for fruitful discussions on classical and quantum nonlinear dynamics. The authors especially thank Ranjith Nair for critically reading the manuscript and suggesting improvements.

\appendix

\section{Literature review}
\label{LiteratureReview}

Arguably the most enduring and influential approach to the quantization of non-Hamiltonian systems is that inspired by Bateman, who, in 1931, proposed two different Lagrangians for the classical damped harmonic oscillator \cite{Bat31}. Assuming the oscillator to have unit mass and damping coefficient $\gamma$, its position $x$ satisfies the second-order differential equation
\begin{align}
\label{ClassicalDHO}
	x'' + \gamma \, x' + \wo^2 \, x = 0  \; .
\end{align}
Here $\wo$ is the undamped (angular) frequency of the oscillator. Bateman's publication of Lagrangians leading to \eqref{ClassicalDHO} seems to have been triggered by a rather innocuous remark (at least measured by our current knowledge of open systems), on the impossibility of dissipative dynamical systems arising from variational principles \cite{Bau31}. He proposed two Lagrangians that ignited an intense discussion. In one version, the Lagrangian involves an ancillary oscillator, and is commonly referred to as the Bateman dual-oscillator, or simply time-independent, model. In the other version, the Lagrangian is time dependent, but does not contain any ancillary system. We will briefly review both approaches below. Bateman's analyses were entirely classical, but spurred a great deal of interest for the corresponding quantum-mechanical description.

\subsection{Origins and early work}
\label{BatemanOrigin}

\subsubsection{Bateman, Caldirola, and Kanai}
\label{BatemanLiterature}

We start with Bateman's dual-oscillator model. Let us denote by $(x,x')$ the variables for the damped harmonic oscillator, and $(q,q')$ the variables for the ancillary system. Bateman's dual Lagrangian may then be stated as\footnote{Bateman originally proposed the Lagrangian $L=x'q'-\gamma\,x'q-\wo^2\,x\,q$ \cite{Bat31}, which omits the $xq'$ term as in \eqref{BatemanLdual}. Its \el\ equation can be easily seen to recover \eqref{ClassicalDHO}. However, the Lagrangian in the form of \eqref{BatemanLdual} is commonly found in more modern literature.} 
\begin{align}
\label{BatemanLdual}
    L = x' \, q' - \frac{\gamma}{2} \, (x'\,q - x \, q') - \wo^2\,x\,q \; .
\end{align}
The Hamiltonian corresponding to this Lagrangian is then defined by the Legendre transform $H=y\,x'+p\,q'-L$, where $y=\partial L/\partial x'$ and $p=\partial L/\partial q'$ are the canonical momenta for $x$ and $q$ respectively. These relations are also used to express $x'$ and $q'$ as functions of $x,y,q,p$. This gives
\begin{align}
\label{BatemanTimeIndepH}
	H = y\,p + \frac{\gamma}{2} \, ( q \, p - x \, y ) + \bigg( \wo^2 - \frac{\gamma^2}{2} \bigg) \, x \, q   \; .
\end{align}
It is then simple to derive Hamilton's equations from \eqref{BatemanTimeIndepH} and see that they are different to
\begin{align}
\label{x'Newton}
	x' = {}& y  \; ,   \\
\label{y'Newton}
	y' = {}& - \gamma \, y - \wo^2 \, x  \; ,
\end{align}
which are based on Newton's second law. Despite \eqref{BatemanTimeIndepH} being inconsistent with \eqref{x'Newton} and \eqref{y'Newton}, one may check that it reproduces \eqref{ClassicalDHO}. Interestingly, the second-order equation for the ancillary system shows that it oscillates with negative damping:
\begin{align}
\label{ClassicalAHO}
	q'' - \gamma \, q' + \wo^2 \, q = 0   \; .
\end{align}
Thus, the coordinate $q$ behaves as a linear amplifier, absorbing the energy lost from the damped oscillator. As is well known, \eqref{ClassicalDHO} and \eqref{ClassicalAHO} may be derived directly from the Lagrangian. But since it is the Hamiltonian that is used for quantization in the literature, and since we are more interested in the first-order dynamics given by $(x',y')$, our primary focus will be on the Hamiltonian.

The Hamiltonian \eqref{BatemanTimeIndepH} was reinvented much later by Morse and Feshbach \cite{MF53}.\footnote{See also Ref.~\cite{Tik78} which refers to \eqref{BatemanTimeIndepH} as Feshbach's Hamiltonian.} The quantization of the damped harmonic oscillator based on \eqref{BatemanTimeIndepH} was considered first by Bopp \cite{Bop73}, and then again by Feshbach and Tikochinsky \cite{FT77,Tik78}. However, in the literature it is the analysis due to Feshbach and Tikochinsky that is most often discussed. In short, Feshbach and Tikochinsky solved the time-independent \sch\ equation governed by the operator form of $H$. The eigenvalues of the quantum Hamiltonian turn out to be complex, and have a real part that is unbounded from below \cite{FT77}. The associated eigenstates are unnormalizable, and thus, not in the Hilbert space of square-integrable functions. We will come back to these problems again in Sec.~\ref{BackToBCK} in light of recent developments.

Bateman noticed that one can eliminate the ancillary oscillator in the above model by noting that $q$ and $x$ are related by $q=x\,\exp(\gamma t)$. This observation led him to propose the time-dependent Lagrangian.\footnote{It is simple to derive \eqref{BatemanLtimedep} by substituting $q=x\,\exp(\gamma\,t)$ into Bateman's original version of the dual Lagrangian, $L=-x'q'+\gamma\,x'q+\wo^2\,x\,q$, and noting that $L$ and $L/2$ have the same \el\ equation. The factor of $1/2$ is chosen so that both $x^2$ and $y^2$ are multiplied by $1/2$ in $H$, as it appears in \eqref{BatemanTimeDepH}.} 
\begin{align}
\label{BatemanLtimedep}
    L =  \frac{e^{\gamma t}}{2} \, ( x'^2 - \wo^2 \, x^2 ) \;.
\end{align}
As before, a Legendre transformation gives a time-dependent Hamiltonian
\begin{align}
\label{BatemanTimeDepH}
	H = \frac{y^2}{2} \, e^{-\gamma\,t} + \frac{\wo^2 \, x^2}{2} \, e^{\gamma\,t}   \; .
\end{align}
Despite the explicit time dependence of $H$ (and subsequently Hamilton's equations), the time-independent second-order equation \eqref{ClassicalDHO} is recovered. The Hamiltonian in \eqref{BatemanTimeDepH} was rediscovered by Caldirola \cite{Cal41} and Kanai \cite{Kan48} in their attempts to quantize the damped harmonic oscillator. A Hamiltonian with a similar form to \eqref{BatemanTimeDepH} was also obtained by Stevens to describe the damped oscillations of electric charges in an LCR circuit \cite{Ste58}. On turning \eqref{BatemanTimeDepH} into an operator for quantization, we no longer have a time-independent \sch\ equation. However, the general time-dependent \sch\ equation can still be solved. Stevens and Kerner independently obtained a discrete set of solutions known in the literature as pseudostationary states, which are normalized for all time \cite{Ste58,Ker58}. In such a state, the mechanical-energy operator $\hat{E}$ (representing the sum of kinetic and potential energies) has the mean
\begin{align}
\label{MechEnergySK}
	\an{\hat{E}\;\!}_n = \frac{\hbar \, \wo^2}{\rt{\wo^2-(\gamma/2)^2}} \, \bigg( n + \frac{1}{2} \bigg) \, e^{-\gamma t}  \; ,
\end{align}
where $n-0,1,2,\ldots$ indexes the pseudostationary state, and we have restored $\hbar$. As can be seen, the mean energy diverges if the oscillator becomes critically damped, i.e.~when $\wo^2=\gamma^2/4$. Recent work has attempted to remedy this issue and we will return to this in Sec.~\ref{BackToBCK}. There has also been a generalization of \eqref{BatemanTimeDepH} to two spatial dimensions, including the solution of the corresponding \sch\ equation in quantum theory \cite{LAG18}.

\subsubsection{Critiques}
\label{CriticismsOfBateman}

Aside from the problems already pointed out above, here we mention two other prominent criticisms associated with \eqref{BatemanTimeIndepH} and \eqref{BatemanTimeDepH}. One major problem is that both Hamiltonians proposed above fail to respect \hei's uncertainty principle for position and momentum. Direct quantum-mechanical calculations have shown that \eqref{BatemanTimeIndepH} leads to an exponentially decaying uncertainty product for the position and canonical momentum operators of the damped oscillator, i.e.~$\xhat$ and $\yhat$. If instead \eqref{BatemanTimeDepH} is used for quantization, then one has to distinguish between the mechanical energy of the damped oscillator from the Hamiltonian. If we take the Hamiltonian simply as the generator of time evolution (i.e.~simply as a mathematical construct) then the physics of the system should refer to the mechanical variables [as in \eqref{MechEnergySK}]. In this case, imposing the canonical commutator $[\hat{x},\hat{y}]=i\,\hbar\,\hat{1}$ leads to an exponentially decaying commutator for position and mechanical momentum, given by $[\hat{x},\hat{x}']=i\,\hbar\,\exp(-\gamma\,t)$.\footnote{This is a simple consequence of the relationship between the canonical and mechanical momentum already present in the classical theory, given by $x'=y\,\exp(-\gamma\,t)$ \cite{Kan48}.} Consequently the uncertainty product for position and mechanical momentum is also exponentially damped. This point appears to have been first highlighted by Brittin a year after Kanai's publication of \eqref{BatemanTimeDepH} \cite{Bri49}.

The alternative of interpreting the Hamiltonian to be physical, as the oscillator's energy, leads to a second critique in the literature. Greenberger has noted that \eqref{BatemanTimeDepH} (or rather the Lagrangian from which it originates to be precise) actually describes an oscillator whose mass is increasing exponentially in time \cite{Gre78}. Classically, \eqref{ClassicalDHO} can be interpreted in two ways---either as an oscillator with constant mass and damping, or as an oscillator with a growing mass but no damping. However, Greenberger has argued that only the time-dependent mass interpretation is correct in quantum theory \cite{Gre78}. The same interpretation of \eqref{BatemanTimeDepH} as a time-dependent mass oscillator was simultaneously advocated by Ray \cite{Ray79}, who also emphasized the difference between standard and non-standard Lagrangians.\footnote{A Lagrangian is standard if it has the form of kinetic energy minus the potential energy. Note there is also an ensuing critique of Ray by Kobe and colleagues in Ref.~\cite{KRS86}.}

Other issues arising from using \eqref{BatemanTimeIndepH} and \eqref{BatemanTimeDepH} for quantization will be deferred to Sec.~\ref{BackToBCK}, where they will be discussed in light of recent developments. For now it suffices to point out that such problems associated with \eqref{BatemanTimeIndepH} and \eqref{BatemanTimeDepH}, and those discussed above, have motivated other phenomenological approaches. A relatively prominent example is the use of state-dependent potentials first proposed by Kostin, which amounts to an effective nonlinear \sch\ equation \cite{Kos72,Kos75,Alb75,Has75}. A similar result as \eqref{MechEnergySK} was also obtained within this paradigm \cite{Has75}. Another approach is based on the observation that a term proportional to $d^{2n}x/dt^{2n}$ appears in the \el\ equation for the system coordinate $x$ if the Lagrangian has a term proportional to $(d^{n}x/dt^n)^2$. Hence, a velocity-dependent force can be obtained if one were to naively set $n=1/2$. Motivated by this, Riewe has generalized Lagrangian and Hamiltonian mechanics to permit fractional derivatives \cite{Rie96,Rie97}.

\subsection{Open systems and related results}
\label{OpenSystems}

What is now known as an open-systems approach to dissipation was not formulated until the early 1960s. By an open-systems approach, we mean coupling the system of interest to infinitely many ancillary degrees of freedom, collectively called the bath (or reservoir), which are then to be eliminated from the model later. Thus, a dash of open-systems thinking was already present in Bateman's dual Hamiltonian, if not earlier.

Sticking to the damped harmonic oscillator as a testbed, Senitzky enunciated how one could arrive at a quantum master equation in Lindblad form [recall \eqref{Lindbladian}] by coupling an undamped system to a bath \cite{Sen60,Sen61}. At around the same time, and pressing on with circuits, Stevens proposed a similar idea by coupling an LC-circuit to a semi-infinite transmission line, which was modelled by chaining units of inductors and capacitors \cite{Ste61}. An important contribution to the quantization literature by Senitzky and Stevens is the preservation of the correct canonical commutation relation in time. They elucidated the role of the bath as a source of noise for the system, and that it was essential for the harmonic oscillator to have the right canonical commutator. Heisenberg's uncertainty principle is thus maintained in their work. Other than the original publications in Refs.~\cite{Sen60,Ste61}, Stevens has also written a reflective comment on this issue \cite{Ste80}.

Given the significance of the Lindblad-form master equation for non-Hamiltonian systems, it is worth pointing out at least one approach in the literature using the Lagrange--Hamilton framework that has acknowledged it. Here we refer to the results of Dekker, who showed how the dynamics of the classical damped harmonic oscillator [given by \eqref{x'Newton} and \eqref{y'Newton}] can be generated on adoption of a complex-valued Hamiltonian \cite{Dek75}. To quantize the damped harmonic oscillator, Dekker then appeals to the classical Liouville equation, except now with a non-Hermitian Hamiltonian. To quantize, Dekker turns the classical Liouville equation into an operator equation (where the classical phase-space density is regarded as the classical analog of the density operator), and imposing canonical commutation relations. The most interesting aspect of Dekker's procedure for us is that it yields a Lindblad-form equation of motion for the density operator of the damped oscillator \cite{Dek79}.\footnote{To preempt our readers who are familiar with quantum-trajectory theory \cite{CSVR89,Car93}, Dekker's approach does not correspond to the non-Hermitian evolution prescribed by the jump trajectories. The evolution under a non-Hermitian Hamiltonian in quantum trajectories is conditioned on null measurements, and therefore does not have a Lindblad form.} Despite this, such a phase-space approach to quantization is not without drawbacks. In view of the results by Senitzky and Stevens, we find Dekker's equation of motion for the density operator only reproduces the case of a zero-temperature bath from an open-systems perspective. Furthermore, this approach is limited to linear systems \cite{Dek79}. A thorough discussion of noise operators in Dekker's approach and its relation to Senitzky's work has been explored in Refs.~\cite{Dek77a,Dek77b}.

\subsection{More recent attempts}

\subsubsection{Generalisations of Hamilton's principle}

One suggestion put forth by the more modern approaches to non-Hamiltonian systems is to modify Hamilton's principle of stationary action. In Ref.~\cite{Gal13}, Galley critiques the subtle pitfall in the conventional formulation of Hamilton's principle, which assumes fixed end points for all varied paths in generalized coordinates. This amounts to imposing boundary conditions on the dynamical system which are actually solved with initial conditions. In particular dissipative systems must be specified as initial-value problems. Galley thus formulates what he referred to as Hamilton's principle with initial data \cite{Gal13}. To do so, the number of generalized coordinates and velocities is doubled. A new action integral is then introduced, which defines a new Lagrangian (now a function of the doubled coordinates and velocities). The end result being new \el\ equations containing nonconservative forces. Equipped with the new formalism, Mart\'{i}nez-P\'{e}rez and Ram\'{i}rez then considered the application of Noether's theorem to dissipative systems \cite{MPR18}. At around the same time as Galley, a formulation of Hamilton's principle as an initial-value problem was also proposed by Polonyi, but based on ideas from quantum field theory \cite{Pol14}. The Bateman Lagrangian from Sec.~\ref{BatemanLiterature} can then be shown to be consistent with Galley's approach \cite{MPR18}. That is, the Bateman Lagrangian has the form stipulated by Galley, and not surprisingly, \eqref{ClassicalDHO} and \eqref{ClassicalAHO} can also be derived using Galley's new \el\ equations. Interestingly, Bateman's dual Langrangian has motivated another modification of the principle of stationary action to include multiple degrees of freedom and nonlinear systems \cite{Lan21}.

There has also been an extension of the principle of least action in Riewe's fractional-calculus version of variational mechanics that we mentioned at the end of Sec.~\ref{CriticismsOfBateman}. In particular, Lazo and Kumreich fixed certain mathematical inconsistencies in Riewe's work and proposed a new principle of stationary action which generalizes Riewe's original version \cite{LK14}.

\subsubsection{Back to Bateman, Caldirola, and Kanai}
\label{BackToBCK}

Instead of tweaking fundamental principles as above, other authors have stayed with the programme initiated by Bateman, Caldirola, and Kanai, where one tries to find a working Hamiltonian by first finding a Lagrangian. Some recent work in the latter category has returned to Bateman's dual-oscillator model \eqref{BatemanTimeIndepH} \cite{DFN19,BGR19,DF19,BGR20,Bag20}.

We said in Sec.~\ref{BatemanLiterature} that the quantum Hamiltonian analyzed by Feshbach and Tikochinsky has complex eigenvalues. A detailed study of the spectral properties of this Hamiltonian was undertaken by Chru\'{s}ci\'{n}ski and Jurkowski \cite{CJ06}. It is beyond the scope of our paper to delve into the mathematics of linear operators on infinite-dimensional Hilbert spaces, but since a Hermitian operator with complex eigenvalues might appear paradoxical to physicists, we need to at least mention that an arbitrary operator $\hat{A}$ has real eigenvalues if and only if it is self-adjoint, and that self-adjoint operators are only a subset of Hermitian operators (i.e.~self-adjointness implies Hermiticity but not vice versa) \cite{Gie00}.\footnote{An operator $\hat{A}$ is defined by both its domain $\mathbb{D}(\hat{A})$, i.e.~the set of states $\hat{A}$ is prescribed to act on, in addition to its action on such states. Hence, two operators $\hat{A}$ and $\hat{B}$ are defined to be equivalent, written as $\hat{A}=\hat{B}$, if and only if $\mathbb{D}(\hat{A})=\mathbb{D}(\hat{B})$, and provided they act identically on any $\ket{\psi}\in\mathbb{D}(\hat{A})$. By the same token, $\hat{A}=\hat{A}\dg$, requires $\hat{A}$ and $\hat{A}\dg$ to have same domain and the same action on their domain. Note the latter condition defines $\hat{A}$ to be Hermitian, and in physics, it is usually taken to be the same as self-adjointness. But mathematically, self-adjointness is a stronger condition than Hermiticity (since it also requires $\hat{A}$ and $\hat{A}\dg$ to have the same domain). An example of an operator in quantum mechanics that is Hermitian but not self-adjoint is the momentum operator of a particle confined to the interval $[0,1]$ by an infinite potential function \cite{Gie00}.} The properties of self-adjointness and Hermiticity are equivalent only for operators acting on a finite-dimensional Hilbert space (which is not the case here). The eigenvalue problem solved by Feshbach and Tikochinsky has been revisited again recently, giving rise to a debate \cite{DFN19,BGR19,DF19,BGR20,Bag20}.\footnote{In a closely related paper, Zhu and Klauder investigated the occurrence of non-self-adjoint Hamiltonians in the quantization of nonlinear Hamiltonian systems \cite{ZK92}. By way of case studies, they examined when a prescribed classical dynamical system will lead to a self-adjoint quantum Hamiltonian, and if not, whether it could be made self-adjoint \cite{ZK92}.} Using a Bogoliubov-like transformation of the usual bosonic annihilation and creation operators, Deguchi and coworkers claim to have found a normalizable vacuum state corresponding to Bateman's dual-oscillator system \cite{DFN19,DF19}. From this, they construct a complete set of number states for the quantum damped harmonic oscillator by acting on the new vacuum state with their transformed creation operator. However, in a counterclaim, Bagarello and colleagues proved that a nonzero vacuum state for the transformed annihilation and creation operators of Refs.~\cite{DFN19,DF19} cannot exist \cite{BGR19,BGR20,Bag20}.

The Hamiltonians \eqref{BatemanTimeIndepH} and \eqref{BatemanTimeDepH} have also inspired other new results. By starting with the Bateman dual Lagrangian, Deguchi and Fujiwara introduced new ancillary variables to the original damped-amplified variables \cite{DF20}. From this they obtained a Hamiltonian that had a similar form as \eqref{BatemanTimeDepH}, but time independent. Upon quantization, the eigenvalues of the oscillator's mechanical energy operator $\hat{E}$ were shown to be
\begin{align}
\label{MechEnergyDF}
	E_n(t) =  \hbar \, \omega_0 \bigg( n + \frac{1}{2} \bigg) \, e^{-\gamma\,t}  \; ,  \quad n=0,1,2,\ldots \; .
\end{align}
This expression should be compared to \eqref{MechEnergySK}, where the classical condition for critical damping, i.e.~$\gamma=2\,\wo$, leads to infinite energy in the oscillator. This divergence in energy is avoided in Ref.~\cite{DF20}. Deguchi and Fujiwara identified a different condition for critical damping in the quantum oscillator, given by $\gamma=(\rt{5}-1)\,\wo$ \cite{DF20}. Soon after, these results were reconsidered by Blacker and Tilbrook \cite{BT21}. They proposed a time-dependent Hamiltonian but without any ancillary variables. Their model shares similar features to Ref.~\cite{DF20}, such as the energy \eqref{MechEnergyDF}, but reproduced the classical condition for critical damping.

Very recently, Javier Valdez and collaborators re-examined the Bateman dual-oscillator model using a semiclassical theory known as momentous quantum mechanics \cite{JVHHCA23}. This framework prescribes classical equations of motion for the moments of phase-space variables that approximate quantum dynamics. Such equations were derived for Bateman's dual-oscillator system and the resulting physics was compared to the usual Lindbladian approach with a thermal bath \cite{JVHHCA23}. The average energy of the damped oscillator from momentous quantum mechanics was shown to be equivalent to that obtained from the Lindblad equation provided the bath has zero temperature.

\section{Variational quantization of nonlinear systems}
\label{VarAppNLSys}

\subsection{Van der Pol oscillator}
\label{AppendixVdP}

\subsubsection{Classical mechanics}

As the variational approach to quantization relies on classical mechanics, we first recall the dual-oscillator Hamiltonian for the \vdp\ oscillator proposed by Shah and colleagues in Ref.~\cite{SCVC15}:
\begin{align}
\label{ShahHAppendix}
	H =  y \, p + x \, q + \mu\; ( x^2 - 1 ) \, q \, p   \; ,       
\end{align}
where the free-oscillator angular frequency has been set to one. We shall also find it convenient to recall from the main text the Hamilton equations generated by \eqref{ShahH}:
\begin{align}
\label{xOscAppendix}
	x' = {}& \frac{\partial H}{\partial y} = p  \; ,  \\
\label{yOscAppendix}
	y' = {}& - \frac{\partial H}{\partial x} = - q - 2 \, \mu \, x \, q \, p  \; , \\
\label{qOscAppendix}	
	q' = {}& \frac{\partial H}{\partial p} = y + \mu \, ( x^2 - 1 ) \, q \; ,  \\
\label{pOscAppendix}
	p' =  {}& - \frac{\partial H}{\partial q} = - x - \mu \, ( x^2 - 1 ) \, p \; .
\end{align}
Clearly, taking the derivative of \eqref{xOscAppendix} and then using \eqref{pOscAppendix} we recover 
\begin{align}
	x'' = - x - \mu \, (x^2 - 1) \, x'  \; .
\end{align}

\subsubsection{Quantum mechanics}

To quantize this system we need to first turn $H$ into a Hermitian operator. This entails that we introduce the Hilbert spaces $\mathbb{H}_{\rm pri}$ and $\mathbb{H}_{\rm anc}$ for the primary and ancillary operators respectively. We then turn $(x,y)$ into $(\xhat,\yhat)$, and $(q,p)$ into $(\qhat,\phat)$, which are operators acting on $\mathbb{H}_{\rm pri}$ and $\mathbb{H}_{\rm anc}$ respectively. As $(x,y)$ and $(q,p)$ are canonically conjugate pairs, we demand that in quantum theory they satisfy, for $\hbar\equiv1$,
\begin{align}
\label{[x,y]}
	[\xhat,\yhat] = i \, \hat{1}  \; ,  \quad        [\qhat,\phat\,] = i \, \hat{1}   \; ,
\end{align}
while all other commutators vanish,
\begin{align}
\label{[x,q]}
	[\xhat,\qhat] = [\xhat,\phat] = 0  \; ,   \quad    [\yhat,\qhat] = [\yhat,\phat] = 0   \; .
\end{align}
The form of \eqref{ShahH} then suggests the following Hamiltonian operator,
\begin{align}
\label{ShahHQ}
	\hat{H} = \yhat \, \phat + \xhat \, \qhat + \frac{\mu}{2} \, ( \xhat^2 - \hat{1} ) \, ( \qhat \, \phat + \phat \, \qhat )  \; .
\end{align}
To ensure $\Hhat$ is Hermitian, we have symmetrized the product of $\qhat$ and $\phat$. We have also omitted the symbol for tensor products. An operator like $\xhat\otimes\hat{1}$, or $\hat{1}\otimes\qhat$, will respectively be denoted simply as $\xhat$, or $\qhat$. To see that \eqref{ShahHQ} makes sense we show that it leads to quantum analogs of \eqref{xOscAppendix}--\eqref{pOscAppendix}. Using \eqref{ShahHQ} we find the following \hei\ equations of motion on $\mathbb{H}_{\rm pri}\otimes\mathbb{H}_{\rm anc}$\,:
\begin{align}
\label{xOscQ}
	\xhat' = {}& i \, [ \hat{H} , \xhat] =  \phat  \; ,  \\
\label{yOscQ}
	\yhat' = {}& i \, [ \hat{H} , \yhat] =  - \qhat -  \mu \, \xhat \, ( \, \qhat \, \phat + \phat \, \qhat )  \; ,  \\
\label{qOscQ}	
	\qhat' = {}& i \, [ \hat{H} , \qhat] = \yhat  + \mu \, ( \xhat^2 - \hat{1} ) \, \qhat \; ,  \\
\label{pOscQ}
	\phat' = {}& i \, [ \hat{H} , \phat] =  - \xhat - \mu \, ( \xhat^2 - \hat{1} ) \, \phat \; .
\end{align}
As with the classical formulation, the \vdp\ oscillator is actually captured by \eqref{xOscQ} and \eqref{pOscQ}. Differentiating \eqref{xOscQ} and using \eqref{pOscQ}, we arrive at the quantized version of $x'' + \mu \, (x^2 - 1) \, x' + x = 0$,
\begin{align}
\label{Sec2vdPQ}
	\xhat'' + \mu \, (\xhat^2 - \hat{1})  \, \xhat' + \xhat = 0   \; .
\end{align}
Again, this is in fact an operator equation on $\mathbb{H}_{\rm pri}\otimes\mathbb{H}_{\rm anc}$. It can be shown explicitly that \eqref{xOscQ}--\eqref{pOscQ} preserves the canonical commutation relations in \eqref{[x,y]} and \eqref{[x,q]}. Assuming that \eqref{[x,y]} and \eqref{[x,q]} hold initially at time $t$, then an infinitesimal interval $dt$ later, we find
\begin{align}
	[\xhat(t+dt),\yhat(t+dt)] = {}& [\xhat(t)+\xhat'(t)\,dt, \yhat(t)+\yhat'(t)\,dt]  \\
	                                        = {}& [\xhat(t), \yhat(t)] + [\xhat(t), \yhat'(t)] \, dt  \nn \\
	                                              & + [\xhat'(t), \yhat(t)] \, dt   \, ,      
\end{align}
where we have neglected terms on the order of $dt^2$. Using \eqref{xOscQ} and \eqref{yOscQ} we find 
\begin{align}
	[\xhat'(t), \yhat(t)] = [\xhat(t),\yhat'(t)] = 0 \; ,
\end{align}
since $\yhat(t)$ commutes with $\phat(t)$, and $\xhat(t)$ commutes with $\qhat(t)$ and $\phat(t)$. Therefore we have
\begin{align}
	[\xhat(t+dt),\yhat(t+dt)] = [\xhat(t), \yhat(t)] = i \, \hat{1}  \; .
\end{align}
The preservation of $[\qhat, \phat] = i \, \hat{1}$ can also be shown in a similar fashion. Next we consider \eqref{[x,q]},
\begin{align}
	[\xhat(t+dt), \qhat(t+dt)] = {}& [\xhat(t)+\xhat'(t) \, dt, \qhat(t)+\qhat'(t) \, dt]  \\
	                            = {}& [\xhat(t),\qhat(t)] + [\xhat(t), \qhat'(t)] \, dt  \nn \\
	                                  & + [\xhat'(t), \qhat(t)] \, dt   \; ,     
\end{align}
where we have again dropped terms on the order of $dt^2$. Substituting in \eqref{xOscQ} and \eqref{qOscQ}, we find
\begin{align}
	[\xhat(t), \qhat'(t)] + [\xhat'(t), \qhat(t)] = {}& [\xhat(t), \yhat(t)] + [\phat(t), \qhat(t)] = 0   \; .                                     
\end{align}
Hence we have shown
\begin{align}
	[\xhat(t+dt),\qhat(t+dt)] = [\xhat(t), \qhat(t)] = 0  \; .
\end{align}
The preservation of the remaining commutators in \eqref{[x,q]} can also be demonstrated by a similar means.

While the above results seem quite sensible, what is ultimately of interest in quantum mechanics, is how a given $\rho(0) \in \mathbb{V}(\mathbb{H}_{\rm pri}\otimes\mathbb{H}_{\rm anc})$ would appear at some future time $t$. Recall from Sec.~\ref{LindbladianDefn} that we have defined $\mathbb{V}(\mathbb{H})$ to be the set of all valid density operators on $\mathbb{H}$. How states evolve in the \sch\ picture is also an important practical question, since it is generally difficult, if not impossible to solve the \hei\ equations of motion. Thus we are interested in 
\begin{align}
\label{rho(t)Shah}
	\rho(t) = e^{-i\Hhat t} \, \rho(0) \, e^{i\Hhat t}  \; .
\end{align}
However, the instabilities in the original Hamiltonian discussed in Sec.~\ref{VdPComparison} make simulations of \eqref{rho(t)Shah} virtually impossible. Furthermore, neither the ancillary or primary systems can be traced out to provide a simpler description. That is, suppose we only cared about the position of the \vdp\ oscillator. Then only density operators in $\mathbb{V}(\mathbb{H}_{\rm pri})$ are required. But since such states are given by a partial trace of \eqref{rho(t)Shah} over $\mathbb{H}_{\rm anc}$, there is still no possibility of extracting the quantum dynamics of just one degree of freedom.

\subsection{Unusual \lie\ oscillator}
\label{unLienHeiEOM}

\subsubsection{Classical mechanics}

As with the \vdp\ oscillator, it helps to first work out the classical mechanics of the unusual Li\'{e}nard oscillator before we quantize it within the variational approach. Recall the classical Hamiltonian for the \lie-type oscillator was given in \eqref{UnusualOscHam},
\begin{align}
\label{UnusualOscHamApp}
	H = - \frac{9}{k^2} \, S^{\frac{1}{2}}(y)  + \frac{1}{2}\,x^2 - \frac{3}{k}\,y - \frac{k}{3} \, x^2 \, y  \; ,
\end{align}
where for ease of writing we have defined
\begin{align}
\label{Sbeta(y)}
	S^\beta(y) = \bigg( 1 - \frac{2k}{3} \, y \bigg)^\beta \; .
\end{align}
The associated Hamilton equations of motion are then,
\begin{align}
\label{xUnusualOscApp}
	x' = {}& \frac{\partial H}{\partial y} = \frac{3}{k} \, S^{-\frac{1}{2}}(y) - \frac{3}{k} - \frac{k}{3} \, x^2 \; ,  \\
\label{yUnusualOscApp}
	y' = {}& - \frac{\partial H}{\partial x} = - x + \frac{2k}{3} \, x \, y  \; .
\end{align}
Note that \eqref{xUnusualOscApp} and \eqref{yUnusualOscApp} reproduce harmonic-oscillator dynamics (i.e.~$x'=y$ and $y'=-x$) in the limit of $k\longrightarrow0$, as they should. We can also show that \eqref{xUnusualOscApp} and \eqref{yUnusualOscApp} decouple to the correct second-order equation for the unusual \lie\ oscillator. Differentiating \eqref{xUnusualOscApp} with respect to time we get 
\begin{align}
\label{x''Step}
	x'' = S^{-\frac{3}{2}}(y) \, y' - \frac{2k}{3} \, x \, x'   \; .
\end{align}
We can now use \eqref{xUnusualOscApp} and \eqref{yUnusualOscApp} to simplify \eqref{x''Step}, giving
\begin{align}
	x'' = {}& - 3\, x \, S^{\!-\frac{1}{2}}(y) + 2 \, x +  \frac{2k^2}{9} \, x^3  \\
	      = {}& - 3\, x \bigg( 1 + \frac{k}{3} \, x' + \frac{k^2}{9} x^2 \bigg)+ 2 \, x +  \frac{2k^2}{9} \, x^3  \\
\label{x''HamEOM}
	      = {}& - k \, x \,x' - \frac{k^2}{9} \, x - x   \; .
\end{align}

\subsubsection{Quantum mechanics}

To obtain the quantum Hamiltonian based on \eqref{UnusualOscHamApp} we first let $(x,y) \longrightarrow (\xhat,\yhat)$ where $[\xhat,\yhat]=i\hat{1}$. To ensure Hermiticity we then impose Weyl ordering on $\xhat$ and $\yhat$ to arrive at
\begin{align}
\label{HunsualLienardApp} 
	\Hhat = {}& - \frac{9}{k^2} \, S^{\frac{1}{2}}(\yhat) + \frac{1}{2}\,\xhat^2 - \frac{3}{k}\,\yhat  \nn \\
			  & - \frac{k}{9} \, \big( \xhat^2 \, \yhat + \xhat \, \yhat \, \xhat + \yhat \, \xhat^2 \big)  \, .
\end{align}
Note that $S^\beta(\yhat)$ is defined by letting $1\longrightarrow\hat{1}$ and $y\longrightarrow\yhat$ in \eqref{Sbeta(y)}. The \hei\ equation of motion for $\xhat$ is then
\begin{align}
	\xhat' = i \, [\Hhat, \xhat ]   
\label{xLienardHeiEOM1}
	          = {}& - i \, \frac{9}{k^2} \, \big[ S^{\frac{1}{2}}(\yhat), \xhat \big] - i \, \frac{3}{k} \, \big[ \yhat, \xhat \big]    \nn \\
	              & - i \, \frac{k}{9} \, \big[ \xhat^2 \, \yhat + \xhat \, \yhat \, \xhat + \yhat \, \xhat^2, \xhat \big] \;.
\end{align}
The first commutator with the square root may be computed by applying 
\begin{align}
\label{[F(A),B]}
	[F(\hat{A}),\hat{B}]= [\hat{A},\hat{B}] \, \frac{dF}{d\hat{A}}   \; .
\end{align}
This identity is valid so long as $\hat{A}$ and $\hat{B}$ are such that $\big[\hat{A}, [\hat{A},\hat{B}]\big]=\big[\hat{B}, [\hat{A},\hat{B}]\big]=0$, and for which $F(\hat{A})$ has a power series in $\hat{A}$. This gives,
\begin{align}
\label{ComId1}
	 \big[ S^{\frac{1}{2}}(\yhat), \xhat \big] = i \, \frac{k}{3} \, S^{-\frac{1}{2}}(\yhat)   \; .
\end{align}
We also have,
\begin{align}
	\big[ \xhat^2 \, \yhat + \xhat \, \yhat \, \xhat + \yhat \, \xhat^2, \xhat \big] 
	= {}& \xhat^2 \, \big[ \yhat, \xhat \big] + \xhat \big[  \yhat,\xhat \big] \xhat + \big[ \yhat, \xhat \big] \xhat^2   \nn  \\
	= {}& - 3 \, i \, \xhat^2 \, .                                                    
\end{align}
Substituting these into \eqref{xLienardHeiEOM1} and simplifying gives
\begin{align}
\label{xHeiEOMLie}
	\xhat' =  \frac{3}{k} \, S^{-\frac{1}{2}}(\yhat) - \frac{3}{k} \, \hat{1} - \frac{k}{3} \, \xhat^2 \; .
\end{align}
Turning now to the \hei\ equation of motion for $\yhat$, we have
\begin{align}
\label{yLienardHeiEOM1}
	\yhat' = i \, [\Hhat, \yhat] = \frac{i}{2} \, \big[\xhat^2, \yhat \big] - i \, \frac{k}{9} \, \big[ \xhat^2 \, \yhat + \xhat \, \yhat \, \xhat + \yhat \, \xhat^2, \yhat \big]   \; ,	          
\end{align}
where
\begin{align}
	\big[ \xhat^2 \, \yhat + \xhat \, \yhat \, \xhat + \yhat \, \xhat^2, \yhat \big] = \big[ \xhat^2, \yhat \big] \yhat + \big[ \xhat \, \yhat \, \xhat, \yhat \big] +  \yhat \big[ \xhat^2, \yhat \big]  .
\end{align}
The second commutator in the right-hand side may be simplified 
\begin{align}
	\big[ \xhat \, \yhat \, \xhat, \yhat \big] = {}& \xhat \, \yhat \, \big[ \xhat, \yhat \big] + \big[ \xhat, \yhat \big] \, \yhat \, \xhat  \nn \\
	                                                             = {}& i \, (\xhat \, \yhat + \yhat \, \xhat)  \, .
\end{align}
Then noting that $[\xhat^2,\yhat]=2\,i\,\xhat$, we have
\begin{align}
	\big[ \xhat^2 \, \yhat + \xhat \, \yhat \, \xhat + \yhat \, \xhat^2, \yhat \big] = 3 \, i \, (\xhat \, \yhat + \yhat \, \xhat)  \; .
\end{align}
Equation \eqref{yLienardHeiEOM1} then becomes, 
\begin{align}
	\yhat' = - \xhat + \frac{2k}{6} \, (\xhat \, \yhat + \yhat \, \xhat)   \; .
\end{align}
The second-order equation for $\xhat$ is now much more difficult to derive because of the ambiguity in operator ordering. Simply differentiating \eqref{xHeiEOMLie} with respect to time and applying the chain rule to $(\hat{1}-2k\yhat/3)^{-1/2}$ produces a factor of $\yhat'$, which does not commute with $\yhat$. However, we can still compute $\xhat''$ by treating $\xhat'$ as a time-dependent operator in the \hei\ equation of motion. Using \eqref{HunsualLienardApp} and \eqref{xHeiEOMLie} we get, 
\begin{align}
	\xhat'' = {}& i \big[ \Hhat, \xhat' \big]    \\
	           = {}& i \, \frac{3}{k} \, \big[ S^{\frac{1}{2}}(\yhat) , \xhat^2 \big] + i \, \frac{3}{2k} \, \big[\xhat^2 , S^{-\frac{1}{2}}(\yhat) \big]  \nn \\
	                 & - \frac{i}{3} \, \big[ \xhat^2 \, \yhat + \xhat\,\yhat\,\xhat + \yhat\,\xhat^2, S^{-\frac{1}{2}}(\yhat) \big]   \nn \\
	                 & + i \, \frac{k^2}{27} \, \big[ \xhat^2 \, \yhat + \xhat\,\yhat\,\xhat + \yhat\,\xhat^2, \xhat^2 \big] + i \, \big[ \yhat, \xhat^2 \big]     \\
\label{x''HeiEOM1}	                 
	           = {}& i \, \frac{3}{k} \, \Big\{ \xhat \, \big[ S^{\frac{1}{2}}(\yhat) , \xhat \big] + \big[ S^{\frac{1}{2}}(\yhat) , \xhat \big] \, \xhat  \Big\}  \nn \\
	                 &  + i \, \frac{3}{2k} \, \Big\{ \xhat \, \big[ \xhat, S^{-\frac{1}{2}}(\yhat) \big] + \big[ \xhat, S^{-\frac{1}{2}}(\yhat) \big] \, \xhat \Big\}   \nn \\
	                 & - \frac{i}{3} \, \xhat \, \big[ \xhat, S^{-\frac{1}{2}}(\yhat) \big] \, \yhat - \frac{i}{3} \, \big[ \xhat, S^{-\frac{1}{2}}(\yhat) \big] \, \xhat \, \yhat    \nn \\
	                 & - \frac{i}{3} \, \xhat \, \yhat\, \big[\xhat, S^{-\frac{1}{2}}(\yhat) \big] - \frac{i}{3} \, \big[\xhat, S^{-\frac{1}{2}}(\yhat) \big] \yhat \, \xhat   \nn \\
	                 & - \frac{i}{3} \,\yhat \,  \xhat \, \big[ \xhat, S^{-\frac{1}{2}}(\yhat) \big]  - \frac{i}{3} \, \yhat \, \big[ \xhat, S^{-\frac{1}{2}}(\yhat) \big] \, \xhat   \nn \\
                        & + i \, \frac{k^2}{27} \,  \big( \xhat^2 \, \big[\yhat, \xhat^2 \big] + \xhat \, \big[ \yhat, \xhat^2 \big] \, \xhat + \big[ \yhat, \xhat^2 \big] \, \xhat^2 \big)   \nn \\
                  & + i \, \big[ \yhat, \xhat^2 \big]  \; .  
\end{align}
As before, the commutators involving the square root of an operator may be simplified using \eqref{[F(A),B]}. This time we also need 
\begin{align}
\label{ComId2}
	 \big[ S^{-\frac{1}{2}}(\yhat), \xhat \big] = - \, i \, \frac{k}{3} \, S^{-\frac{3}{2}}(\yhat)   \; .
\end{align}
Then using \eqref{ComId1} and \eqref{ComId2} in \eqref{x''HeiEOM1}, we have
\begin{align}
\label{x''HeiEOM2}
	\xhat'' = {}& - \,  \xhat \, S^{-\frac{1}{2}}(\yhat) - S^{-\frac{1}{2}}(\yhat) \, \xhat   \\
	                 & - \frac{1}{2} \,  \xhat \, S^{-\frac{3}{2}}(\yhat) - \frac{1}{2} \, S^{-\frac{3}{2}}(\yhat) \, \xhat    \nn \\
	                 & + \frac{2k}{9} \, \xhat \, \yhat \, S^{-\frac{3}{2}}(\yhat) + \frac{2k}{9} \, S^{-\frac{3}{2}}(\yhat) \, \yhat \, \xhat  \nn \\
	                 & + \frac{k}{9} \, S^{-\frac{3}{2}}(\yhat) \, \xhat \, \yhat + \frac{k}{9} \,\yhat \,  \xhat \, S^{-\frac{3}{2}}(\yhat)  
	                    + 2 \, \xhat + \frac{2k^2}{9} \, \xhat^3   \; .  \nn
\end{align}
Now using $[\xhat,\yhat]=i\,\hat{1}$, we have, for any $F(\yhat)$,
\begin{align}
\label{OpId}
	\yhat \, \xhat \, F(\yhat) + F(\yhat) \, \xhat \, \yhat = \xhat \, \yhat \, F(\yhat) + F(\yhat) \, \yhat \, \xhat   \; .
\end{align}
Applying \eqref{OpId} to \eqref{x''HeiEOM2} we get
\begin{align}
	\xhat'' = {}& - \,  \xhat \, S^{-\frac{1}{2}}(\yhat) - S^{-\frac{1}{2}}(\yhat) \, \xhat      \\
	                 & - \frac{1}{2} \,  \xhat \, S^{-\frac{3}{2}}(\yhat) - \frac{1}{2} \, S^{-\frac{3}{2}}(\yhat) \, \xhat       \nn \\
	                 & + \frac{k}{3} \, \xhat \, \yhat\, S^{-\frac{3}{2}}(\yhat) + \frac{k}{3} \, S^{-\frac{3}{2}}(\yhat) \, \yhat \, \xhat   
	                    + 2 \, \xhat + \frac{2k^2}{9} \, \xhat^3  \; .  \nn
\end{align}
Now factorizing terms containing $S^{-3/2}(\yhat)$ and simplifying
\begin{align}
\label{x''HeiEOM3}
	\xhat'' = {}& - \,  \xhat \, S^{-\frac{1}{2}}(\yhat) - S^{-\frac{1}{2}}(\yhat) \, \xhat     \nn \\
	                 & - \frac{1}{2} \, \xhat \, S(\yhat) \, S^{-\frac{3}{2}}(\yhat) - \frac{1}{2} \, S^{-\frac{3}{2}}(\yhat) \, S(\yhat) \, \xhat  \nn \\
	                 & + 2 \, \xhat + \frac{2k^2}{9} \, \xhat^3   \\
		    = {}& - \frac{3}{2} \, \xhat \, S^{-\frac{1}{2}}(\yhat) - \frac{3}{2} \, S^{-\frac{1}{2}}(\yhat) \, \xhat + 2 \, \xhat + \frac{2k^2}{9} \, \xhat^3   \; .
\end{align}
We can now use \eqref{xHeiEOMLie} to write 
\begin{align}
	S^{-\frac{1}{2}}(\yhat) = \frac{k}{3} \,\xhat' + \hat{1} + \frac{k^2}{9} \, \xhat^2     \; .
\end{align}
Substituting this into \eqref{x''HeiEOM3},
\begin{align}
	\xhat'' = {}& - \frac{3}{2} \,  \xhat \, \bigg( \frac{k}{3} \,\xhat' + \hat{1} + \frac{k^2}{9} \, \xhat^2 \bigg)  \nn \\
	                 & - \frac{3}{2} \, \bigg( \frac{k}{3} \,\xhat' + \hat{1} + \frac{k^2}{9} \, \xhat^2 \bigg) \, \xhat + 2 \, \xhat + \frac{2k^2}{9} \, \xhat^3    \\
		    = {}&  - \frac{k}{2} \, \big( \xhat \, \xhat' + \xhat' \xhat \big) - 3 \xhat - \frac{k^2}{3} \, \xhat^3 + 2 \xhat + \frac{2k^2}{9} \, \xhat^3    \\
\label{x''HeiEOM4}	
		    = {}&  - \frac{k}{2} \, \big( \xhat \, \xhat' + \xhat' \xhat \big) - \frac{k^2}{9} \, \xhat^3 - \xhat   \; .	    
\end{align}
Equation \eqref{x''HeiEOM4} can now be seen as the quantum analog of \eqref{x''HamEOM}, with the product $\xhat \xhat'$ symmetrized. This is in contrast to the variational formulation of the \vdp\ oscillator where the nonlinear damping term in \eqref{Sec2vdPQ} did not have to be symmetrized because $\xhat'$ was given by the momentum operator of an ancillary oscillator.


\begin{widetext}
\section{Constructive proof of cascade quantization}
\label{ConsProof}

For ease of reference, let us recall from Sec.~\ref{QuantStrategy} that our strategy to quantize a system given by $\alpha'=h(\alpha,\alphastar)$, where $h(\alpha,\alphastar)$ is a polynomial of degree $m$, is to first write it as a sum of homogeneous polynomials of degree $n$, 
\begin{align}
	h(\alpha,\alphastar) = \sum_{n=0}^m \, h_n(\alpha,\alphastar)  \; .
\end{align}
Thus, to quantize a general polynomial system, we only need to quantize an arbitrary homogeneous polynomial. And to achieve this we futher decompose $h_n(\alpha,\alphastar)$ as
\begin{align}
\label{hn(a,a*)SumApp}
	h_n(\alpha,\alphastar) = \lambda_n \, \alphastar{}^n + (1-\delta_{n,0}) \sum_{k=0}^{K(n)} \, h_{n,k}(\alpha,\alphastar) \; ,
\end{align}
where for $n \ge 1$ we have defined, 
\begin{align}
\label{hnk(a,a*)App}
	h_{n,k}(\alpha,\alphastar) = \mu_{n,k} \, \alphastar{}^k \alpha^{n-k} + \nu_{n,k} \, \alphastar{}^{n-k-1} \alpha^{k+1}  \; ,
\end{align}
with $\lambda_n$, $\mu_{n,k}$, and $\nu_{n,k}$ being arbitrary complex coefficients. We have also defined the top limit of the sum as
\begin{align}	
\label{K(n)DefnApp}
	 K(n) = \begin{cases}  \frac{n}{2} - 1  \; ,  \quad    \text{$n$ even}    \\   
	                                   \frac{n-1}{2}  \; ,  \quad   \text{$n$ odd}         \end{cases}  .
\end{align}

We prove constructively in this Appendix that there is always a Lindbladian $\Lin_{n,k}$, i.e.~for all values of $n$ and $k$, such that it quantizes, 
\begin{align}
\label{<a>=In+NotinApp}
	\alpha' = h_{n,k}(\alpha,\alphastar) + h^{\notin}_{n,k}(\alpha,\alphastar)    \; ,
\end{align}
where $h^{\notin}_{n,k}(\alpha,\alphastar)$ are byproduct terms not in $h_{n,k}(\alpha,\alphastar)$, but have lower degree than $h_{n,k}(\alpha,\alphastar)$. Therefore we can always construct another Lindbladian $\Lnot_{n,k}$ such that it quantizes
\begin{align}
	\alpha' = - \, h^{\notin}_{n,k}(\alpha,\alphastar)     \; .
\end{align}
The Lindbladian that quantizes $h_{n,k}(\alpha,\alphastar)$ is then given by 
\begin{align}
\label{L=Lin+LnotApp}
	\Lcal_{n,k} = \Lin_{n,k} + \Lnot_{n,k}     \; . 
\end{align}

\subsection{$\Lin_{n,k}$ for even $n\ge2$}
\label{nEvenApp}

To make our proof easier to read, we will suppress the subscripts $n$ and $k$ throughout most of our detailed calculations. The dependence on $n$ and $k$ will be restored at the end of the subsection, when we come to summarize our results. To begin with, let us assume a Lindbladian of the form 
\begin{align}
\label{SimpleL}
	\Lin = - i \, [\Hhat,\supopdot] + \Dcal[\hat{c}]   \; .
\end{align}
It is then simple show that the evolution of $\an{\ann}$, as generated by \eqref{SimpleL}, can be written as
\begin{align}
\label{General<a>}
	\an{\ann}' = i \, \an{\,[\Hhat,\ann]\,} + \frac{1}{2} \, \an{\,[\hat{c}\dg, \ann] \, \hat{c}\,} + \frac{1}{2} \, \an{\,\hat{c}\dg [\ann, \hat{c}]\,}    \; .
\end{align}
We then seek an $\Hhat$ and a $\hat{c}$ such that \eqref{General<a>} evaluates to 
\begin{align}
\label{a'=hbar_nk}
	\an{\ann}' = \mu \, \an{\adg{}^k \, \ann^{n-k}} + \nu \, \an{\adg{}^{n-k-1} \, \ann^{k+1}}  + \an{\,\normord{h^{\notin}(\ann,\adg)}\,}   \; ,
\end{align}
where $h^{\notin}(\alpha,\alphastar)$ will be determined by our construction of $\Lin$. To this end, we make the ansatz that $\Hhat$ can be put in the form
\begin{align}
\label{EvenHamiltonian}
	\Hhat = \chi \, \adg{}^{k+1} \ann^{n-k} + \chi^* \, \adg{}^{n-k} \ann^{k+1}  \; ,
\end{align}
while the Lindblad operator $\hat{c}$ can be written as
\begin{align}
\label{EvenLindbladOpc}
	\hat{c} = \adg{}^{\frac{n}{2}-k-1} \, \ann^{k+1} + \sigma \, \ann^{\frac{n}{2}+1}   \; .
\end{align}
Here $\chi$ and $\sigma$ are constants which depend on $\mu$ and $\nu$, and whose forms are to be determined. We now use \eqref{EvenHamiltonian} and \eqref{EvenLindbladOpc} to evaluate each term in \eqref{General<a>}. First we have
\begin{align}
	i \, [\Hhat,\ann] = {}& i \, \chi \, [\adg{}^{k+1}\ann^{n-k}, \ann] + i \, \chi^* \, [\adg{}^{n-k}\ann^{k+1}, \ann]   \\
	                     = {}& i \, \chi \, [\adg{}^{k+1}, \ann] \, \ann^{n-k} + i \, \chi^* \, [\adg{}^{n-k}, \ann] \, \ann^{k+1}   \\
\label{i[H,a]}                    
                      = {}& - i \, \chi \, (k+1) \, \adg{}^k \, \ann^{n-k} - i \, \chi^* \, (n-k) \, \adg{}^{n-k-1} \, \ann^{k+1}   \; .
\end{align}
Now using \eqref{EvenLindbladOpc},
\begin{align}
	[\hat{c}\dg,\ann] = [\adg{}^{k+1}, \ann] \, \ann^{\frac{n}{2}-k-1} + \sigma^* \, [\adg{}^{\frac{n}{2}+1}, \ann]   
                                  = -(k+1) \, \adg{}^k \, \ann^{\frac{n}{2}-k-1} - \sigma^* \, \bigg(\frac{n}{2}+1\bigg) \, \adg{}^{\frac{n}{2}}  \; .
\end{align}
Multiplying this on the right by $\hat{c}$ we get
\begin{align}
	[\hat{c}\dg,\ann] \, \hat{c} = {}& - \bigg[ (k+1) \, \adg{}^k \, \ann^{\frac{n}{2}-k-1} + \sigma^* \, \bigg(\frac{n}{2}+1\bigg) \, \adg{}^{\frac{n}{2}} \bigg]  
	                                                    \big( \adg{}^{\frac{n}{2}-k-1} \, \ann^{k+1} + \sigma \, \ann^{\frac{n}{2}+1}  \big)  \\
\label{[cdg,a]c1}	
						        = {}& - (k+1) \, \adg{}^k \, \ann^{\frac{n}{2}-k-1} \, \adg{}^{\frac{n}{2}-k-1} \, \ann^{k+1}  
						                 - \sigma \, (k+1) \, \adg{}^k \, \ann^{\frac{n}{2}-k-1} \, \ann^{\frac{n}{2}+1}   \nn  \\
	                                                & - \sigma^* \, \bigg(\frac{n}{2}+1\bigg) \, \adg{}^{\frac{n}{2}} \, \adg{}^{\frac{n}{2}-k-1} \, \ann^{k+1}  
	                                                   - |\sigma|^2 \, \bigg(\frac{n}{2}+1\bigg) \, \adg{}^{\frac{n}{2}} \, \ann^{\frac{n}{2}+1}    \; .
\end{align}
Every term in \eqref{[cdg,a]c1} except the first is already in normal order. We may reorder $\ann^{n/2-k-1} \adg{}^{n/2-k-1}$ by using the identity \cite{SH15,BPS03}
\begin{align}
\label{OrderingId}
	\ann^r \adg{}^s = \sum_{p=0}^{{\rm min}\{r,s\}} \, \tbo{r}{p} \, \frac{s!}{(s-p)!} \, \adg{}^{s-p} \, \ann^{r-p}   \; ,
\end{align}
where we have used the standard notation for binomial coefficients, defined by
\begin{align}
	\tbo{r}{p} = \frac{r!}{p! \, (r-p)!}   \; .
\end{align}
Applying this to \eqref{[cdg,a]c1} we thus obtain
\begin{align}
\label{[cdg,a]c2}
	[\hat{c}\dg,\ann] \, \hat{c} = {}& - (k+1) \,   \adg{}^k \left[ \sum_{p=0}^{\frac{n}{2}-k-1}
	                                                    \tbo{\frac{n}{2}-k-1}{p} \frac{\big[(n/2) - k - 1\big]!}{\big[(n/2) - k - 1 - p\big]!} \, \adg{}^{\frac{n}{2}-k-1-p} \,  \ann^{\frac{n}{2}-k-1-p} \right] \ann^{k+1}   \nn  \\
	                                                 & - \sigma \, (k+1) \, \adg{}^k \, \ann^{n-k} - \sigma^* \, \bigg( \frac{n}{2}+1 \bigg) \, \adg{}^{n-k-1} \, \ann^{k+1}  
                                                           - |\sigma|^2 \, \bigg(\frac{n}{2}+1\bigg) \, \adg{}^{\frac{n}{2}} \, \ann^{\frac{n}{2}+1}    \; .
\end{align}
Note the first line of \eqref{[cdg,a]c2} is a polynomial in $\ann$ and $\adg$ of degree $n-1$. Thus on multiplying \eqref{[cdg,a]c2} by one half, it helps to write the resulting expression as,
\begin{align}
\label{[cdg,a]c3}
	\frac{1}{2} \, [\hat{c}\dg,\ann] \, \hat{c} = - \frac{\sigma}{2} \, (k+1) \, \adg{}^{k} \, \ann^{n-k} - \frac{\sigma^*}{2} \, \bigg( \frac{n}{2}+1 \bigg) \, \adg{}^{n-k-1} \, \ann^{k+1}  
	                                                                  - \frac{|\sigma|^2}{2} \, \bigg( \frac{n}{2}+1 \bigg) \, \adg{}^{\frac{n}{2}} \, \ann^{\frac{n}{2}+1} + \, \normord{R^{\notin,1}(\ann,\adg)}   \; ,                                          
\end{align}
where we have defined
\begin{align}
\label{Rhat(1)Even}
	R^{\notin,1}(\alpha,\alphastar) = - \frac{k+1}{2} \, \sum_{p=0}^{\frac{n}{2}-k-1} \tbo{\frac{n}{2}-k-1}{p} \frac{\big[(n/2) - k - 1\big]!}{\big[(n/2) - k - 1 - p\big]!} \; \alphastar{}^{\frac{n}{2}-p-1} \,  \alpha^{\frac{n}{2}-p}  \; .
\end{align}
Since $R^{\notin,1}(\alpha,\alphastar)$ has degree $n-1$, we may assume $-R^{\notin,1}(\alpha,\alphastar)$ to be quantized. However, the last term in \eqref{[cdg,a]c3} has degree $n+1$ so its Lindbladian will need to be found explicitly. To do so, consider the Lindblad operator and coefficient 
\begin{align}
\label{EvenLindbladOpb}
	\hat{b} = \adg{}^{\frac{n}{2}+1}  \; ,   \quad      \kappa = |\sigma|^2   \; .
\end{align}
This gives
\begin{align}
	\frac{\kappa}{2} \, [\hat{b}\dg,\ann]\,\hat{b} + \frac{\kappa}{2} \, \hat{b}\dg [\ann, \hat{b}] = \frac{|\sigma|^2}{2} \, \ann^{\frac{n}{2}+1} \, [\ann, \adg{}^{\frac{n}{2}+1}]  
\label{EvenDeg<a>2}	        
	         = \frac{|\sigma|^2}{2} \bigg( \frac{n}{2} + 1 \bigg) \, \ann^{\frac{n}{2}+1} \, \adg{}^{\frac{n}{2}}   \; .
\end{align}
We can again use \eqref{OrderingId} to reorder $\ann^{n/2+1} \, \adg{}^{n/2}$, giving,
\begin{align}
    \frac{\kappa}{2} \, [\hat{b}\dg,\ann]\,\hat{b} + \frac{\kappa}{2} \, \hat{b}\dg [\ann, \hat{b}]
	   = {}& \frac{|\sigma|^2}{2} \bigg( \frac{n}{2} + 1 \bigg) \, \sum_{p=0}^{\frac{n}{2}} \tbo{\frac{n}{2}+1}{p} \frac{(n/2)!}{[(n/2)-p]!} \, \adg{}^{\frac{n}{2}-p} \, \ann^{\frac{n}{2}+1-p}   \\
\label{EvenDeg<a>3}
	   = {}& \frac{|\sigma|^2}{2} \bigg( \frac{n}{2} + 1 \bigg) \, \adg{}^{\frac{n}{2}} \, \ann^{\frac{n}{2}+1} + \normord{R^{\notin,2}(\ann,\adg)}   \; ,
\end{align}
where we have picked out the $p=0$ term in the sum, and defined another polynomial in $\alpha$ and $\alphastar$ of degree $n-1$,
\begin{align}
\label{Rhat(3)Even}
	R^{\notin,2}(\alpha,\alphastar) = \frac{|\sigma|^2}{2} \, \bigg( \frac{n}{2} + 1 \bigg) \, \sum_{p=1}^{\frac{n}{2}} \, \tbo{\frac{n}{2}+1}{p} \,  \frac{(n/2)!}{[(n/2)-p]!} \, \alphastar{}^{\frac{n}{2}-p} \, \alpha^{\frac{n}{2}+1-p}   \; .
\end{align}
As $R^{\notin,2}(\alpha,\alphastar)$ is also of degree $n-1$, we are free to assume $-R^{\notin,2}(\alpha,\alphastar)$ to be quantized.

The last term in \eqref{General<a>} remains to be evaluated. Again using \eqref{EvenLindbladOpc}, we find
\begin{align}
	[\ann,\hat{c}\,] = [\ann, \adg{}^{\frac{n}{2}-k-1} ] \, \ann^{k+1}   
	                         = \bigg( \frac{n}{2}-k-1 \bigg) \, \adg{}^{\frac{n}{2}-k-2} \, \ann^{k+1}   \; .
\end{align}
Now multiplying this on the left by $\hat{c}\dg$,
\begin{align}
	\hat{c}\dg [\ann, \hat{c}] = {}& \big( \adg{}^{k+1} \, \ann^{\frac{n}{2}-k-1} + \sigma^* \, \adg{}^{\frac{n}{2}+1}  \big) \bigg( \frac{n}{2}-k-1 \bigg) \, \adg{}^{\frac{n}{2}-k-2} \, \ann^{k+1}   \\
\label{cdg[a,c]1}
							= {}& \bigg( \frac{n}{2}-k-1 \bigg) \adg{}^{k+1} \, \ann^{\frac{n}{2}-k-1} \, \, \adg{}^{\frac{n}{2}-k-2} \, \ann^{k+1}  
							         + \sigma^* \, \bigg( \frac{n}{2}-k-1 \bigg) \, \adg{}^{n-k-1} \, \ann^{k+1}   \; .
\end{align}
Using \eqref{OrderingId} on the first term of \eqref{cdg[a,c]1},
\begin{align}
	\hat{c}\dg \,[\ann,\hat{c}] = {}& \bigg( \frac{n}{2}-k-1 \bigg) \adg{}^{k+1}
                                                          \left[ \sum_{p=0}^{\frac{n}{2}-k} \tbo{\frac{n}{2}-k-1}{p} \frac{\big[(n/2) - k -1\big]!}{\big[(n/2) - k - 2- p\big]!} \, \adg{}^{\frac{n}{2}-k-2-p} \,  \ann^{\frac{n}{2}-k-1-p} \right] \ann^{k+1}   \nn  \\
	                                                & + \sigma^* \, \bigg( \frac{n}{2}-k-1 \bigg) \, \adg{}^{n-k-1} \, \ann^{k+1}   \; .
\end{align}
The first line here has degree $n-1$. Hence we write, on multiplying across by one half,
\begin{align}
\label{cdg[a,c]2}
	\frac{1}{2} \, \hat{c}\dg [\ann,\hat{c}] = \frac{\sigma^*}{2} \, \bigg( \frac{n}{2}-k-1 \bigg) \, \adg{}^{n-k-1} \, \ann^{k+1} + \, \normord{R^{\notin,3}(\ann,\adg)}  \; ,
\end{align}
where we have again defined the terms of degree $n-1$ by
\begin{align}
\label{Rhat(2)Even}
	R^{\notin,3}(\alpha,\alphastar) 
	= \frac{1}{2} \, \bigg( \frac{n}{2}-k-1 \bigg) \sum_{p=0}^{\frac{n}{2}-k} \tbo{\frac{n}{2}-k-1}{p} \frac{\big[(n/2) - k -1\big]!}{\big[(n/2) - k - 2- p\big]!} \, \alphastar{}^{\frac{n}{2}-1-p} \,  \alpha^{\frac{n}{2}-p}  \; .
\end{align}

We thus have, on using \eqref{i[H,a]}, \eqref{[cdg,a]c3}, \eqref{Rhat(1)Even},  \eqref{cdg[a,c]2}, \eqref{Rhat(2)Even}, \eqref{EvenDeg<a>3}, and \eqref{Rhat(3)Even}, 
\begin{align}
	\an{\ann}' = {}& i \, \an{[\Hhat, \ann]} + \frac{1}{2} \, \an{[\hat{c}\dg,\ann] \, \hat{c}} + \frac{1}{2} \, \an{\hat{c}\dg \, [\ann,\hat{c}]}  
	                         +  \frac{\kappa}{2} \, \an{[\hat{b}\dg,\ann]\,\hat{b}} + \frac{\kappa}{2} \, \an{\hat{b}\dg [\ann, \hat{b}]}   \\
\label{a'=hbar_nkDerived}
		  = {}& \bigg( -i \, \chi - \frac{\sigma}{2} \bigg) \, (k+1) \, \an{\adg{}^k \, \ann^{n-k}} + \bigg[ -i \, \chi^* \, (n-k) - \frac{\sigma^*}{2} \, (k+2) \bigg] \an{\adg{}^{n-k-1} \, \ann^{k+1}} 
		           + \an{\,\normord{h^{\notin}(\ann,\adg)}\,}    
\end{align}
where 
\begin{align}
\label{EvenNotinh}
	 h^{\notin}(\alpha,\alphastar) = R^{\notin,1}(\alpha,\alphastar) + R^{\notin,2}(\alpha,\alphastar) + R^{\notin,3}(\alpha,\alphastar)   \; .
\end{align}
Note that \eqref{EvenNotinh} now defines $\Lnot$, which quantizes $-h^{\notin}(\alpha,\alphastar)$, and which always exists. It now remains for us to determine a closed form for $\Lin$. To do this, we simply match \eqref{a'=hbar_nkDerived} to \eqref{a'=hbar_nk}. Recall for convenience that  \eqref{a'=hbar_nk} is given by
\begin{align}
\label{a'=hbar_nkRecall}
	\an{\ann}' = \mu \, \an{\adg{}^k \, \ann^{n-k}} + \nu \, \an{\adg{}^{n-k-1} \, \ann^{k+1}} + \an{\,\normord{h^{\notin}(\ann,\adg)}\,}   \; .
\end{align}
The appropriate $\chi$ and $\sigma$ that will generate \eqref{a'=hbar_nkRecall} can thus be found by equating coefficients, giving,
\begin{align}
	 \mu = - \bigg( i \, \chi + \frac{\sigma}{2} \bigg) (k+1) \; ,   \quad   \nu = - i \, \chi^* \, (n-k) - \frac{\sigma^*}{2} \, (k+2)  \; .
\end{align}
Solving these gives
\begin{align}
\label{chiSolution}
	\chi = i \, \frac{\big[ (k+2) \mu - (k+1) \nu^* \big]}{(n+2)(k+1)}   \; ,   \quad    \sigma = - \frac{2\,\big[ (n-k) \mu + (k+1) \nu^* \big]}{(n+2)(k+1)}  \; .
\end{align}
Equation \eqref{chiSolution} now determines the ansatz equations \eqref{EvenHamiltonian}, \eqref{EvenLindbladOpc}, and \eqref{EvenLindbladOpb} completely.

Let us now summarize our findings with the indices written out explicitly. The system
\begin{align}
	\alpha' = h_{n,k}(\alpha,\alphastar)   \; ,    \quad n=2,4,6,\ldots,  \quad   k=0,1,\ldots, K  \; ,
\end{align}
where $K=n/2-1$, is quantized exactly by 
\begin{align}
	\Lcal_{n,k} = \Lin_{n,k} + \Lnot_{n,k}   \; .
\end{align}
The Lindbladian $\Lin_{n,k}$ is given by
\begin{align}
	\Lin_{n,k} = - \, i \, [\Hhat_{n,k}\,, \supopdot\,] + \Dcal[\hat{c}_{n,k}] + \kappa_{n,k} \, \Dcal[\hat{b}_{n,k}]  \; .
\end{align}
From \eqref{EvenHamiltonian} and \eqref{chiSolution} we have
\begin{align}
\label{HnkEven}
	\Hhat_{n,k} =  \chi_{n,k} \, \adg{}^{k+1} \ann^{n-k} + \chi^*_{n,k}\, \adg{}^{n-k} \ann^{k+1} \; ,   \quad \chi_{n,k} = i \, \frac{\big[ (k+2) \mu_{n,k} - (k+1) \nu^*_{n,k} \big]}{(n+2)(k+1)}  \; ,
\end{align}
while from \eqref{EvenLindbladOpc}, \eqref{EvenLindbladOpb}, and \eqref{chiSolution},
\begin{gather}
\label{cnkEven}
	\hat{c}_{n,k} = \adg{}^{\frac{n}{2}-k-1} \, \ann^{k+1} + \sigma_{n,k} \, \ann^{\frac{n}{2}+1}  \; ,   \quad  \sigma_{n,k} = -\frac{2\,\big[ (n-k) \mu_{n,k} + (k+1) \nu^*_{n,k} \big]}{(n+2)(k+1)} \; ,   \\ 
\label{bnkEven}
	\hat{b}_{n,k} = \adg{}^{\frac{n}{2}+1} \; ,   \quad   \kappa_{n,k} = |  \sigma_{n,k} |^2  \; .
\end{gather} 
The Lindbladian $\Lnot_{n,k}$ quantizes $-h^{\notin}_{n,k}(\alpha,\alphastar)$ and is thus prescribed by the form of $h^{\notin}_{n,k}(\alpha,\alphastar)$. Substituting \eqref{Rhat(1)Even}, \eqref{Rhat(3)Even}, and \eqref{Rhat(2)Even} into \eqref{EvenNotinh} we get
\begin{align}
\label{rTermsEven}
	h^{\notin}_{n,k}(\alpha,\alphastar) 
	= {}& - \frac{k+1}{2} \, \sum_{p=0}^{\frac{n}{2}-k-1} \tbo{\frac{n}{2}-k-1}{p} \frac{\big[(n/2) - k - 1\big]!}{\big[(n/2) - k - 1 - p\big]!} \; \alpha^*{}^{\frac{n}{2}-p-1} \,  \alpha^{\frac{n}{2}-p}   \nn \\
	      & + \frac{|\sigma_{n,k}|^2}{2} \, \bigg( \frac{n}{2} + 1 \bigg) \, \sum_{p=1}^{\frac{n}{2}} \, \tbo{\frac{n}{2}+1}{p} \,  \frac{(n/2)!}{[(n/2)-p]!} \, \alpha^*{}^{\frac{n}{2}-p} \, \alpha^{\frac{n}{2}+1-p}   \nn \\
	      & + \frac{1}{2} \, \bigg( \frac{n}{2}-k-1 \bigg) \sum_{p=0}^{\frac{n}{2}-k} \tbo{\frac{n}{2}-k-1}{p} \frac{\big[(n/2) - k -1\big]!}{\big[(n/2) - k - 2- p\big]!} \, \alpha^*{}^{\frac{n}{2}-1-p} \,  \alpha^{\frac{n}{2}-p}   \; .	      
\end{align}

\subsection{$\Lin_{n,k}$ for odd $n\ge3$}

Finding a valid quantization for odd $n$ follows the same steps as even $n$. We first consider the case of $h_{n,k}(\alpha,\alpha^*)$ for $k=0,1,\ldots,K-1$ in Appendix~\ref{nOddk<KApp}, followed by $h_{n,K}(\alpha,\alpha^*)$ in Appendix~\ref{nOddk=KApp}.

\subsubsection{$0 \le k < K$}
\label{nOddk<KApp}

We again start by assuming
\begin{align}
	\Lin = - i \, [\Hhat,\supopdot] + \Dcal[\hat{c}]   \; ,
\end{align}
and proceed to find an $\Hhat$ and a $\hat{c}$ such that
\begin{align}
\label{General<a>Odd}
	\an{\ann}' = i \, \an{ [\Hhat,\ann] } + \frac{1}{2} \, \an{[\hat{c}\dg, \ann] \, \hat{c}} + \frac{1}{2} \, \an{\hat{c}\dg [\ann, \hat{c}]}   
\end{align}
generates
\begin{align}
\label{Odd<a>'}
	\an{\ann}' = \mu \, \an{\adg{}^k \, \ann^{n-k}} + \nu \, \an{\adg{}^{n-k-1} \, \ann^{k+1}} + \an{\,\normord{h^{\notin}(\ann,\adg)}\,}   \; .
\end{align}
This can be accomplished using similar ansatz as before. In particular, $\Hhat$ still has the form
\begin{align}
\label{OddHamiltonian}
	\Hhat = \chi \, \adg{}^{k+1} \ann^{n-k} + \chi^* \, \adg{}^{n-k} \ann^{k+1}  \; ,
\end{align}
but now the Lindblad operator $\hat{c}$ is
\begin{align}
\label{OddLindbladOp}
	\hat{c} = \adg{}^{\frac{n-1}{2}-k} \, \ann^{k+1} + \sigma \, \ann^{\frac{n+1}{2}}   \; .
\end{align}
Since there is no change in $\Hhat$ we have again, from \eqref{i[H,a]},
\begin{align}
\label{Oddi[H,a]}
	i \, [\Hhat,\ann] = {}& - i \, \chi \, (k+1) \, \adg{}^k \, \ann^{n-k}  - i \, \chi^* \, (n-k) \, \adg{}^{n-k-1} \, \ann^{k+1}   \; .
\end{align}
Now using \eqref{OddLindbladOp},
\begin{align}
	[\hat{c}\dg,\ann] = {}& [\adg{}^{k+1}, \ann] \, \ann^{\frac{n-1}{2}-k} - \sigma^* \, [\adg{}^{\frac{n+1}{2}}, \ann]    \\
                                   = {}& -(k+1) \, \adg{}^k \, \ann^{\frac{n-1}{2}-k} - \sigma^* \, \bigg(\frac{n+1}{2}\bigg) \, \adg{}^{\frac{n-1}{2}}  \; .
\end{align}
Multiplying this on the right by $\hat{c}$ we get
\begin{align}
\label{Odd[cdg,a]c1}
	[\hat{c}\dg,\ann] \, \hat{c} = {}& -\bigg[ (k+1) \, \adg{}^k \, \ann^{\frac{n-1}{2}-k} + \sigma^* \, \bigg(\frac{n+1}{2}\bigg) \, \adg{}^{\frac{n-1}{2}} \bigg] \big( \adg{}^{\frac{n-1}{2}-k} \, \ann^{k+1} + \sigma \, \ann^{\frac{n+1}{2}} \big)   \\[0.25cm]
	                                           = {}& - (k+1) \, \adg{}^k \, \ann^{\frac{n-1}{2}-k} \, \adg{}^{\frac{n-1}{2}-k} \, \ann^{k+1}  - \sigma \, (k+1) \, \adg{}^k \, \ann^{\frac{n-1}{2}-k} \, \ann^{\frac{n+1}{2}}   \nn  \\
	                                                 & - \frac{\sigma^*}{2} \, (n+1) \, \adg{}^{\frac{n-1}{2}} \, \adg{}^{\frac{n-1}{2}-k} \, \ann^{k+1}  - \frac{|\sigma|^2}{2} \, (n+1) \, \adg{}^{\frac{n-1}{2}} \, \ann^{\frac{n+1}{2}}    \; .
\end{align}
Every term except the first is already in normal order. This can be reordered again using \eqref{OrderingId} which we recall here for ease of reference,
\begin{align}
\label{OrderingIdRecall}
	\ann^r \adg{}^s = \sum_{p=0}^{{\rm min}\{r,s\}} \, \tbo{r}{p} \, \frac{s!}{(s-p)!} \, \adg{}^{s-p} \, \ann^{r-p}   \; ,    \quad      \tbo{r}{p} = \frac{r!}{p! \, (r-p)!}   \; .
\end{align}
Using the identity \eqref{OrderingIdRecall},
\begin{align}
\label{Odd[cdg,a]c2}
	[\hat{c}\dg,\ann] \, \hat{c} = {}& - (k+1) \, \adg{}^k \left[ \sum_{p=0}^{\frac{n-1}{2}-k}
	                                                    \tbo{\frac{n-1}{2}-k}{p} \frac{\big[(n-1)/2 - k\big]!}{\big[(n-1)/2 - k - p\big]!} \, \adg{}^{\frac{n-1}{2}-k-p} \,  \ann^{\frac{n-1}{2}-k-p} \right] \ann^{k+1}   \nn  \\
	                                                 & - \sigma \, (k+1) \, \adg{}^k \, \ann^{\frac{n-1}{2}-k} \, \ann^{\frac{n+1}{2}} - \frac{\sigma^*}{2} \, (n+1) \, \adg{}^{\frac{n-1}{2}} \, \adg{}^{\frac{n-1}{2}-k} \, \ann^{k+1}  
                                                           - \frac{|\sigma|^2}{2} \, (n+1) \, \adg{}^{\frac{n-1}{2}} \, \ann^{\frac{n+1}{2}}    \; .
\end{align}
The first line of \eqref{Odd[cdg,a]c2} comprises of terms with degree $n-2p$, while the second line consists of only terms of degree $n$. Hence the $p=0$ term in the sum in the first line of \eqref{Odd[cdg,a]c2} has degree $n$, while every other term in the sum are of degree $n-2$ or less. We thus write,
\begin{align}
\label{Odd[cdg,a]c3}
	\frac{1}{2} \, [\hat{c}\dg,\ann] \, \hat{c} = {}& - \frac{k+1}{2} \, \adg{}^{\frac{n-1}{2}} \, \ann^{\frac{n+1}{2}}  - \frac{\sigma}{2} \, (k+1) \, \adg{}^k \, \ann^{n-k} 
	                                                                        - \frac{\sigma^*}{4} \, (n+1) \, \adg{}^{n-k-1} \, \ann^{k+1}  - \frac{|\sigma|^2}{4} \, (n+1) \, \adg{}^{\frac{n-1}{2}} \, \ann^{\frac{n+1}{2}}  \nn \\
	                                                                    & + \, \normord{R^{\notin,1}(\ann,\adg)}   \; ,                                          
\end{align}
where we have defined,
\begin{align}        
\label{Rhat(1)Odd}
	R^{\notin,1}(\alpha,\alphastar) = - \frac{k+1}{2} \sum_{p=1}^{\frac{n-1}{2}-k}  \tbo{\frac{n-1}{2}-k}{p} \frac{\big[(n-1)/2 - k\big]!}{\big[(n-1)/2 - k - p\big]!} \, \alphastar{}^{\frac{n-1}{2}-p} \,  \alpha^{\frac{n+1}{2}-p}   \; .                                          
\end{align}
Note that $R^{\notin,1}(\alpha,\alphastar)$ has only odd-degree terms [i.e.~the sum of powers of $\alpha$ and $\alphastar$ in \eqref{Rhat(1)Odd} is always odd].

Now applying the same procedure to $\hat{c}\dg [\ann, \hat{c}]$ gives, 
\begin{align}
	[\ann, \hat{c}] = [\ann, \adg{}^{\frac{n-1}{2}-k}]  \, \ann^{k+1} = \bigg( \frac{n-1}{2} - k \bigg) \adg{}^{\frac{n-1}{2}-k-1}  \, \ann^{k+1}  \; .
\end{align}
Multiplying on the left by $\hat{c}\dg$,
\begin{align}
	\hat{c}\dg [\ann, \hat{c}] = {}& \big( \adg{}^{k+1} \, \ann^{\frac{n-1}{2}-k} + \sigma^* \, \adg{}^{\frac{n+1}{2}} \big)  \bigg( \frac{n-1}{2} - k \bigg) \adg{}^{\frac{n-1}{2}-k-1}  \, \ann^{k+1}  \\
\label{Oddcdg[a,c]1}	
	                                         = {}& \bigg( \frac{n-1}{2}-k \bigg) \adg{}^{k+1} \, \ann^{\frac{n-1}{2}-k} \, \adg{}^{\frac{n-3}{2}-k} \, \ann^{k+1} + \sigma^* \, \bigg( \frac{n-1}{2}-k \bigg) \, \adg{}^{n-k-1} \, \ann^{k+1}   \; .  
\end{align}
Normally ordering the first term in \eqref{Oddcdg[a,c]1} using \eqref{OrderingIdRecall} then gives,
\begin{align}
	\hat{c}\dg [\ann, \hat{c}] = {}& \bigg( \frac{n-1}{2}-k \bigg) \adg{}^{k+1}
                                                         \left[ \sum_{p=0}^{\frac{n-3}{2}-k} \tbo{\frac{n-1}{2}-k}{p} \frac{\big[(n-3)/2 - k\big]!}{\big[(n-3)/2 - k - p\big]!} \, \adg{}^{\frac{n-3}{2}-k-p} \,  \ann^{\frac{n-1}{2}-k-p} \right] \ann^{k+1}   \nn  \\
	                                               & + \sigma^* \, \bigg( \frac{n-1}{2}-k \bigg) \, \adg{}^{n-k-1} \, \ann^{k+1}   \; .
\end{align}
As with \eqref{Odd[cdg,a]c3}, we will group terms of degree $n-2$ into one operator, writing
\begin{align}
\label{Oddcdg[a,c]2}
	\frac{1}{2} \, \hat{c}\dg [\ann, \hat{c}] = \frac{1}{2} \bigg( \frac{n-1}{2}-k \bigg) \adg{}^{\frac{n-1}{2}} \, \ann^{\frac{n+1}{2}} + \frac{\sigma^*}{2} \, \bigg( \frac{n-1}{2}-k \bigg) \, \adg{}^{n-k-1} \, \ann^{k+1} 
	                                                                + \, \normord{R^{\notin,2}(\ann,\adg)}   \; ,
\end{align}
where $R^{\notin,2}(\alpha,\alphastar)$ has only odd-degree terms, defined by
\begin{align}
\label{Rhat(2)Odd}
	R^{\notin,2}(\alpha,\alphastar) 
	= \frac{1}{2} \bigg( \frac{n-1}{2}-k \bigg) \sum_{p=1}^{\frac{n-3}{2}-k} \tbo{\frac{n-1}{2}-k}{p} \frac{\big[(n-3)/2 - k\big]!}{\big[(n-3)/2 - k - p\big]!} \, \alphastar{}^{\frac{n-1}{2}-p} \,  \alpha^{\frac{n+1}{2}-p}  \; .
\end{align}
Using \eqref{Oddi[H,a]}, \eqref{Odd[cdg,a]c3}, and \eqref{Oddcdg[a,c]2} in \eqref{General<a>Odd} we thus have
\begin{align}
	i \, [\Hhat,\ann] + \frac{1}{2} \, [\hat{c}\dg, \ann] \, \hat{c} + \frac{1}{2} \, \hat{c}\dg [\ann, \hat{c}]   
\label{OddDeg<a>1}
		  = {}& - \bigg( i \, \chi + \frac{\sigma}{2} \bigg) (k+1) \, \adg{}^k \, \ann^{n-k}  + \bigg[ \! - i \, \chi^* \, (n-k) + \sigma^* \, \frac{(k+1)}{2} \bigg] \, \adg{}^{n-k-1} \, \ann^{k+1}    \nn \\
		        & + \bigg[ \frac{n-3}{4} - |\sigma|^2\frac{(n+1)}{4} - k \bigg] \adg{}^{\frac{n-1}{2}} \, \ann^{\frac{n+1}{2}} + \, \normord{R^{\notin,1}(\ann,\adg) + R^{\notin,2}(\ann,\adg)}   \; .
\end{align}
Note the first term of the second line in \eqref{OddDeg<a>1} is an unwanted term since it has the wrong powers of $\ann$ and $\adg$ [compared to \eqref{Odd<a>'}]. We thus need to cancel this term, depending on the sign of
\begin{align}
\label{zetaDefn}
	\zeta= \frac{(n-3)-(n+1) |\sigma|^2}{4} - k  \; .
\end{align}
If $\zeta>0$ we may consider the Lindblad operator $\hat{b}=\ann^{(n+1)/2}$, with a coefficient $\kappa$ to be determined:
\begin{align}
\label{EvenDeg<a>1b}
	\frac{\kappa}{2} \; [\hat{b}\dg,\ann]\,\hat{b} + \frac{\kappa}{2} \; \hat{b}\dg \, [\ann, \hat{b}]    
	= \frac{\kappa}{2} \; [\adg{}^{\frac{n+1}{2}},\ann] \, \ann^{\frac{n+1}{2}} = - \, \kappa \; \frac{(n+1)}{4} \, \adg{}^{\frac{n-1}{2}} \, \ann^{\frac{n+1}{2}}   \; .
\end{align}
Comparing this to the first term on the second line in \eqref{OddDeg<a>1} shows that we ought to choose
\begin{align}
\label{kappa+Odd}
	\kappa = \frac{4}{(n+1)} \,  \zeta   \; ,
\end{align}
If on the other hand, $\zeta<0$, then we may consider $\hat{b}=\adg{}^{(n+1)/2}$, for which
\begin{align}
	\frac{\kappa}{2} \; [\hat{b}\dg,\ann]\,\hat{b} + \frac{\kappa}{2} \; \hat{b}\dg \, [\ann, \hat{b}] = {}&  \frac{\kappa}{2} \; \ann^{\frac{n+1}{2}} \, [\ann, \adg{}^{\frac{n+1}{2}}]   	\\        
	         = {}& \kappa \; \frac{(n+1)}{4} \, \ann^{\frac{n+1}{2}} \, \adg{}^{\frac{n-1}{2}}  
\label{OddDeg<a>3}
	         = \kappa \; \frac{(n+1)}{4} \, \adg{}^{\frac{n-1}{2}} \, \ann^{\frac{n+1}{2}} + \, \normord{\hat{R}^{\notin,3}(\ann,\adg)}    \; .
\end{align}
Note that we have used \eqref{OrderingIdRecall} to normally order \eqref{OddDeg<a>3}, and introduced 
\begin{align}
\label{Rhat(3)Odd}
	R^{\notin,3}(\alpha,\alphastar) = \kappa \, \frac{(n+1)}{4} \, \sum_{p=1}^{\frac{n-1}{2}} \tbo{\frac{n+1}{2}}{p} \frac{\big[(n-1)/2\big]!}{\big[(n-1)/2-p\big]!} \, \alphastar{}^{\frac{n-1}{2}-p} \, \alpha^{\frac{n+1}{2}-p}   \; .
\end{align}
Our choice of $\kappa$ in this case is then the negative of \eqref{kappa+Odd}:
\begin{align}
\label{kappa-Odd}
	\kappa = -\frac{4}{(n+1)} \,  \zeta  \; .
\end{align}
Using \eqref{OddDeg<a>1} and whichever choice of $\hat{b}$ above depending on the sign of $\zeta$, we get
\begin{align}
	\an{\ann}' = {}& i \, \an{[\Hhat, \ann]} + \frac{1}{2} \, \an{[\hat{c}\dg,\ann] \, \hat{c}} + \frac{1}{2} \, \an{\hat{c}\dg \, [\ann,\hat{c}]}  
	                         + \frac{\kappa}{2} \, \an{[\hat{b}\dg,\ann]\,\hat{b}} + \frac{\kappa}{2} \, \an{\hat{b}\dg [\ann, \hat{b}]}   \\
		          = {}& - \bigg( i \, \chi + \frac{\sigma}{2} \bigg) (k+1) \, \an{\adg{}^k \, \ann^{n-k}}  
                      - \bigg[  i \, \chi^* \, (n-k) - \sigma^* \, \frac{(k+1)}{2} \bigg] \, \an{\adg{}^{n-k-1} \, \ann^{k+1}} 
                      + \an{\,\normord{h^{\notin}(\ann,\adg)}\,}  \; , 
\end{align}
where 
\begin{align}
\label{OddNotinhk<K}
	h^{\notin}(\alpha,\alphastar) = R^{\notin,1}(\alpha,\alphastar) + R^{\notin,2}(\alpha,\alphastar) + \theta(-\zeta) \, R^{\notin,3}(\alpha,\alphastar)
\end{align}
is now a polynomial of degree $n-2$. Note that we have used the Heaviside step function for $R^{\notin,3}(\alpha,\alphastar)$ as it is only nonzero if $\zeta<0$. Recall from the main text in \eqref{HeavisideDefn} that the step function is defined by
\begin{align}
\label{HeavisideDefnRecall3}
	\theta(x) = \begin{cases} 0 , \; x \le 0   \\
	                                         1 , \; x > 1 \end{cases} . 
\end{align}
Equation \eqref{OddNotinhk<K} now defines $\Lnot$, which quantizes $-h^{\notin}(\alpha,\alphastar)$, and which can always be constructed. Since we are trying to replicate
\begin{align}
	\an{\ann}'= \mu \, \an{\adg{}^k \, \ann^{n-k}} + \nu \, \an{\adg{}^{n-k-1} \, \ann^{k+1}}  \; ,
\end{align}
we require
\begin{align}
	 \mu = - \bigg( i \, \chi + \frac{\sigma}{2} \bigg) (k+1)  \; ,   \quad        \nu = - i \, \chi^* \, (n-k) + \sigma^* \, \frac{(k+1)}{2}  \; .
\end{align}
Solving these equations simultaneously for $\chi$ and $\sigma$ gives
\begin{align}
\label{chiOddDeg}
	\chi = \frac{i}{n+1} \, (\mu - \nu^*)   \; ,   \quad      \sigma = -\frac{2\big[ (n-k) \mu + (k+1) \nu^* \big]}{(n+1)(k+1)}  \; .
\end{align}
Substituting \eqref{chiOddDeg} back into \eqref{OddHamiltonian} and \eqref{OddLindbladOp} then gives us the desired Hamiltonian and Lindblad operator.

Let us now summarize our results with the indices written out explicitly. The system
\begin{align}
	\alpha' = h_{n,k}(\alpha,\alphastar)   \; ,    \quad n=3,5,7,\ldots,  \quad   k=0,1,\ldots, K-1  \; ,
\end{align}
where $K=(n-1)/2$, is quantized exactly by 
\begin{align}
	\Lcal_{n,k} = \Lin_{n,k} + \Lnot_{n,k}   \; .
\end{align}
We found $\Lin_{n,k}$ to have the form 
\begin{align}
	\Lin_{n,k} = - \, i \, [\Hhat_{n,k}\,, \supopdot\,] + \Dcal[\hat{c}_{n,k}] + \kappa^-_{n,k} \, \Dcal[\hat{b}^-_{n,k}] + \kappa^+_{n,k} \, \Dcal[\hat{b}^+_{n,k}]   \; ,
\end{align}
for which \eqref{OddHamiltonian} and \eqref{chiOddDeg} give
\begin{align}
\label{HnkOdd}
	\Hhat_{n,k} = \chi_{n,k} \, \adg{}^{k+1} \, \ann^{n-k} + \chi^*_{n,k} \, \adg{}^{n-k} \, \ann^{k+1}   \; ,   \quad    \chi_{n,k} = i \, \frac{( \mu_{n,k} - \nu_{n,k}^* )}{n+1}   \; ,
\end{align}
and which \eqref{OddLindbladOp}, \eqref{chiOddDeg}, give
\begin{align}
\label{cnkOddApp}
	\hat{c}_{n,k} = \adg{}^{\frac{n-1}{2}-k} \, \ann^{k+1} + \sigma_{n,k} \, \ann^{\frac{n+1}{2}}    \; ,   \quad      \sigma_{n,k} = -\frac{2\,\big[ (n-k) \mu_{n,k} + (k+1) \nu^*_{n,k} \big]}{(n+1)(k+1)}   \; .
\end{align}
We also found from \eqref{zetaDefn}--\eqref{kappa-Odd}, 
\begin{gather}	
\label{bnk-OddApp}
	\hat{b}^-_{n,k} = \adg{}^{\frac{n+1}{2}} \; ,   \quad   \kappa^-_{n,k} = - \frac{4}{n+1}  \, \theta(-\zeta_{n,k}) \, \zeta_{n,k}    \; ,    \\
\label{bnk+OddApp}
	\hat{b}^+_{n,k} = \ann^{\frac{n+1}{2}} \; ,   \quad   \kappa^+_{n,k} = \frac{4}{n+1}  \, \theta(\zeta_{n,k}) \, \zeta_{n,k}    \; ,
\end{gather} 
where 
\begin{align}
\label{sigmaOddApp} 
	\zeta_{n,k} = {}& \frac{(n-3)-(n+1) |\sigma_{n,k}|^2}{4} - k \; .
\end{align}
The form of $\Lnot_{n,k}$ on the other hand, is determined by $h^{\notin}_{n,k}(\alpha,\alphastar)$, as it quantizes $-h^{\notin}_{n,k}(\alpha,\alphastar)$. We can now provide the explicit form of $h^{\notin}_{n,k}(\alpha,\alphastar)$ by substituting \eqref{Rhat(1)Odd}, \eqref{Rhat(2)Odd}, and \eqref{Rhat(3)Odd} in \eqref{OddNotinhk<K}. This gives
\begin{align}
\label{rTermsOdd}
	h^{\notin}_{n,k}(\alpha,\alpha^*) 
	= {}& - \frac{k+1}{2} \sum_{p=1}^{\frac{n-1}{2}-k}  \tbo{\frac{n-1}{2}-k}{p} \frac{\big[(n-1)/2 - k\big]!}{\big[(n-1)/2 - k - p\big]!} \, \alpha^*{}^{\frac{n-1}{2}-p} \,  \alpha^{\frac{n+1}{2}-p}     \nn \\
	      & + \frac{1}{2} \; \theta\bigg( \frac{n-3}{2}-k \bigg) \bigg( \frac{n-1}{2}-k \bigg) 
	             \sum_{p=1}^{\frac{n-3}{2}-k} \tbo{\frac{n-1}{2}-k}{p} \frac{\big[(n-3)/2 - k\big]!}{\big[(n-3)/2 - k - p\big]!} \, \alpha^*{}^{\frac{n-1}{2}-p} \,  \alpha^{\frac{n+1}{2}-p}    \nn \\
	      & - \theta(-\zeta_{n,k}) \, \zeta_{n,k}  \sum_{p=1}^{\frac{n-1}{2}} \tbo{\frac{n+1}{2}}{p} \frac{\big[(n-1)/2\big]!}{\big[(n-1)/2-p\big]!} \, \alpha^*{}^{\frac{n-1}{2}-p} \, \alpha^{\frac{n+1}{2}-p}  \; .
\end{align}

\subsubsection{$k=K$}
\label{nOddk=KApp}

As before, we begin with the Lindbladian 
\begin{align}
	\Lin = - i \, [\Hhat,\supopdot] + \gamma \, \Dcal[\hat{c}]   \; ,
\end{align}
whose equation of motion for $\an{\ann}$ is
\begin{align}
\label{<a>'Oddk=K}
	\an{\ann}' = i \, \an{ [\Hhat,\ann] } + \frac{\gamma}{2} \, \an{[\hat{c}\dg, \ann] \, \hat{c}} + \frac{\gamma}{2} \, \an{\hat{c}\dg [\ann, \hat{c}]}   \; .
\end{align}
Our goal is then to find an $\Hhat$ and a $\hat{c}$ so that
\begin{align}
	\an{\ann}' = \epsilon \, \an{\adg{}^{\frac{n-1}{2}} \, \ann^{\frac{n+1}{2}}} + \an{\,\normord{h^{\notin}(\ann,\adg)}\,}   
\end{align}
is reproduced for an arbitrary complex constant $\epsilon$. Writing $\epsilon$ in terms of its real and imaginary parts we get
\begin{align}
\label{a'=hbarnK(a,adg)}
	\an{\ann}' = \Re[\epsilon] \, \an{\adg{}^{\frac{n-1}{2}} \, \ann^{\frac{n+1}{2}}} + i \, \Im[\epsilon] \, \an{\adg{}^{\frac{n-1}{2}} \, \ann^{\frac{n+1}{2}}}  \; .
\end{align}
We first try to generate the second term in \eqref{a'=hbarnK(a,adg)} by using the Hamiltonian 
\begin{align}
\label{HnKApp}
	\Hhat = -\frac{2\,\Im[\epsilon]}{n+1} \, \adg{}^{\frac{n+1}{2}} \, \ann^{\frac{n+1}{2}}    \; .
\end{align}
This gives
\begin{align}
	i \, [\Hhat, \ann] = {}& - i \, \frac{2\,\Im[\epsilon]}{n+1} \, [\adg{}^{\frac{n+1}{2}},\ann] \, \ann^{\frac{n+1}{2}} = i \, \Im[\epsilon] \, \adg{}^{\frac{n-1}{2}} \, \ann^{\frac{n+1}{2}}    \; .
\end{align}
To generate the first term in \eqref{a'=hbarnK(a,adg)} we note that the appropriate Lindblad operator and coefficient depends on the sign of $\Re[\epsilon]$. If $\Re[\epsilon]<0$, we can use
\begin{align}
\label{cnK-App}
	\hat{c} = \ann^{\frac{n+1}{2}} \; ,  \quad   \gamma = -\,\frac{4\,\Re[\epsilon]}{n+1}   \; ,
\end{align}
which produces 
\begin{align}
	\frac{\gamma}{2} \; [\hat{c}\dg,\ann]\,\hat{c} + \frac{\gamma}{2} \; \hat{c}\dg \, [\ann, \hat{c}]  = -\frac{2\,\Re[\epsilon]}{n+1} \; [\adg{}^{\frac{n+1}{2}},\ann] \, \ann^{\frac{n+1}{2}}    
	= \Re[\epsilon] \, \adg{}^{\frac{n-1}{2}} \, \ann^{\frac{n+1}{2}}   \; .
\end{align}
If instead $\Re[\epsilon]>0$, then we should use the Lindblad operator and coefficient
\begin{align}
\label{cnK+App}
	\hat{c} = \adg{}^{\frac{n+1}{2}}  \; ,   \quad  \gamma = \frac{4\,\Re[\epsilon]}{n+1}   \; .
\end{align}
This gives
\begin{align}
	\frac{\gamma}{2} \; [\hat{c}\dg,\ann]\,\hat{c} + \frac{\gamma}{2} \; \hat{c}\dg \, [\ann, \hat{c}]  = {}& \frac{2\,\Re[\epsilon]}{n+1}  \; \ann^{\frac{n+1}{2}} \, [\ann, \adg{}^{\frac{n+1}{2}}]    \\
	= {}& \Re[\epsilon] \, \ann^{\frac{n+1}{2}} \, \adg{}^{\frac{n-1}{2}} = \Re[\epsilon] \, \adg{}^{\frac{n-1}{2}} \, \ann^{\frac{n+1}{2}} + \, \normord{R^{\notin}(\ann,\adg)}   \; ,
\end{align}
where we have used \eqref{OrderingIdRecall} and defined
\begin{align}
\label{Rhat(4)Even}
	R^{\notin}(\alpha,\alphastar) 
	=  \Re[\epsilon] \, \sum_{p=1}^{\frac{n-1}{2}} \tbo{\frac{n+1}{2}}{p} \frac{\big[(n-1)/2\big]!}{\big[(n-1)/2-p\big]!} \, \alphastar{}^{\frac{n-1}{2}-p} \, \alpha^{\frac{n+1}{2}-p}   \; ,
\end{align}
which is a polynomial of degree $n-2$.

In summary, we are able to quantize exactly
\begin{align}
\label{a'Odd2}
	\alpha' = h_{n,K}(\alpha,\alphastar) = \epsilon_n \, \alphastar{}^{\frac{n-1}{2}}  \alpha^{\frac{n+1}{2}}  \;,   \quad   K=\frac{n-1}{2}  \; ,  \quad    n=3,5,7,\dots  \; ,
\end{align}
by the Lindbladian
\begin{align}
	\Lcal_{n,K} = \Lin_{n,K} + \Lnot_{n,K}  \; .
\end{align}
We found $\Lin_{n,K}$ to be given by
\begin{align}
	\Lin_{n,K} = -i \, [\Hhat_{n,K}\,, \supopdot\,] + \gamma^-_{n,K} \, \Dcal[\hat{c}^-_{n,K}] + \gamma^+_{n,K} \, \Dcal[\hat{c}^+_{n,K}]   \; ,
\end{align}
where the Hamiltonian has the form given in \eqref{HnKApp},
\begin{align}
	\Hhat_{n,K} = -\frac{2\,\Im[\epsilon_{n,K}]}{n+1} \, \adg{}^{\frac{n+1}{2}} \, \ann^{\frac{n+1}{2}}   \; ,
\end{align}
while from \eqref{cnK-App} and \eqref{cnK+App}, we have
\begin{gather}
	\hat{c}^-_{n,K} = \ann^{\frac{n+1}{2}} \; ,   \quad   \gamma^-_{n,K} = -\,\frac{4\,\Re[\epsilon_{n}]}{n+1}   \, \theta(-\Re[\epsilon_{n}])    \; ,    \\
	\hat{c}^+_{n,K} = \adg{}^{\frac{n+1}{2}} \; ,   \quad   \gamma^+_{n,K} = \frac{4\,\Re[\epsilon_{n}]}{n+1}   \, \theta(\Re[\epsilon_{n}])    \; .
\end{gather} 
Finally, $\Lnot_{n,K}$ is defined by the unwanted terms captured by \eqref{Rhat(4)Even}. As these terms only exist for $\Re[\epsilon_n]>0$, we can use the step function to define
\begin{align}
	h^{\notin}_{n,K}(\alpha,\alphastar) 
	= \theta(\Re[\epsilon_n]) \, \Re[\epsilon_n] \, \sum_{p=1}^{\frac{n-1}{2}} \tbo{\frac{n+1}{2}}{p} \frac{\big[(n-1)/2\big]!}{\big[(n-1)/2-p\big]!} \, \alphastar{}^{\frac{n-1}{2}-p} \, \alpha^{\frac{n+1}{2}-p}   \; .
\end{align}
This now conditions the form of $\Lnot_{n,K}$, which quantizes $-h^{\notin}_{n,K}(\alpha,\alphastar)$. 

\end{widetext}


\end{document}